\documentclass[a4paper,fleqn,usenatbib,english]{mnras}
\usepackage[T1]{fontenc}
\usepackage{ae,aecompl}
\usepackage{amssymb}
\usepackage{amsmath}
\usepackage{babel}
\usepackage{graphicx}
\usepackage{amsmath}
\usepackage{amssymb}
\usepackage{subfig}
\usepackage{rotating}
\usepackage{longtable}
\usepackage{bigints}

\makeatletter

\newcommand{\sn}[2]{\ensuremath{#1 \times 10^{#2}}}

\newcommand{\rom}[2]{\ensuremath{#1_{\textrm{#2}}}}

\newcommand{\ebv}{\ensuremath{E(B\,\textrm{--}\,V)}}
\newcommand{\um}{$\mu$m}
\newcommand{\ium}{$\mu$m$^{-1}$}

\DeclareMathAlphabet\mathbfcal{OMS}{cmsy}{b}{n}

\title[Photo-z's, Fuzzy Templates, and SOMs. II.]{Deriving Photometric Redshifts using Fuzzy Templates and Self-Organizing Maps. II. Comparing Sampling Techniques Using Mock Data}

\author[Speagle and Eisenstein]{
	Joshua S. Speagle$^{1,2}$\thanks{E-mail: jspeagle@cfa.harvard.edu}
	and Daniel J. Eisenstein$^{1}$
	\\
	$^{1}$Harvard University Department of Astronomy, 60 Garden St., MS 46, Cambridge, MA 02138, USA\\
	$^{2}$Kavli IPMU (WPI), UTIAS, The University of Tokyo, Kashiwanoha 5-1-5, Kashiwa, Chiba, Japan\\
}

\date{Accepted XXX. Received YYY; in original form ZZZ}

\pubyear{2015}

\begin{document}
\label{firstpage}
\pagerange{\pageref{firstpage}--\pageref{lastpage}}
\maketitle
	
\begin{abstract}
In a companion paper (Speagle \& Eisenstein 2015a), we proposed combining large numbers of ``fuzzy archetypes'' with Self-Organizing Maps (SOMs) to derive photometric redshifts in a data-driven way. In this paper, we investigate the performance of several sampling approaches that build on this general idea using a mock catalog designed to approximately simulate LSST ($ugrizY$) and \textit{Euclid} ($YJH$) data from $z=0$\,--\,$6$ at fixed LSST $Y=24$\,mag. We test eight different approaches: two brute-force methods, two Markov Chain Monte Carlo (MCMC)-based methods, two hierarchical sampling methods, and two ``quick-search'' methods based on quantities derived during the initial SOM training process. We find most methods perform reasonably well with small catastrophic outlier fractions and are able to robustly identify redshift probability distribution functions that are multi-modal and/or poorly constrained. Once these insecure objects are removed, the results are generally in good agreement with the strict accuracy requirements necessary to meet \textit{Euclid} weak lensing goals for most redshifts above $z \sim 0.8$. These results demonstrate the utility of our data clustering-based approach and highlight its effectiveness to derive quick and accurate photo-z's using large numbers of templates.
\end{abstract}
	



\section{Introduction}
\label{sec:intro}

Future large-scale surveys such as \textit{Euclid} \citep{laureijs+11} and the Large Synoptic Survey Telescope \citep{ivezic+08} will rely on accurate redshifts ($z$) measured to a significant portion of observed galaxies in order to measure the dark energy equation-of-state to target goals of a few percent accuracy using weak lensing \citep{albrecht+06,bordoloi+12,newman+15}. Due to the challenge of obtaining spectroscopic redshifts (spec-z's) to the majority galaxies in these samples, many of these studies will heavily rely on accurate photometric redshifts (photo-z's) derived from galaxy broad- and/or narrow-band spectral energy distributions (SEDs).

Two main techniques are used to derive photo-z's and the associated redshift probability distribution function (PDF) $P(z|\rom{\mathbf{F}}{obs})$ for a given set of observed fluxes $\rom{\mathbf{F}}{obs}$. Template fitting approaches \citep[e.g.,][]{benitez00,bolzonella+00,feldmann+06,ilbert+06,cool+13,johnson+13,speagle+15} rely on deriving a set of forward mappings and their associated likelihoods from a collection of templates and associated model parameters to observed color space. Machine learning approaches \citep[e.g.,][]{collisterlahav04,abazajian+09,liyee08,gerdes+10,carrascokindbrunner14,bonnett15,hoyle15,elliott+15,sadeh+15,almosallam+15}, on the other hand, use training data to derive the best inverse mapping from observed color space to redshift.\footnote{See \citet{hildebrandt+10}, \citet{dahlen+13}, and \citet{sanchez+14} for recent overviews and comparisons, and \citet{menard+13} as well as related work for new applications of clustering-based redshifts (which will not be discussed here).}

The majority of photo-z developments in the last few years has been dominated by the introduction of numerous machine learning-based approaches. These are attractive relative to template fitting methods for several reasons:
\begin{enumerate}
	\item \textit{Single parameter estimation}: While template-fitting approaches are physically motivated, they involve fitting a number of extraneous parameters that must be marginalized over afterwards to derive $P(z|\rom{\mathbf{F}}{obs})$. Machine learning methods by design tend to focus on single-parameter estimation, making them somewhat more attractive.
	\item \textit{Computational speed}: Most template codes either resort to brute-force sampling \citep{ilbert+06,cool+13} or Markov Chain Monte Carlo (MCMC) techniques \citep{johnson+13,speagle+15} run over a large pre-generated grid of model photometry to derive $P(z|\rom{\mathbf{F}}{obs})$, both of which tend to take significantly more time than the majority of machine learning architectures.
	\item \textit{Empirically driven}: Most template fitting codes suffer from serious systematic uncertainties involved with modeling the impact of dust attenuation and emission line variation on an often limited set of templates \citep[e.g.,][]{coleman+80,kinney+96,polletta+07} along with the need for good relative photometric calibrations. Machine learning methods are not subject to these uncertainties.\footnote{If photo-z's are used to fit stellar population synthesis models at a later period, however, then good relative photometric calibrations still are required to avoid introducing biases.}
\end{enumerate}

While machine learning approaches have been shown to be effective within the bounds of their training sets \citep{sanchez+14}, attempts to extend their predictions beyond them have only been partially successful \citep{hoyle+15}. Due to the limited regions of color space spanned by spec-z's training sets today \citep{masters+15} and the difficulty of obtaining reliable spectra for higher redshift sources, it is likely that template-based photo-z's will continue to serve a crucial role in upcoming imaging surveys.

This paper is the latest in a series that attempt to make template fitting-based approaches more competitive in order to keep pace with the rapid development in machine learning applications.
In \citet{speagle+15}, we investigated the general features of typical pre-generated model ``grids'' and were able to develop a simulated annealing and ensemble MCMC \citep{goodmanweare10,foremanmackey+13} driven approach to exploring those grids more efficiently in an attempt to improve computational speed and efficiency. However, this did not affect the way that these grids of model photometry were created or improve on the modeling assumptions involved.

In an attempt to fix this glaring drawback, we introduced a new framework in Speagle \& Eisenstein (2015a) (henceforth Paper I) that allows an algorithm to explore a large number of ``fuzzy archetypes'' using clustering methods such as Self Organizing Maps (SOMs). This approach significantly reduces systematics from dust and emission-line modeling and allows for an adaptive exploration of the space in a computationally efficient manner.

In this paper, we investigate the performance of several extensions of this new hybrid approach against a mock LSST and \textit{Euclid} catalog. As the relevant likelihood surface is challenging to search, with many potential local minima that are often widely-separated \citep{speagle+15}, we limit our focus in this paper to the case the same model parameters has been used to generate both the templates and observed galaxies. This enables us to better investigate the relative performance of each algorithm and avoid issues relating to template-mismatch systematics.

The paper is organized as follows. In \S\ref{sec:formalism}, we briefly summarize the basics of template-fitting approaches and the framework outlined in Paper I. In \S\ref{sec:methods}, we outline the eight basic methods we use to derive photo-z's using our fuzzy templates and SOMs. In \S\ref{sec:results}, we describe our mock catalog and examine the performance of each of our individual approaches. We conclude in \S\ref{sec:conc}.

Throughout this work, boldface font ($\mathbf{x}$, $\mathbf{\Theta}$) is used to represent vectors/matrices while normal font ($x$, $\Theta$) is used for singular variables. All observations are generated from 9 bands of data (LSST $ugrizY$ and \textit{Euclid} $YJH$) with error properties based on $5\sigma$ imaging depths of 26.5\,mag (24.5 in $u$) for LSST and 24\,mag for \textit{Euclid} along with an assumed calibration uncertainty of 0.01\,mag ($\approx$\,1\%).

\section{Formalism}
\label{sec:formalism}

\subsection{Fuzzy Templates}
\label{subsec:fuzzy}

Model photometric fluxes $\rom{\mathbf{F}}{model}$ for a given series of templates are generated via
\begin{equation}\label{eq:phot}
\rom{\mathbf{F}}{model}=\frac{\int_{\nu_z}  S_{\nu,\textrm{model}}(\nu) R_{\textrm{model}}(\nu) \mathbf{T}(\nu)\nu^{-1}d\nu}{\int_{\nu_z} \mathbf{T}(\nu) \nu^{-1} d\nu},
\end{equation}
where the model template
\begin{equation}
S_{\nu,\textrm{model}}(\nu)=\sum_{\textrm{gal}} c_{\textrm{gal}} \times  S_{\nu,\textrm{gal}}(\nu) + \sum_{\textrm{lines}} \textrm{EW}_{\textrm{lines}} \times  S_{\nu,\textrm{lines}}(\nu)
\end{equation}
is a linear combination of ``basis'' galaxy \citep[see, e.g.,][]{coleman+80,kinney+96,polletta+07} and emission line templates \citep[e.g.,][]{kennicuttevans12} scaled by their equivalent widths (EWs), the dust (i.e. reddening) model
\begin{equation}
-2.5\log \left(R_{\textrm{model}}(\nu)\right) = \sum_{\textrm{dust}}\rom{\ebv}{dust}\rom{k}{dust}^\prime(\nu) + \rom{A}{IGM}(\nu,z)
\end{equation}
is the sum of a linear combination of dust templates for a given $\mathbf{E}(B-V)$ that are applied as a uniform screen, with the rest-frame dust screen $\rom{k}{dust}^\prime(\nu)=\rom{k}{dust}+c_{b,\textrm{dust}}\rom{k}{bump}$ composed of a combination of an underlying dust continuum and the 2175\,{\AA} bump \citep{fitzpatrickmassa07} and the IGM reddening template given by the parameterization outlined in \citet{madau95}. This combined template is convolved with the corresponding set of filter curves $\mathbf{T}(\nu)$ and over the redshifted set of frequencies $\nu_z$ probed by the respective filters.

As many template-fitting approaches only explore a single basis galaxy and dust template at a given time, we henceforth ignore linear combinations of those respective components. In addition, for convenience we will re-parameterize our full set of parameters into two vectors: $\boldsymbol{\theta}=\left\lbrace z,\textrm{gal},\textrm{dust} \right\rbrace$, which contains the parameters surrounding the baseline archetype, and $\boldsymbol{\phi}=\left\lbrace \ebv, c_b^\prime\equiv c_b \times \ebv, \boldsymbol{\Delta} \textrm{\textbf{EW}} \right\rbrace$, which contains the associated ``fuzzy'' parameters.

For computational reasons \citep[see][]{speagle+15}, most codes opt to use large \textit{pre-generated grids} of photometry from combinations of parameters sampled at a given granularity in each dimension during the fitting process. However, most of these grids are constructed using only a small collection of galaxy ($\lesssim$\,$30$) and dust ($\lesssim$\,$5$) templates taken from low-$z$ observations and then subsequently applied to the high-$z$ universe with large modifications.

Instead of starting with a small number of templates in well-understood regions of color space and trying to simulate less-understood regions of color space, we instead attempt to sample the color-space manifold occupied by galaxies directly using a much larger set of galaxy spectral ``archetypes''. In order to allow for possible variation away from our discrete set of starting archetypes, we generate a small level of intrinsic ``fuzziness'' using sets of individualized, independent priors on dust attenuation and emission line strengths that allows for small deviations away from each baseline archetype. This transforms an individual galaxy's position in color space into a multidimensional PDF centered at its original location. By keeping the perturbations relatively small, we remain in the regime where the uniform dust screen approximation is likely to remain valid while also allowing emission line strengths to vary independently of one another.


As Paper I, we use the set of 129 high-quality UV\,--\,IR spectra from \citet{brown+14} to serve as our baseline set of archetypes and measure the corresponding EWs of $\lbrace$H$\alpha$+N{\scriptsize[II]}, H$\beta$, H$\gamma$, O{\scriptsize[II]}, and the O{\scriptsize[III]} doublet to derive appropriate emission line templates. These are combined with a series of 8 dust curves of the form $k(x)=a_x(\beta)(x)^{-\beta}$+$b_x(\beta)+c_b\rom{k}{bump}(x|x_0,\gamma)$, where $\beta=\lbrace 0.3,0.45,0.6,0.75,0.9,1.05,1.2,1.35 \rbrace$, $a_x$ and $b_x$ are chosen such that each dust curve is normalized to $A(V)=3.0$ and $A(2.5\,\textrm{\um})=0$, and the mean amplitude ($c_b$), position ($x_0$), and width ($\gamma$) of the 2175\,{\AA} bump are taken from the mean fits presented in \citet{fitzpatrickmassa07}. We then choose $P(\boldsymbol{\phi})=P[\Delta\ebv] \times P[\Delta c_b^\prime] \times P[\boldsymbol{\Delta} \rom{\textrm{\textbf{EW}}}{line}]$ such that
\begin{eqnarray}
P[\Delta\ebv]=\mathcal{N}[0,\Delta A(1500\textrm{\,{\AA}})=0.05\,\textrm{mag}], \nonumber \\
P[\Delta c_b^\prime]= \mathcal{N}[0,1.5] \times P[\Delta\ebv],\,\,\textrm{and} \\
P[\Delta \rom{\textrm{EW}}{line}]=\mathcal{N}[0,0.2 \times \max(\textrm{EW},0.005\rom{\lambda}{line})], \nonumber
\end{eqnarray}
where $\mathcal{N}(\mu,\sigma)$ indicates a normal (i.e. Gaussian) distribution with mean $\mu$ and standard deviation $\sigma$. These form the underlying set of priors for our fuzzy archetypes. 

Our decision to fix $P[\Delta\ebv]$ to $A(1500\,{\AA})$ limits the amount of dust extinction in the UV to the approximately linear regime ($\sim$\,$5\%$ $1\sigma$ variation) across the entire wavelength range for each dust curve. Likewise, our choice of prior for $\Delta c_b^\prime$ allows fluctuations to $0$ to take place at the $2\sigma$ level, which enables us to fit dust curves without the impact of the 2175\,{\AA} bump while still disfavoring possible non-physical solutions. Finally, the 20\% variation ($\sim 0.1$\,dex) in EW allowed by  $P(\Delta \rom{\textrm{EW}}{line})$ mimics the observed scatter among emission line scaling relations \citep{kennicutt98,kennicuttevans12}, while the EW ``floor'' leaves open the possibility for small variation at the $\lesssim 1$\,--\,$2$\% level.


\subsection{Self-Organizing Maps}
\label{subsec:som}

The Self-Organizing Map \citep[SOM;][]{kohonen82,kohonen01} is an unsupervised machine learning algorithm that projects high-dimensional data onto a lower-dimensional space using competitive training of a large set of ``cells'' in a way that preserves general topological features and correlations present in the higher-dimensional data. Each cell in the grid is assigned a position $\rom{\mathbf{x}}{som}^{\textrm{cell}}$ and contains a cell model $\rom{\mathbf{F}}{som}^{\textrm{cell}}(t)$ that is the same size as a vector from the training data ($\rom{N}{filter}$) and initialized (most often randomly) at iteration $t=0$. Training then proceeds as follows:
\begin{enumerate}
\item Draw (with replacement) a random object $\rom{\mathbf{F}}{obs}^{i}$ from the input dataset.
\item Compute $\chi^2\left(\rom{\mathbf{F}}{obs}^{i},\rom{\mathbf{F}}{som}^{\textrm{cell}}(t),\rom{s}{cell}(t)\right)$ for every cell on the map.
\item Select the best-matching (i.e. lowest $\chi^2$) cell $\rom{\mathbf{x}}{som}^{\textrm{best}}$ from the corresponding set of $\chi^2$'s.
\item Update the cell models within the map according to an evolving \textit{learning rate} $\mathcal{A}(t)$ and \textit{neighborhood function} (i.e. kernel density) $\mathcal{H}(\rom{\mathbf{x}}{som}^{\textrm{best}},\rom{\mathbf{x}}{som}^{\textrm{cell}}|t)$ such that
\begingroup\makeatletter\def\f@size{8.5}\check@mathfonts
\begin{equation}
\rom{\mathbf{F}}{som}^{\textrm{cell}}(t+1)=\rom{s}{cell}(t)\rom{\mathbf{F}}{som}^{\textrm{cell}}(t)+\mathcal{A}(t)\mathcal{H}(t)\left[\rom{\mathbf{F}}{obs}^{i}-\rom{s}{cell}(t)\rom{\mathbf{F}}{som}^{\textrm{cell}}\right].
\end{equation}
\endgroup
\item Increment $t$ and repeat steps (i) through (iv) while $t<\rom{N}{iter}$.
\end{enumerate}
After training, objects are ``sorted'' onto the map by repeating steps (ii) and (iii) for every object in the input dataset and assigning them to the best-matching cell.

Following Paper I, we set
\begin{eqnarray}
\mathcal{A}(t)=a_0\left(\frac{a_1}{a_0}\right)^{t/\rom{N}{iter}}, \nonumber \\
\mathcal{H}(t|\rom{\mathbf{x}}{som}^{\textrm{best}},\rom{\mathbf{x}}{som}^{\textrm{cell}}) = \exp\left(-\frac{||\rom{\mathbf{x}}{som}^{\textrm{best}}-\rom{\mathbf{x}}{som}^{\textrm{cell}}||^2}{\sigma^2(t)}\right), \textrm{ and} \\
\sigma(t)=\sigma_0\left(\frac{1}{\sigma_0}\right)^{t/\rom{N}{iter}}, \nonumber
\end{eqnarray}
where $\mathcal{A}(t)$ and $\mathcal{H}(t)$ are now functions of the hyper-parameters $a_0$, $a_1$, and $\sigma_0$.

We use the SOM to organize both rest-frame and observed-frame colors using ``model catalogs'' generated from numerous Monte Carlo realizations of our fuzzy templates and a given $P(\boldsymbol{\theta})$. These are generated as follows:
\begin{enumerate}
	\item \textit{Rest-frame SOM } (2-D): We construct a model catalog of 9,675 objects using 75 Monte Carlo realizations from $P(\boldsymbol{\phi})$ and uniform sampling of $\rom{k}{dust}$ for each fuzzy archetype, which are assigned $1$\% errors to in order to ensure the SOM training proceeds in logarithmic rather than linear space. We construct a two-dimensional $30 \times 30$ SOM and initialize the cell models randomly on the uniform interval $[0,1]$. We set $[a_0,a_1,\sigma_0]=[1,0.5,30]$ and allow the map to train for $\rom{N}{iter}=10,000$ iterations. The corresponding set of mappings from the model catalog onto the SOM are then recorded.
	\item \textit{Observed-frame SOM} (3-D): We construct a model catalog of 3,228,225 objects using 25 Monte Carlo realizations from $P(\boldsymbol{\phi})$ with uniform sampling from $\rom{k}{dust}$ for each fuzzy archetype sampled on every point of an input $z=0$\,--\,$10$ ($\Delta z=0.01$) redshift grid. Each object is assigned $1$\% errors as well as a baseline error floor of $0.01$, which are added in quadrature. We construct a three-dimensional $30 \times 30 \times 30$ SOM and initialize the cell models according to pre-defined color gradients in $u-r$, $r-z$, and $z-H$. We set $[a_0,a_1,\sigma_0]=[0.5,0.2,10]$ and allow the map to train for $\rom{N}{iter}=1,000,000$ iterations. The corresponding set of mappings from the model catalog onto the SOM are then recorded.
\end{enumerate}



\section{Methods}
\label{sec:methods}


To test whether our use of (linearized) fuzzy templates and Self-Organizing Maps (SOMs) can be used to derive photo-z's, we decide to compare the following eight different methods:
\begin{enumerate}
	\item {\tt{BruteForce}}: A standard brute-force approach that involves comparing against all 3,228,225 models used to create the 3-D observed-frame SOM.
	\item {\tt{BruteForce\_LinearFuzzy}}: A modified brute-force approach that explores all $N_z \times \rom{N}{gal} \times \rom{N}{dust} = 1,033,032$ possible combinations of our non-linear parameters $\boldsymbol{\theta}$, where at each step we optimize over all linear parameters $\boldsymbol{\phi}$.
	\item {\tt{SOM\_MCMC\_RestFrame}}: An MCMC-based approach that explores a 2-D SOM constructed with rest-frame realizations of our fuzzy templates along with changes in photometry as a function of redshift. 
	\item {\tt{SOM\_MCMC\_ObservedFrame}}: An MCMC-based approach that explores a 3-D SOM constructed from \textit{observed-frame} realizations (i.e. redshift already included) of our fuzzy templates. 
	\item {\tt{SOM\_Hierarchical\_MonteCarlo}}: A hierarchical sampling approach that uses a fixed number of randomly sampled objects drawn from each SOM cell.
	\item {\tt{SOM\_Hierarchical\_ImportanceSampling}}: A hierarchical sampling approach that draws objects randomly from SOM cells based on an adaptive SOM importance density.
	\item {\tt{SOM\_CellModel\_LimitedSum}}: A limited reconstruction using all models contained within a small subset of SOM cells selected based on their SOM cell models.
	\item {\tt{SOM\_CellModel\_Average}}: A weighted redshift average across the SOM based on the SOM cell models.
\end{enumerate}

We discuss each of these in turn.

\subsection{{\tt{BruteForce}} (gold standard)}
\label{subsec:BruteForce}

Deriving $P(z|\rom{\mathbf{F}}{obs})$ for any individual object $\rom{\mathbf{F}}{obs}$ in a brute-force approach involves computing the likelihood $P(\boldsymbol{\theta},\boldsymbol{\phi}|\rom{\mathbf{F}}{obs})$ to the entire collection $\rom{\mathbf{F}}{model}(\boldsymbol{\theta},\boldsymbol{\phi})$'s. Each $\rom{\mathbf{F}}{model}$ has a corresponding redshift attached to it, and the final $P(z|\rom{\mathbf{F}}{obs})$ is constructed by marginalizing over all the other input parameters such that
\begin{equation}\label{eq:pz}
\boxed{P(z|\rom{\mathbf{F}}{obs}) \propto \sum_{\textrm{dust}}\sum_{\textrm{gal}}\sum_{\boldsymbol{\phi}} w(\boldsymbol{\theta},\boldsymbol{\phi}) P(\boldsymbol{\theta},\boldsymbol{\phi}|\rom{\mathbf{F}}{obs}),}
\end{equation}
where $w(\boldsymbol{\theta},\boldsymbol{\phi})$ is an associated model weight (from, e.g., priors on $\boldsymbol{\theta}$ and $\boldsymbol{\phi}$) and $P(\boldsymbol{\theta},\boldsymbol{\phi}|\rom{\mathbf{F}}{obs})$ is the computed likelihood between the data and the model. For the purposes of this paper, we take our collection of model photometry to be the 3,228,225 objects included in our observed-frame model catalog (see \S\ref{subsec:som}) and assign all models equal weight such that $w(\boldsymbol{\theta},\boldsymbol{\phi})=1$.

\subsection{{\tt{BruteForce\_LinearFuzzy}}}
\label{subsec:BruteForce_linear}

In order to analytically marginalize over the ``fuzziness'' in our fuzzy templates, we first need to render them in a more statistically-friendly and computationally-convenient form. As outlined in Paper I, we can accomplish this by approximating the changes in the underlying archetype as being linear in $\boldsymbol{\phi}$ and recasting the fitting procedure as a (conditional) linear algebra problem that can be solved using iterative least squares. We detail the full derivation in Appendix~\ref{app:linear_fuzzy} and give a brief outline below.

Taylor expanding around the baseline archetype and ignoring all terms higher than first order, $\rom{\mathbf{F}}{model}(z)$ can be written as
\begin{eqnarray}
s\rom{\mathbf{F}}{model}(z) = s\rom{\mathbf{F}}{gal}(z) + \Delta\ebv \,s\rom{\mathbf{F}}{dust}(z) \nonumber \\
+ \Delta c_b^\prime \,s\rom{\mathbf{F}}{bump}(z) + \sum_{\textrm{lines}} \Delta \textrm{EW}_{\textrm{lines}}^\textrm{gal} \,s\rom{\mathbf{F}}{lines}^{\textrm{gal}}(z),
\end{eqnarray}
where $\rom{\mathbf{F}}{gal}(z)$ is the flux of the baseline galaxy archetype, $\rom{\mathbf{F}}{dust}(z)$ is the ``dust flux'', $\rom{\mathbf{F}}{bump}(z)$ is the ``2175\,{\AA} bump flux'', and $\rom{\mathbf{F}}{lines}^{\textrm{gal}}(z)$ is the corresponding emission line flux for a given archetype and emission line.

Assuming that our priors are Gaussian, the corresponding log-likehood is
\begingroup\makeatletter\def\f@size{8.2}\check@mathfonts
\begin{equation}\label{eq:chi2_mod}
-2\log P(\boldsymbol{\theta},\boldsymbol{\phi},s|\rom{\mathbf{F}}{obs}) \equiv \sum_i \sigma_{i}^{-2} \left[\Delta F_i(\boldsymbol{\theta},\boldsymbol{\phi},s)\right]^2 + \sum_j \left(\frac{\phi_j}{\sigma_{\phi_j}(\boldsymbol{\theta})}\right)^2
\end{equation}
\endgroup
where $\boldsymbol{\sigma}^2$ is the observed variances (we take all model errors to be 0), $\boldsymbol{\Delta}\mathbf{F}(\boldsymbol{\theta},\boldsymbol{\phi},s) = \mathbf{F}_{\textrm{obs}} - s\mathbf{F}_{\textrm{model}}(\boldsymbol{\theta},\boldsymbol{\phi})$ are the associated flux residuals, $\boldsymbol{\sigma}_{\phi}(\boldsymbol{\theta})$ are the standard deviations of the corresponding Gaussian priors for a given $\boldsymbol{\theta}$, the sum over $i$ is taken over all observed bands, and the sum over $j$ is taken over all relevant nuisance parameters. This allows us to re-write our expression for $\rom{\mathbf{F}}{model}$ as
\begin{equation}\label{eq:linear_phot}
sF_{\textrm{model},i}(\boldsymbol{\theta},\boldsymbol{\phi})=sF_{\textrm{gal},i}(\boldsymbol{\theta})+\sum_j s\mathbf{X}_{ij}(\boldsymbol{\theta}) \boldsymbol{\phi}_j,
\end{equation}
where $\mathbf{X}(\boldsymbol{\theta})$ is an $\rom{N}{filt} \times (2+\rom{N}{lines})$ matrix of pre-computed coefficients for a given $\boldsymbol{\theta}$. 

For fixed $\boldsymbol{\theta}$ and $\boldsymbol{\phi}$, we can marginalize over $s$ to minimize $\rom{\chi^2}{mod}(s|\boldsymbol{\theta},\boldsymbol{\phi})$ via
\begin{equation}\label{eq:chi_s}
s = \left. {\sum_i {\sigma_i^{-2}}} F_{\textrm{obs},i}F_{\textrm{model},i} \middle/ {\sum_i \sigma_i^{-2} F_{\textrm{model},i}F_{\textrm{model},i}} \right. ,
\end{equation}
while for fixed $s$ and $\boldsymbol{\theta}$, we can marginalize over $\boldsymbol{\phi}$ by solving 
\begin{equation}\label{eq:chi_lparams}
\left(\mathbf{X}(\boldsymbol{\theta})^T\rom{\mathbf{W}}{obs}\right)\Delta\rom{\mathbf{F}}{gal}(\boldsymbol{\theta})=\left(\mathbf{X}(\boldsymbol{\theta})^{T}\rom{\mathbf{W}}{obs}\mathbf{X}(\boldsymbol{\theta})+\mathbf{W}_\phi(\boldsymbol{\theta})\right)\boldsymbol{\phi},
\end{equation}
where $\Delta\rom{\mathbf{F}}{gal}(\boldsymbol{\theta}) = \mathbf{F}_{\textrm{obs}} - s\rom{\mathbf{F}}{gal}(\boldsymbol{\theta})$ is the baseline galaxy flux residual, $\rom{\mathbf{W}}{obs}=\textrm{diag}(\dots,\sigma_i^{-2},\dots)$ is the associated observational weight matrix, $\mathbf{W}_\phi=\textrm{diag}(\dots,\sigma_{\phi_j}^{-2},\dots)$ is the prior weight matrix, and $T$ is the transpose operator.

This gives us an simple iterative scheme for computing $\rom{P}{max}(\boldsymbol{\theta}|\rom{\mathbf{F}}{obs})$ with respect to $s$ and $\boldsymbol{\phi}$ for a given choice of $\boldsymbol{\theta}$. Using this approach, we execute a brute-force search over all $N_z \times \rom{N}{gal} \times \rom{N}{dust} = 1,033,032$ possible combinations of our non-linear parameters $\boldsymbol{\theta}$. This gives us
\begin{equation}\label{eq:pz_linear}
\boxed{P(z|\rom{\mathbf{F}}{obs}) \propto \sum_{\textrm{dust}}\sum_{\textrm{gal}} \rom{P}{max}(\boldsymbol{\theta}|\rom{\mathbf{F}}{obs}).}
\end{equation}

\subsection{{\tt{SOM\_MCMC\_RestFrame}}}
\label{subsec:mcmc_restframe}


In Paper I, we introduced a method to derive estimates to $P(z|\rom{\mathbf{F}}{obs})$ in a more exact sense using Markov Chain Monte Carlo (MCMC) methods running ``over'' the SOM whose performance only depends on the ability of the SOM to effectively cluster the space in a topologically smooth fashion. We summarize our approach here and direct the reader to Paper I for additional details.

In brief, we implement a hybrid version of MCMC sampling that transitions to new regions based on comparisons between individual sets of model photometry but with chains that live ``on top of'' the SOM. Chains propose new models in a hierarchical fashion by first proposing a new cell position\footnote{If a proposed cell has no objects, the process is repeated from the proposed cell until a valid cell is reached.} and then selecting a random model within that cell. Each chain then transitions to the proposed cell with an associated transition probability that satisfies detailed balance.

After the chains have converged to the stationary target distribution (the ``burn-in'' phase), they are allowed to sample for a set number of iterations. The final $P(z|\rom{\mathbf{F}}{obs})$ can then be reconstructed via
\begingroup\makeatletter\def\f@size{8.7}\check@mathfonts
\begin{equation}
\boxed{P(z|\rom{\mathbf{F}}{obs})\approx\sum_{\textrm{chain}} \rom{w}{chain} P(z|\textrm{chain},\rom{\mathbf{F}}{obs}).}
\end{equation}
\endgroup
where $\rom{w}{chain} = \rom{P}{max}(\rom{\mathbf{F}}{obs}|\rom{\mathbf{F}}{model},\textrm{chain})$ is a maximum-likelihood weight used to suppress chains that happen to converge to spurious minima and
\begin{equation}
P(z|\textrm{chain},\rom{\mathbf{F}}{obs})={\sum_{\textrm{dust}}}^{\textrm{chain}} {\sum_{\textrm{gal}}}^{\textrm{chain}} {\sum_{\boldsymbol{\phi}}}^{\textrm{chain}} w(\boldsymbol{\theta},\boldsymbol{\phi}|\rom{\mathbf{F}}{obs})
\end{equation}
is the sum over the trials from each individual chain with $w(\boldsymbol{\theta},\boldsymbol{\phi}|\rom{\mathbf{F}}{obs})$ the associated weight of each trial. This is usually taken to be $w=1$ if the trial was accepted and $w=0$ if the trial was rejected, although other weighting schemes are possible \citep{bernton+15}.

We implement this sampling scheme on the 2-D rest-frame SOM with where new cells are proposed according to a multivariate Gaussian with $\boldsymbol{\Sigma} = \textrm{diag}\left[(\rom{\sigma}{som}^\textrm{dim\_1})^2, (\rom{\sigma}{som}^\textrm{dim\_1})^2\right]$ where $\rom{\sigma}{som}^\textrm{dim\_1}=\rom{\sigma}{som}^\textrm{dim\_2}=2$. Once a specific archetype, dust template, and $\boldsymbol{\phi}$ have been chosen, we generate the associated redshifted photometry based on a separate Gaussian proposal with $\sigma_z=2.5\%$ of the corresponding redshift grid.

To improve performance and reduce the dependence on our initial choice of parameters, we include a short learning step during every MCMC run after the burn-in phase is complete where the corresponding size of each component of the proposal ($\lbrace\rom{\sigma}{som}^\textrm{dim\_i}\rbrace_i,\sigma_{z}$) is adjusted to better approximate the underlying shape of the PDF. We set a fixed number of iterations for burn-in, learning, and sampling of $5000$, $1000$, and $5000$ trials, respectively, and sample each object using 32 chains run in parallel.

\subsection{{\tt{SOM\_MCMC\_ObservedFrame}}}
\label{subsec:mcmc_observedframe}

In \S\ref{subsec:mcmc_restframe}, we used a 2-D SOM with 900 cells to sort 9675 rest-frame relations of our fuzzy archetypes that were then allowed to vary as a function of redshift. However, we can instead attempt to reorganize the entire fuzzy archetype-redshift \textit{observed-frame} color space using a single SOM. We use this approach to construct an 3-D observed-frame SOM composed of 27,000 cells based on $\sim$\,$\sn{3}{6}$ individual realizations of our model photometry from $z=0$\,--\,$10$.

To explore this observed-frame SOM, we again implement the MCMC sampling scheme described above on the 3-D observed-frame SOM. New cells are proposed according to the multivariate Gaussian distribution with $\boldsymbol{\Sigma} = \textrm{diag}\left[(\rom{\sigma}{som}^\textrm{dim\_1})^2, (\rom{\sigma}{som}^\textrm{dim\_2})^2,(\rom{\sigma}{som}^\textrm{dim\_3})^2\right]$ where $\rom{\sigma}{som}^\textrm{dim\_1}=\rom{\sigma}{som}^\textrm{dim\_2}=\rom{\sigma}{som}^\textrm{dim\_3}=2$. As before, we set a fixed number of iterations for burn-in, learning, and sampling of $5000$, $1000$, and $5000$ trials, respectively, and sample each object using 32 chains run in parallel.

\subsection{{\tt{SOM\_Hierarchical\_MonteCarlo}}}
\label{subsec:stratifiedMC}

Rather than taking advantage of the topologically smooth nature of the SOM to ``explore'' its corresponding reduced-dimensional manifold, we can instead treat the SOM as simply a way to partition the model space into a series of  clusters (i.e. cells). Rather than using the relationship between these clusters (i.e. the ``graph'' they form) to define some sampling algorithm, we can then treat each cluster as an independent entity. Thus, instead of sampling from a local neighborhood function in SOM space to determine what quasi-contiguous regions of the SOM contain the majority of the likelihood, we instead wish to determine which isolated clusters contain significant amounts of likelihood to merit further exploration and/or sampling.

In other words, we want to break down the problem into a \textit{hierarchical sampling} approach, where we first consider which clusters (stored in some arbitrary, hierarchical structure) to explore before considering the individual models themselves. In the case of the SOM, where the clusters are organized along a single manifold, this is a simple two-tiered problem involving the individual cells and the objects contained within them.

The process of determining how likely a particular $\rom{\mathbf{F}}{obs}$ (along with associated errors $\rom{\boldsymbol{\sigma}}{obs}$) is matched to a particular cluster is an open problem. As clusters can have very different distributions of galaxy colors and occupation rates depending of its location in color space (Paper I), the use of simple summary statistics will often introduce biases depending on the properties data in question (see \S\ref{sec:results}).

Alternately, we can directly sample from the PDF spanned by each cluster using Monte Carlo techniques. Unlike the case of summary statistics, this approach is guaranteed to give a good approximation to the likelihood contained within each cluster when the number of samples $\rom{N}{MC}\sim N(\rom{\mathbf{x}}{som}^{\textrm{cell}})$. 

In addition to giving us information on the likelihood contained within each cluster, such an approach naturally introduces a \textit{stratified Monte Carlo} sampling scheme to uncover the underlying PDF, where random model photometry drawn from partitioned regions of color space can be used to stochastically reconstruct the color-redshift relation probed by $\rom{\mathbf{F}}{obs}$ and established by the individual SOM cells. In the case outlined above where $\rom{N}{MC}\sim N(\rom{\mathbf{x}}{som}^{\textrm{cell}})$ for each cell, this approach simply reduces to a Monte Carlo version of a brute-force approach.

However, the power of this stratified Monte Carlo approach is especially apparent in the sparse-sampling regime where only a small amount of objects is drawn from each cluster. While such an approach might not be effective in determining the precise likelihood contained within individual cluster, it does provide a fair statistical representation of a sample galaxy's distribution over the entire region of color space spanned by the collection of SOM cells. Sparse sampling of this sort using single-draw Monte Carlo approaches has been shown to be quite effective at reconstructing $N(z)$ distributions (and relevant summary statistics) to ensembles of objects, as shown in, e.g., \citet{wittman09}, \citet{carrascokindbrunner14c}, and \citep{masters+15}.

Due to the small number of occupied cells in our observed-frame SOM ($\sim$\,21,600 out of 27,000), we decide to simply draw an equal number of samples from occupied cells on the SOM. The corresponding $P(z|\rom{\mathbf{F}}{obs})$ can then be directly calculated via
\begin{equation}
\boxed{P(z|\rom{\mathbf{F}}{obs})\approx\sum_{\textrm{cell}} \sum_{\mathbfcal{R}(\textrm{cell})} N(\mathbf{x}_{\textrm{som}}^{\textrm{cell}}) P(\rom{\mathbf{F}}{model}^{\mathcal{R}_i}|\rom{\mathbf{F}}{obs}),}
\end{equation}
where $\sum_{\textrm{cell}}$ is taken over all occupied cells, $\mathbfcal{R}(\textrm{cell})$ is a collection of $\rom{N}{MC}$ random numbers $\lbrace\mathcal{R}_i\rbrace_i$ that select the corresponding indices of model fluxes $\rom{\mathbf{F}}{model}^{\mathcal{R}_i}$ contained within a given cell, and $N(\rom{\mathbf{x}}{som}^{\textrm{cell}})$ accounts for the different occupation rates of each cell. We set $\rom{N}{MC}=5$ to allow for more robust sampling than single-draw approaches while still keeping the number of total draws ($\sim$\,110,000) a factor of $\sim$\,30 below the total number of available models.

\subsection{{\tt{SOM\_Hierarchical\_ImportanceSampling}}}
\label{subsec:importancesampling}

While the stratified Monte Carlo approach is easy to implement and relatively resilient against random misses (since every cell on the SOM is sampled evenly), most objects with well-localized colors are strongly associated with a very small set of clusters. As a result, a significant amount of computation time will be wasted evaluating clusters containing model fluxes with $P(\rom{\mathbf{F}}{obs}|\rom{\mathbf{F}}{model})\approx 0$. Rather than drawing an equal number of samples from every SOM cell, we could use preliminary approximations of the likelihood across the SOM to instead take advantage of adaptive \textit{importance sampling} to favor target cells that are more likely to contain the majority of the likelihood. 

In the scheme of the hierarchical interpretation outlined above in \S\ref{subsec:stratifiedMC}, we can approximate the likelihood contained in each individual cluster $P(\rom{\mathbf{x}}{som}^{\textrm{cell}}|\rom{\mathbf{F}}{obs})$ via the identity
\begin{eqnarray}
P(\rom{\mathbf{F}}{obs})=\int P(\rom{\mathbf{x}}{som}^{\textrm{cell}})P(\rom{\mathbf{F}}{obs}|\rom{\mathbf{x}}{som}^{\textrm{cell}})d\rom{\mathbf{x}}{som}^{\textrm{cell}}\nonumber \\
=\int\pi(\rom{\mathbf{x}}{som}^{\textrm{cell}})\frac{P(\rom{\mathbf{x}}{som}^{\textrm{cell}})P(\rom{\mathbf{F}}{obs}|\rom{\mathbf{x}}{som}^{\textrm{cell}})}{\pi(\rom{\mathbf{x}}{som}^{\textrm{cell}})}d\rom{\mathbf{x}}{som}^{\textrm{cell}},
\end{eqnarray}
where $\pi(\rom{\mathbf{x}}{som}^{\textrm{cell}})$ is a generic approximation to $P(\rom{\mathbf{x}}{som}^{\textrm{cell}}|\rom{\mathbf{F}}{obs})$. For a series of $\rom{N}{IS}$ samples drawn from $\pi(\rom{\mathbf{x}}{som}^{\textrm{cell}})$, the $n$th posterior moment of $P(\rom{\mathbf{x}}{som}^{\textrm{cell}}|\rom{\mathbf{F}}{obs})$ can then be estimated as
\begingroup\makeatletter\def\f@size{8.7}\check@mathfonts
\begin{equation}
\int (\rom{\mathbf{x}}{som}^{\textrm{cell}})^nP(\rom{\mathbf{x}}{som}^{\textrm{cell}}|\rom{\mathbf{F}}{obs})d\rom{\mathbf{x}}{som}^{\textrm{cell}}=\frac{\int (\rom{\mathbf{x}}{som}^{\textrm{cell}})^n \pi(\rom{\mathbf{x}}{som}^{\textrm{cell}}) w(\rom{\mathbf{x}}{som}^{\textrm{cell}})d\rom{\mathbf{x}}{som}^{\textrm{cell}}}{\int \pi(\rom{\mathbf{x}}{som}^{\textrm{cell}})w(\rom{\mathbf{x}}{som}^{\textrm{cell}})d\rom{\mathbf{x}}{som}^{\textrm{cell}}}\nonumber
\end{equation}
\endgroup
\begin{equation}
\approx \left. {\sum_{\rom{N}{IS}} (\rom{\mathbf{x}}{som}^{\textrm{cell}})^n  w(\rom{\mathbf{x}}{som}^{\textrm{cell}})} \middle/ {\sum_{\rom{N}{IS}} w(\rom{\mathbf{x}}{som}^{\textrm{cell}})} \right. ,
\end{equation}
where $w(\rom{\mathbf{x}}{som}^{\textrm{cell}})=\frac{P(\rom{\mathbf{x}}{som}^{\textrm{cell}})P(\rom{\mathbf{F}}{obs}|\rom{\mathbf{x}}{som}^{\textrm{cell}})}{\pi(\rom{\mathbf{x}}{som}^{\textrm{cell}})}$ is the associated \textit{importance weight} -- the overall likelihood of a given trial formed by a combination of the initial prior and the computed likelihood normalized by the proposal distribution. Estimates of the relevant marginalized distributions can be constructed by assigning each sampled location an appropriate \textit{kernel density} with an amplitude proportional to its importance weight.

Using the same logic as \S\ref{subsec:stratifiedMC}, implementing a Monte Carlo sampling scheme within each cluster allows us to separate out the proposal distribution $\pi(\rom{\mathbf{x}}{som}^{\textrm{cell}})$ which operates on the set of clusters and the associated importance weights $w(\rom{\mathbf{F}}{model})$ assigned to each particular chosen model. Assuming $\pi(\rom{\mathbf{x}}{som}^{\textrm{cell}})$ is properly normalized, the corresponding importance weight for a random model $\rom{\mathbf{F}}{model}^{\mathcal{R}_i}$ located at cell $\rom{\mathbf{x}}{som}^{\textrm{cell}}$ is then
\begin{equation}
w(\rom{\mathbf{F}}{model}^{\mathcal{R}_i})=N(\rom{\mathbf{x}}{som}^{\textrm{cell}})\frac{P(\rom{\mathbf{F}}{model}^{\mathcal{R}_i}|\rom{\mathbf{F}}{obs})}{\pi(\rom{\mathbf{x}}{som}^{\textrm{cell}})}.
\end{equation}

For a number of $\rom{N}{IS}$ samples drawn randomly from cells drawn from $\pi(\rom{\mathbf{x}}{som}^{\textrm{cell}})$, we can then estimate $P(z|\rom{\mathbf{F}}{obs})$ as
\begin{equation}
\boxed{P(z|\rom{\mathbf{F}}{obs})\approx \sum_{\mathbfcal{R}(\pi(\rom{\mathbf{x}}{som}^{\textrm{cell}}))} w(\rom{\mathbf{F}}{model}^{\mathcal{R}_i}) P(\rom{\mathbf{F}}{model}^{\mathcal{R}_i}|\rom{\mathbf{F}}{obs}).}
\end{equation}
In this paper, we have avoided the use of kernels to avoid additional computational overhead.

For the implementation used in this paper, we further build on this general approach using an adaptive importance sampling scheme equivalent to a crude form of Sequential/Population Monte Carlo \citep{doucet+01} and related methods:
\begin{enumerate}
	\item Compute $\pi(\rom{\mathbf{x}}{som}^{\textrm{cell}})$ using the stratified Monte Carlo scheme outlined in \S\ref{subsec:stratifiedMC} with $\rom{N}{MC}$ random models sampled in each cell.
	\item Draw $\rom{N}{batch}$ cells from $\pi(\rom{\mathbf{x}}{som}^{\textrm{cell}})$.
	\item Draw a random model $\rom{\mathbf{F}}{model}^{\mathcal{R}_i}$ from each cell.
	\item Use the corresponding set of $w(\rom{\mathbf{F}}{model}^{\mathcal{R}_i})$'s to update $\pi(\rom{\mathbf{x}}{som}^{\textrm{cell}})$.
	\item Repeat steps (ii) through (iv) for $\rom{N}{iter}$ iterations.
	\item Use the total number of objects sampled from steps (i) through (v) to compute $P(z|\rom{\mathbf{F}}{obs})$.
\end{enumerate}
We set $\rom{N}{MC}=1$, $\rom{N}{batch}=2500$, and $\rom{N}{iter}=20$ in this paper, giving us a total of $\sim$\,70,000 models comparisons (slightly less than the amount used in \S\ref{subsec:stratifiedMC}). In addition, we impose a ``likelihood floor'' for each computed weight of $10^{-3}$ when updating the proposal distribution both to prevent multi-valued inverse functions when drawing from the proposal and to ensure each cell has a small (but non-negligible) probability of being selected with each draw in order to ensure great robustness during sampling. Note that this likelihood floor is \textit{not} used when constructing $P(z|\rom{\mathbf{F}}{model})$.

\subsection{{\tt{SOM\_CellModel\_LimitedSum}}}
\label{subsec:som_limited}

Instead of sampling over the entire SOM, we can instead attempt to limit the total number of terms included in the sum to only cells which carry significant weight, after which we can just compute the reduced set of $P(z|\rom{\mathbf{x}}{som}^{\textrm{cell}})$'s directly. In other words, after computing the full set of $P(\rom{\mathbf{F}}{obs},\rom{\mathbf{F}}{som}^{\textrm{cell}})$'s across the 3-D observed-frame SOM for a particular object, we select a collection of $k$ cells such that
\begin{equation}
P(\rom{\mathbf{F}}{som}^{\textrm{cell}}|\rom{\mathbf{F}}{obs})>c_\eta \textrm{max}\left[P(\rom{\mathbf{F}}{som}^{\textrm{cell}}|\rom{\mathbf{F}}{obs})\right],
\end{equation}
where $c_\eta$ is a specified acceptance threshold which we take to be $10^{-3}$ and $\rom{\mathbf{F}}{som}^{\textrm{cell}}$ is the associated SOM cell model (see \S\ref{subsec:som}), which we assume to be a reasonable proxy of the data in a given cell. As noted in \S\ref{subsec:stratifiedMC}, while this assumption is not always true (especially in sparely populated and/or highly variable regions of color space), it is a good first-order approximation to the problem.

After using the cell models to select our best-fitting cells, we can then compute $P(z|\rom{\mathbf{F}}{obs})$ as
\begingroup\makeatletter\def\f@size{8.4}\check@mathfonts
\begin{equation}
\boxed{P(z|\rom{\mathbf{F}}{obs}) \approx \sum_k P(z|\mathbf{x}_k) = \sum_k {\sum_{\textrm{dust}}}^{\mathbf{x}_k} {\sum_{\textrm{gal}}}^{\mathbf{x}_k} {\sum_{\boldsymbol{\phi}}}^{\mathbf{x}_k} P(\boldsymbol{\theta},\boldsymbol{\phi}|\rom{\mathbf{F}}{obs}),}
\end{equation}
\endgroup
where ${\sum}^{\mathbf{x}_k}$ indicates the sum is taken over all objects sorted into the $k$th cell. As the number of terms included in the sum over $k$ should only be a small fraction of the total number of cells, it should be reasonably quick to compute.

\subsection{{\tt{SOM\_CellModel\_Average}}}
\label{subsec:som_average}

Instead of comparing individual models and/or fuzzy templates, we now wish to take advatange of the quantization the SOM provides by assigning the models to a discrete number of cells. In particular, every cell in our 3-D observed-frame SOM has an associated redshift number distribution $N(z|\rom{\mathbf{x}}{som}^{\textrm{cell}})$, whose combined sum would be proportional to the input redshift prior which all the objects in our model catalog were drawn from. For any individual object, this $N(z|\rom{\mathbf{x}}{som}^{\textrm{cell}})$ distribution corresponds to a specific redshift probability $P(z|\rom{\mathbf{x}}{som}^{\textrm{cell}},\rom{\mathbf{F}}{obs})$ for the given cell through a redshift-dependent weight function $w(z|\rom{\mathbf{x}}{som}^{\textrm{cell}},\rom{\mathbf{F}}{obs})$. This allows us to write $P(z|\rom{\mathbf{F}}{obs})$ as
\begingroup\makeatletter\def\f@size{8}\check@mathfonts
\begin{equation}
P(z|\rom{\mathbf{F}}{obs})=\sum_{\textrm{cell}}P(z|\rom{\mathbf{x}}{som}^{\textrm{cell}},\rom{\mathbf{F}}{obs})=\sum_{\textrm{cell}} w(z|\rom{\mathbf{F}}{obs},\rom{\mathbf{x}}{som}^{\textrm{cell}})N(z|\rom{\mathbf{x}}{som}^{\textrm{cell}}),
\end{equation}
\endgroup
where unoccupied cells again are excluded from the sum.

Assuming that the SOM has clustered objects efficiently relative to the observed errors on $\rom{\mathbf{F}}{obs}$, we roughly can approximate the likelihood of a given cell as simply being proportional to its intrinsic redshift distribution such that $P(z|\rom{\mathbf{F}}{obs},\rom{\mathbf{x}}{som}^{\textrm{cell}}) \sim N(z|\rom{\mathbf{x}}{som}^{\textrm{cell}})$. If we further assume (as in \S\ref{subsec:som_limited}) that the cell model is a good proxy for the data in a given cell, we can further approximate the cell weight as being the likelihood of the corresponding cell model $P(\rom{\mathbf{F}}{som}^{\textrm{cell}}|\rom{\mathbf{F}}{obs})$. This gives, up to a normalization factor, 
\begin{equation}
P(z|\rom{\mathbf{F}}{obs}) \approx \sum_{\textrm{cell}} P(\rom{\mathbf{F}}{som}^{\textrm{cell}}|\rom{\mathbf{F}}{obs})N(z|\rom{\mathbf{x}}{som}^{\textrm{cell}}).
\end{equation}

As each $N(z|\rom{\mathbf{x}}{som}^{\textrm{cell}})$ is an array of $N_z$ elements, this sum is computationally taxing due to the large number of cells used in the construction of the observed-frame SOM. As a result, we instead opt to use a point estimate for $N(z|\rom{\mathbf{x}}{som}^{\textrm{cell}})$ based on its zeroth moment (i.e. mean) $\bar{z}(\rom{\mathbf{x}}{som}^{\textrm{cell}})$, which can be pre-computed and saved to memory. The mean weighted redshift across the SOM for a given object can then be written as
\begin{equation}
\boxed{\rom{z}{avg}(\rom{\mathbf{F}}{obs}) = \left . \sum_{\textrm{cell}} P(\rom{\mathbf{F}}{som}^{\textrm{cell}}|\rom{\mathbf{F}}{obs})\bar{z}(\rom{\mathbf{x}}{som}^{\textrm{cell}}) \middle / \sum_{\textrm{cell}} P(\rom{\mathbf{F}}{obs},\rom{\mathbf{F}}{som}^{\textrm{cell}}) \right . .}
\end{equation}
While the full $P(z|\rom{\mathbf{F}}{model})$ distribution could be recovered using kernel density estimation centered on the corresponding collection of $\bar{z}(\rom{\mathbf{x}}{som}^{\textrm{cell}})$'s and $P(\rom{\mathbf{F}}{som}^{\textrm{cell}}|\rom{\mathbf{F}}{obs})$'s (see \S\ref{subsec:importancesampling}), for computational convenience we instead opt to compute only the first moment
\begingroup\makeatletter\def\f@size{8.5}\check@mathfonts
\begin{equation}
\sigma_{z,\textrm{avg}}^2(\rom{\mathbf{F}}{obs}|\rom{z}{avg}) = \frac{\sum_{\textrm{cell}}  P(\rom{\mathbf{F}}{obs},\rom{\mathbf{F}}{som}^{\textrm{cell}})\left[\bar{z}(\rom{\mathbf{x}}{som}^{\textrm{cell}})-\rom{z}{avg}(\rom{\mathbf{F}}{obs})\right]^2}{\sum_{\textrm{cell}} P(\rom{\mathbf{F}}{som}^{\textrm{cell}}|\rom{\mathbf{F}}{obs})}.
\end{equation}
\endgroup
Given that this method is the ``crudest'' approach that makes the strongest assumptions concerning the data and the SOM, we take this as a suitable estimate of the shape of the corresponding PDF.

\subsection{Summary}
\label{subsec:pz_list}

We have outlined eight separate methods to derive photo-z's for individual objects. Each pair of methods (brute-force, MCMC, hierarchical, cell model-dependent) represent four different ways to use templates to derive $P(z|\rom{\mathbf{F}}{obs})$. In our brute-force approaches, all available models and/or model parameters are compared to the data in order to derive the appropriate redshift PDF. In our MCMC-based approaches, we attempt to explore the set of models contained within contiguous regions on the SOM. In our hierarchical sampling approaches, we randomly sample galaxies according to the SOM's partitions in color space to stochastically construct estimates to the specific portion of the SOM-based color-redshift relation for each individual galaxy. Finally, in our cell model-dependent approaches, we attempt to ``triage'' the space spanned by the SOM using a series of summary statistics in order to derive limited estimates of $P(z|\rom{\mathbf{F}}{obs})$. Each take advantage of different aspects of our initial template set and the final SOM outputs and represent a variety of ways to take advantage of the SOM in order to maximize the utility of our fuzzy templates.

\begin{figure*}
	\centering
	\captionsetup[subfigure]{labelformat=empty}
	\subfloat[][{\tt{BruteForce}} (gold standard)]{\includegraphics[scale=0.25]{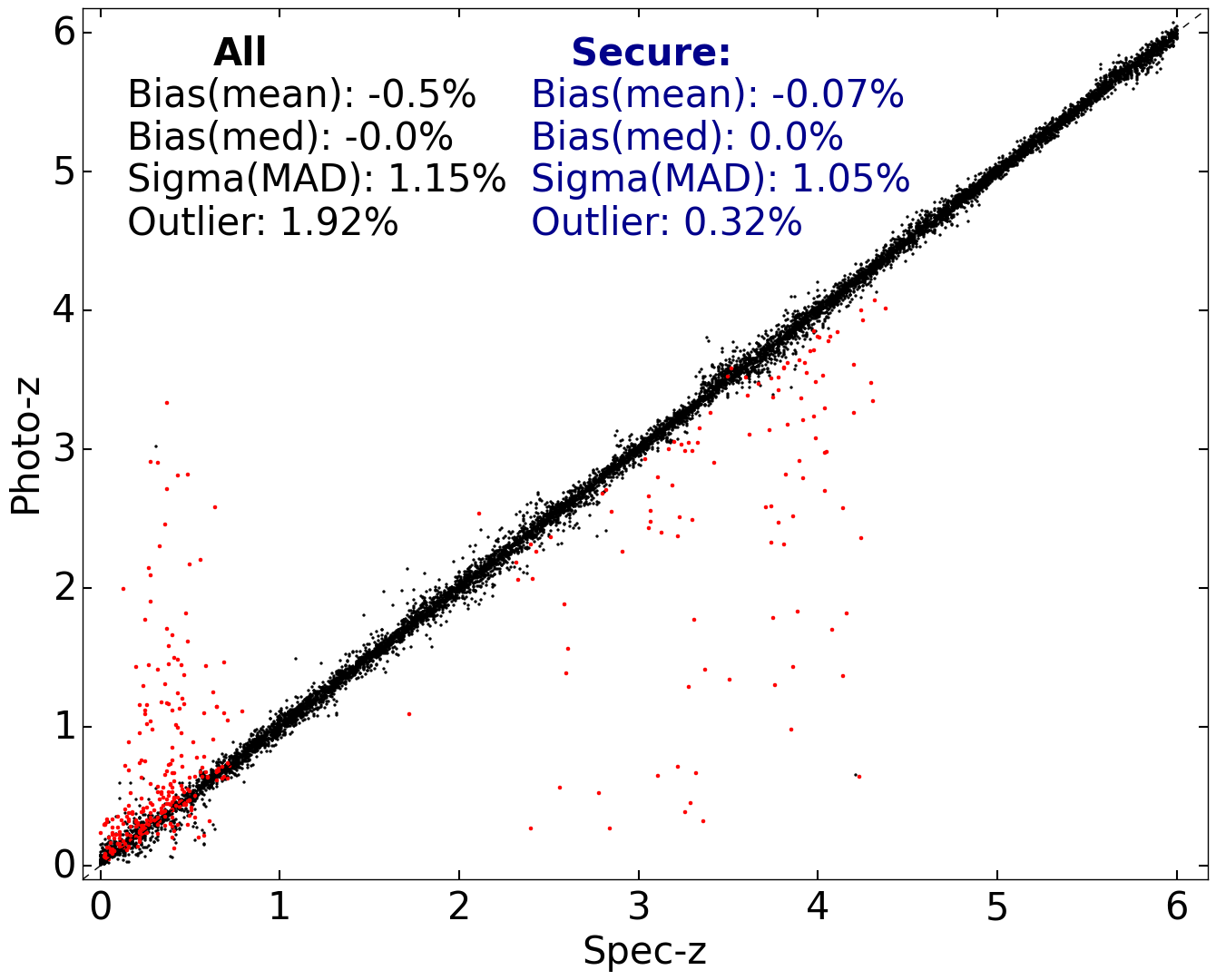}}
	\qquad
	\subfloat[][{\tt{BruteForce\_LinearFuzzy}}]{\includegraphics[scale=0.25]{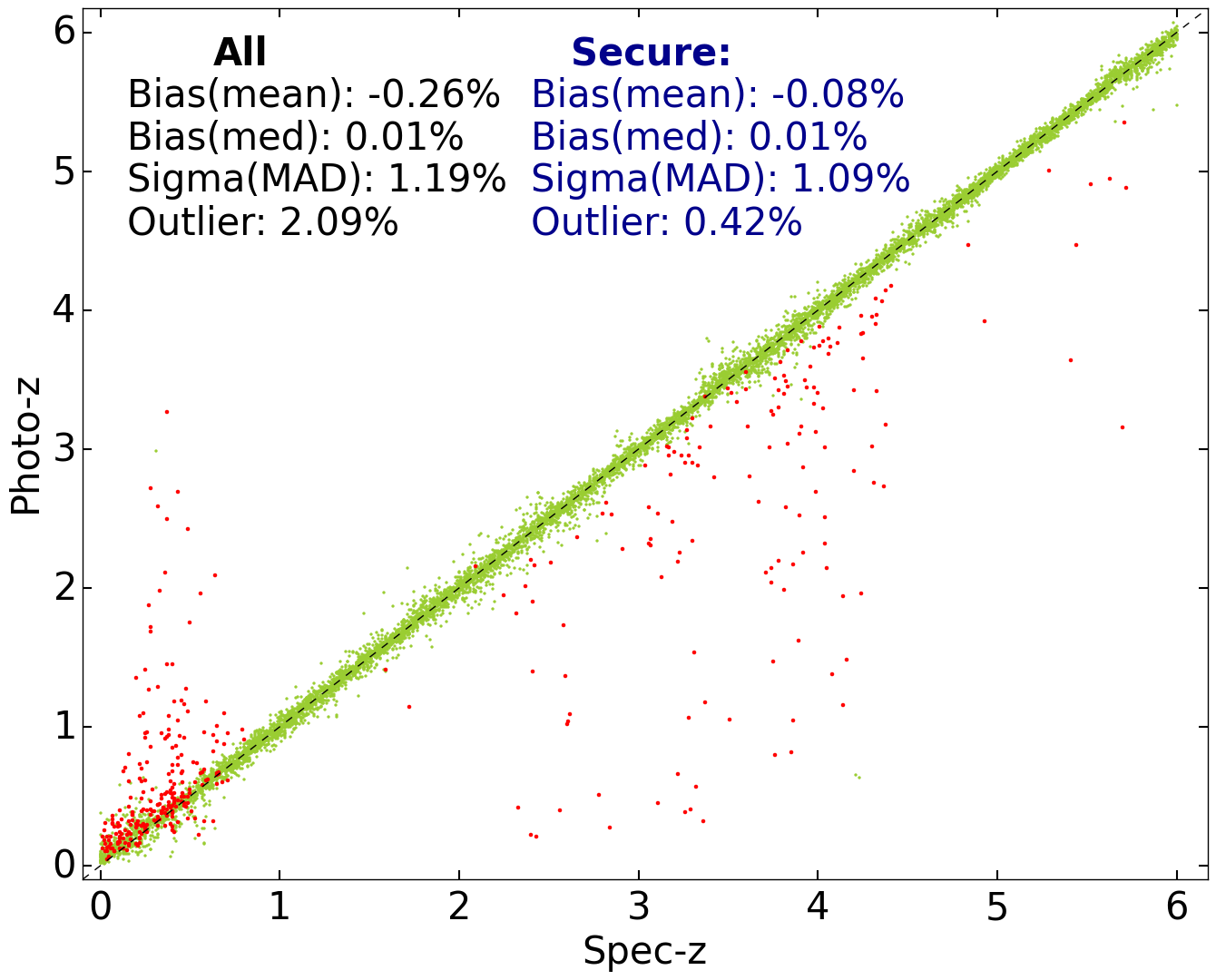}}\\
	\subfloat[][{\tt{SOM\_MCMC\_RestFrame}}]{\includegraphics[scale=0.25]{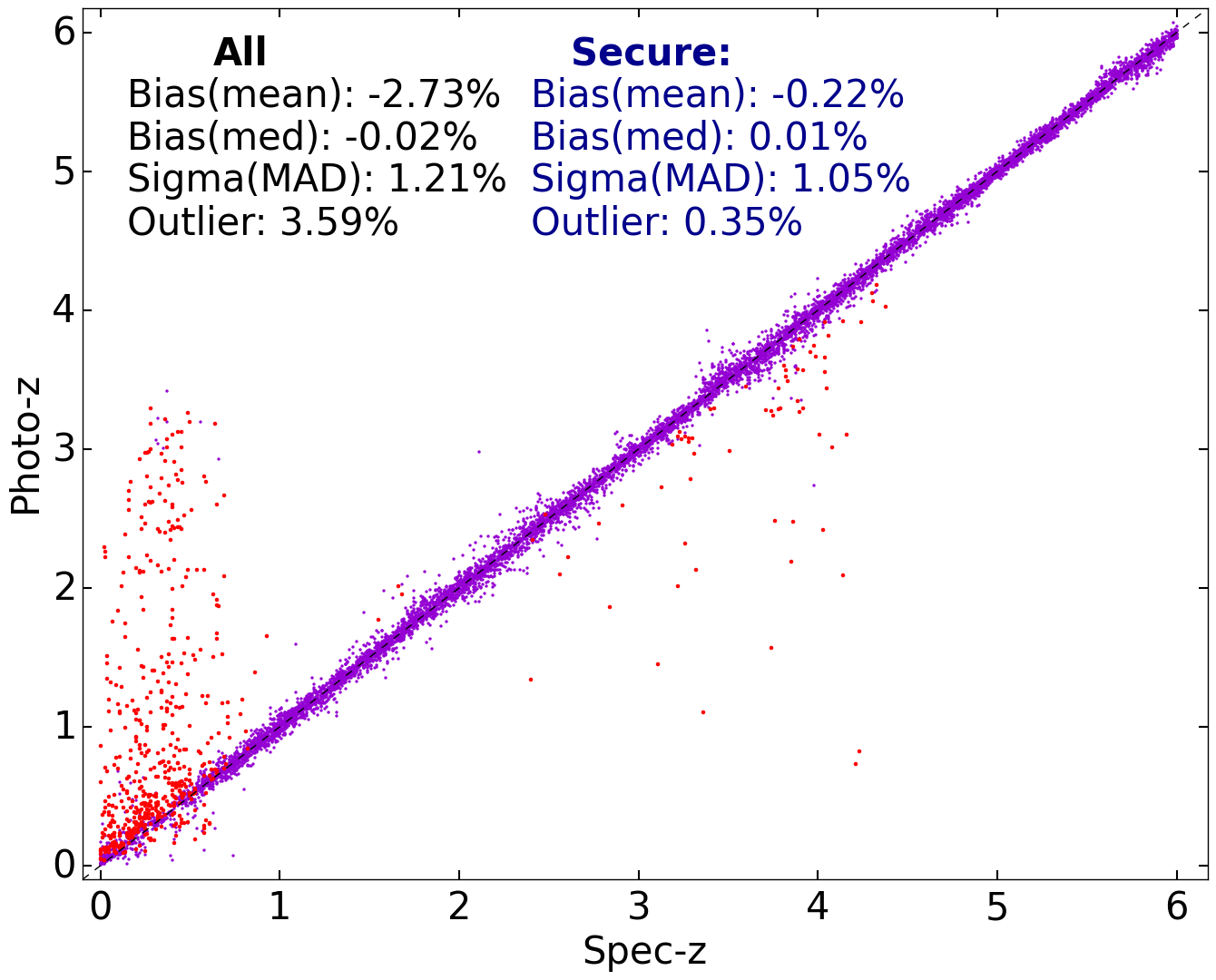}}
	\qquad
	\subfloat[][{\tt{SOM\_MCMC\_ObservedFrame}}]{\includegraphics[scale=0.25]{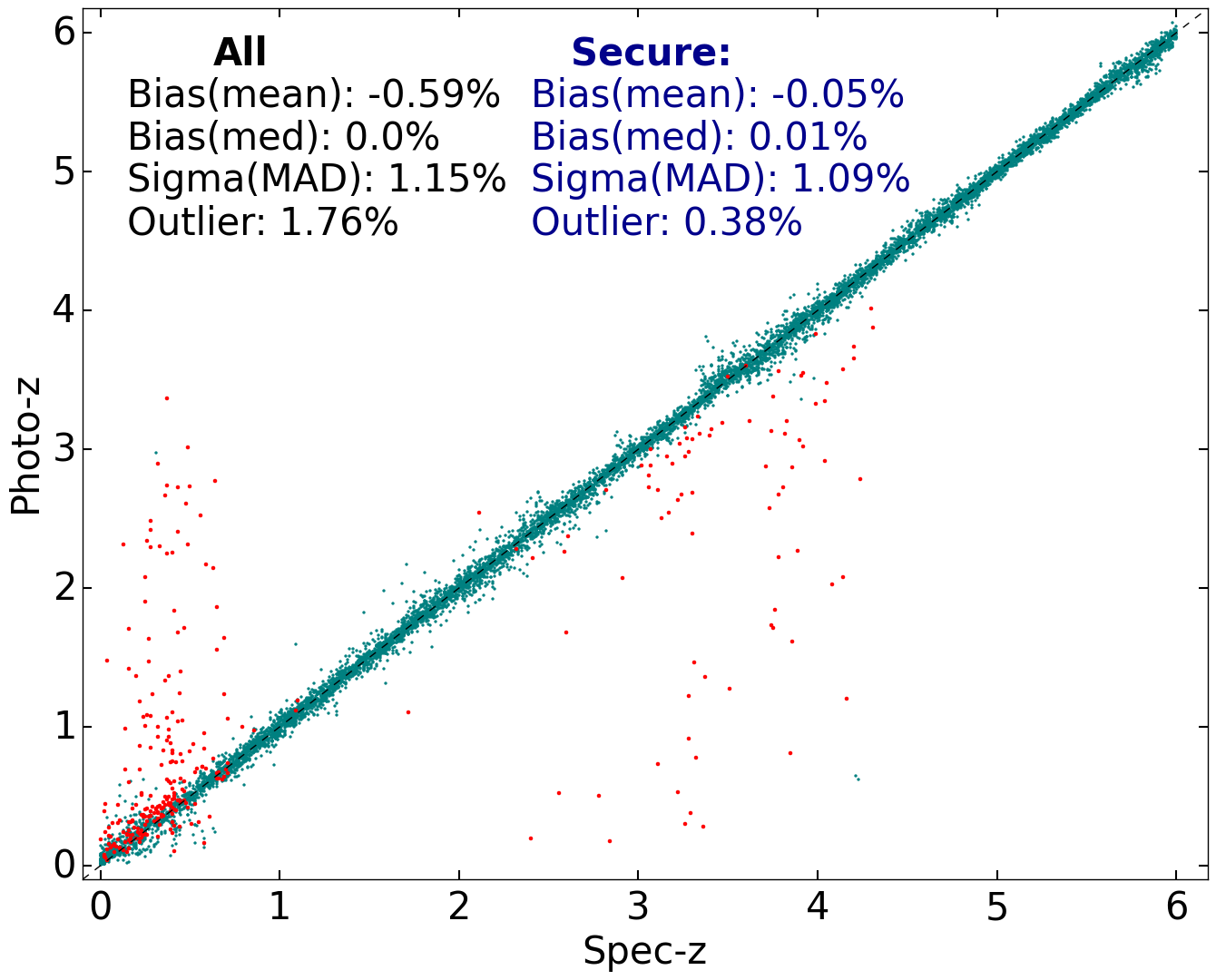}}
	\caption{
		The input ($\rom{z}{spec}$) versus fitted ($\rom{\bar{z}}{phot}$) redshifts for our {\tt{BruteForce}} (our ``gold standard''; black), {\tt{BruteForce\_LinearFuzzy}} (yellow-green), {\tt{SOM\_MCMC\_RestFrame}} (purple), and {\tt{SOM\_MCMC\_ObservedFrame}} (teal) methods (see \S\ref{sec:methods}) to the 10,000 objects in our 9-band LSST $ugrizY$ and \textit{Euclid} $YJH$ mock catalog consisting of $\rom{Y}{LSST}=24$\,mag objects. $P(z|\rom{\mathbf{F}}{obs})$ distributions with redshift-normalized standard deviations $\sigma_{z}^\prime > 0.15$ are flagged as insecure and plotted as red points. The redshift-normalized mean bias $\bar{\Delta z}^\prime$, median bias ${\Delta z}_{50}^\prime$, 1$\sigma$ median absolute deviation (MAD) $\sigma_{z,\textrm{MAD}}^\prime$, and the catastrophic outlier fraction $\rom{f}{cat}$ for the full and secure samples are listed in each plot.
		See Figure~\ref{fig:photz_2} for more details.}
	\label{fig:photz}
\end{figure*}

\begin{figure*}
	\centering
	\captionsetup[subfigure]{labelformat=empty}
	\subfloat[][{\tt{SOM\_Hierarchical\_MonteCarlo}}]{\includegraphics[scale=0.25]{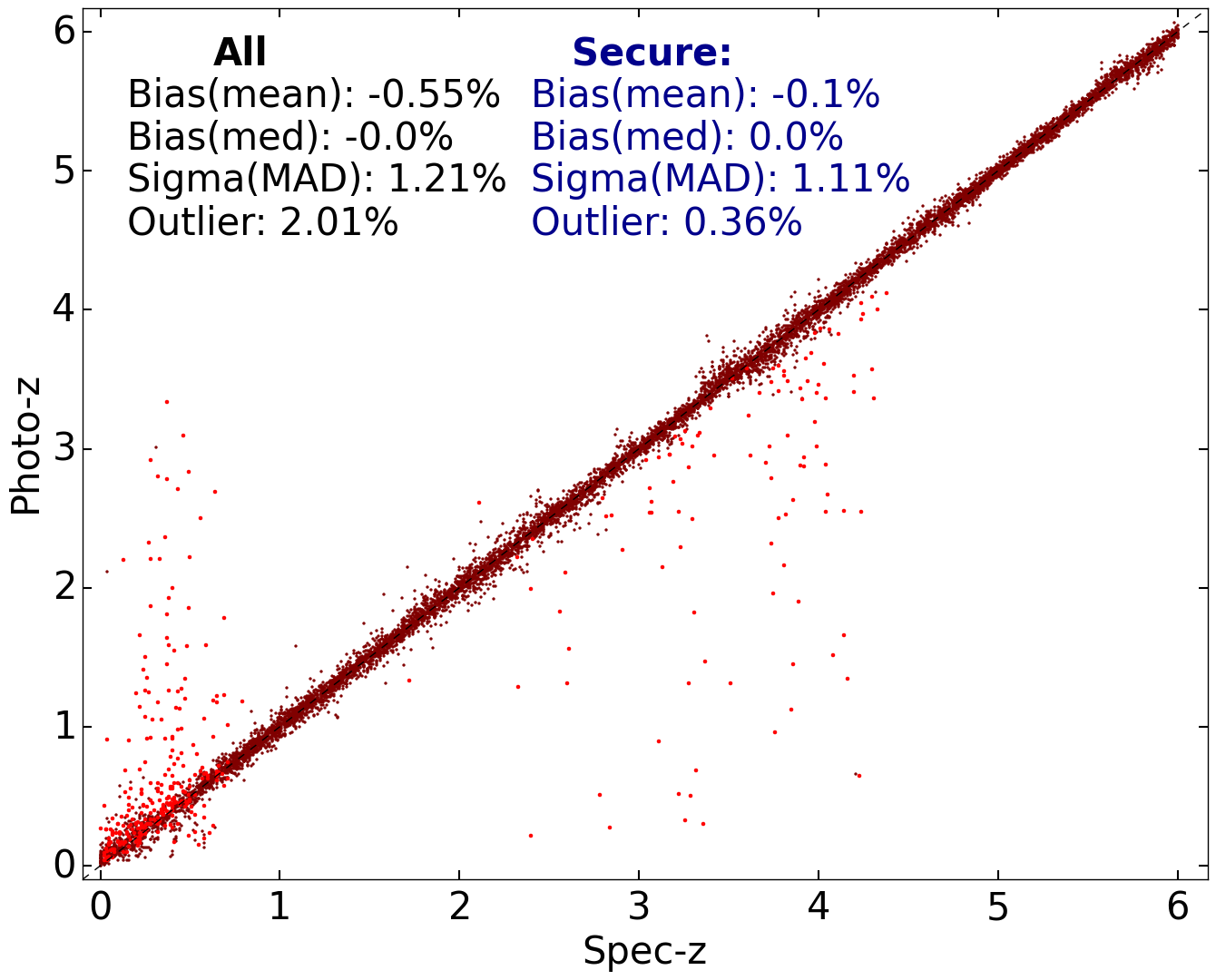}}
	\qquad
	\subfloat[][{\tt{SOM\_Hierarchical\_ImportanceSampling}}]{\includegraphics[scale=0.25]{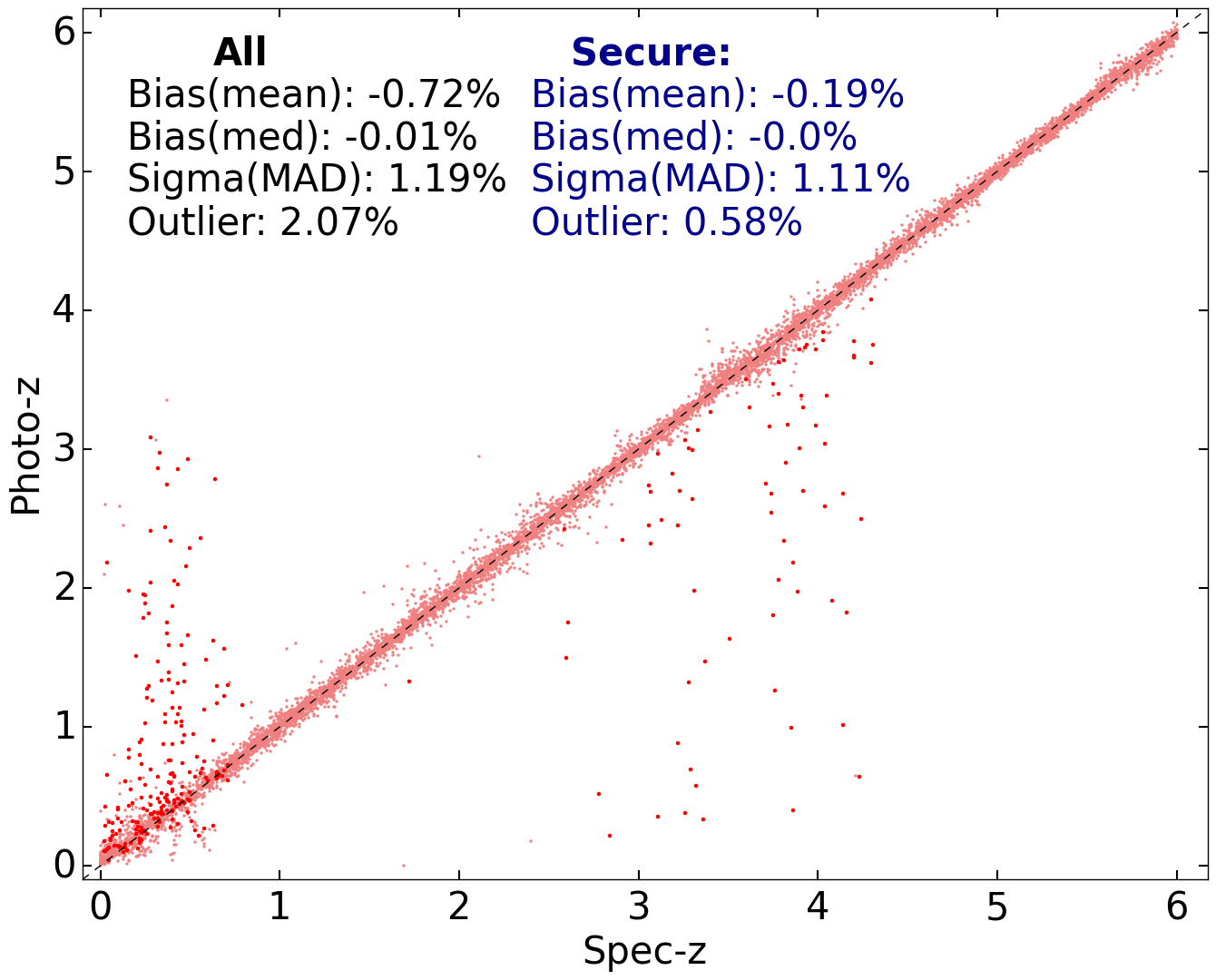}}\\
	\subfloat[][{\tt{SOM\_CellModel\_LimitedSum}}]{\includegraphics[scale=0.25]{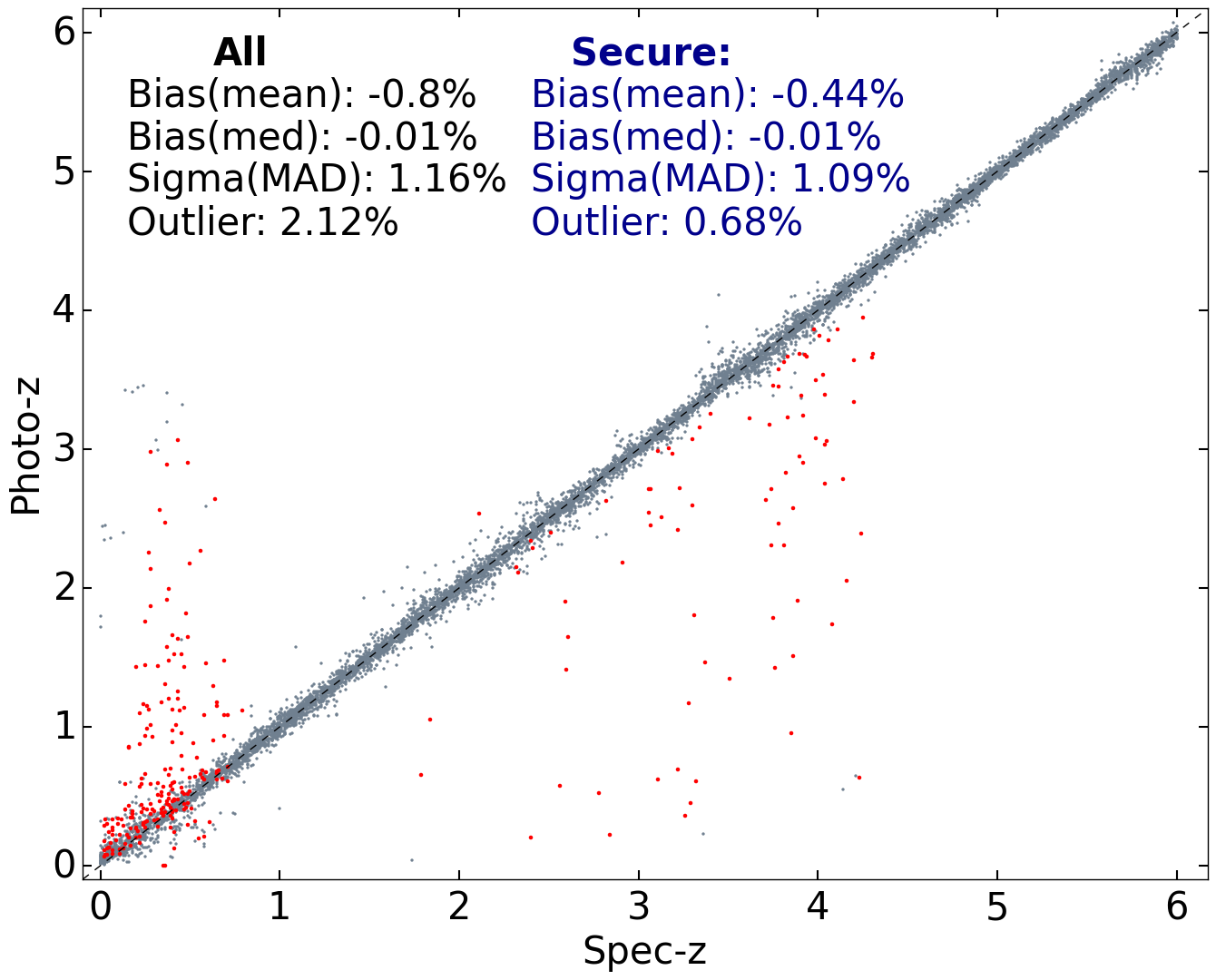}}
	\qquad
	\subfloat[][{\tt{SOM\_CellModel\_Average}}]{\includegraphics[scale=0.25]{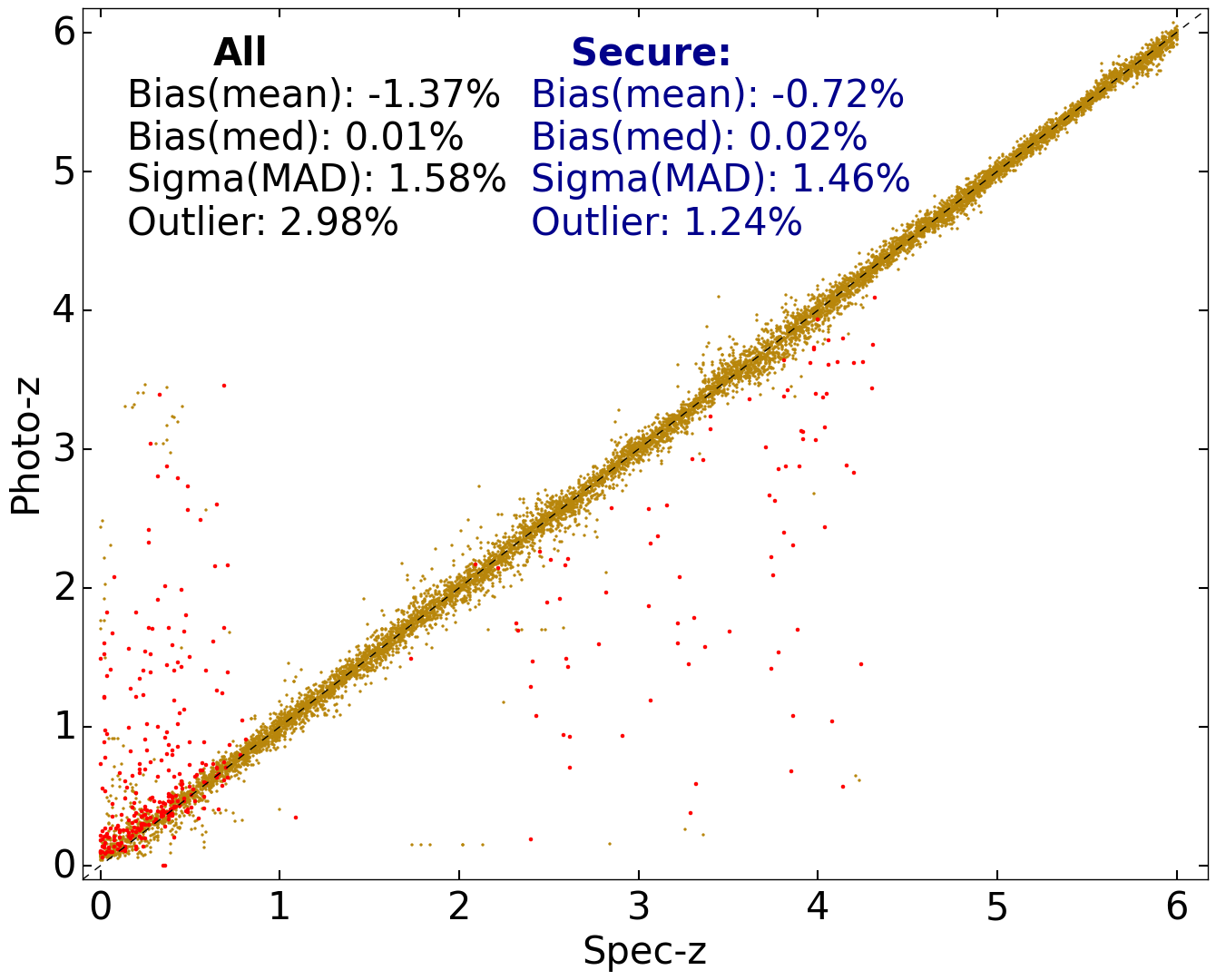}}
	\caption{
		As Figure~\ref{fig:photz}, but for our {\tt{SOM\_Hierarchical\_MonteCarlo}} (dark red), {\tt{SOM\_Hierarchical\_ImportanceSampling}} (pink), {\tt{SOM\_CellModel\_LimitedSum}} (gray), and {\tt{SOM\_CellModel\_Average}} (dark yellow) methods (see \S\ref{sec:methods}). 
		While all of our methods are able to reproduce the one-to-one relationship between spec-z and photo-z quite well with $\sigma_{z,\textrm{MAD}}^\prime\lesssim0.015$, our  {\tt{SOM\_CellModel\_LimitedSum}} and {\tt{SOM\_CellModel\_Average}} approaches consistently fail to identify unreliable photo-z's due to subtle biases introduced from their reliance on SOM cell models. Once insecure photo-z's have been removed, we find the majority of methods achieve accuracies comparable to our brute-force approaches with $|\bar{\Delta z}^\prime|\approx0.5\%$, $\sigma_{z,\textrm{MAD}}^\prime\approx1.1\%$, and catastrophic outlier rates of $\lesssim 0.4$\%, with our {\tt{BruteForce\_LinearFuzzy}}, {\tt{SOM\_MCMC\_ObservedFrame}}, and {\tt{SOM\_Hierarchical\_MonteCarlo}} giving the best overall results.}
	\label{fig:photz_2}
\end{figure*}

\begin{figure*}
	\centering
	\captionsetup[subfigure]{labelformat=empty}
	\subfloat[][{\tt{BruteForce}} (gold standard)]{\includegraphics[scale=0.25]{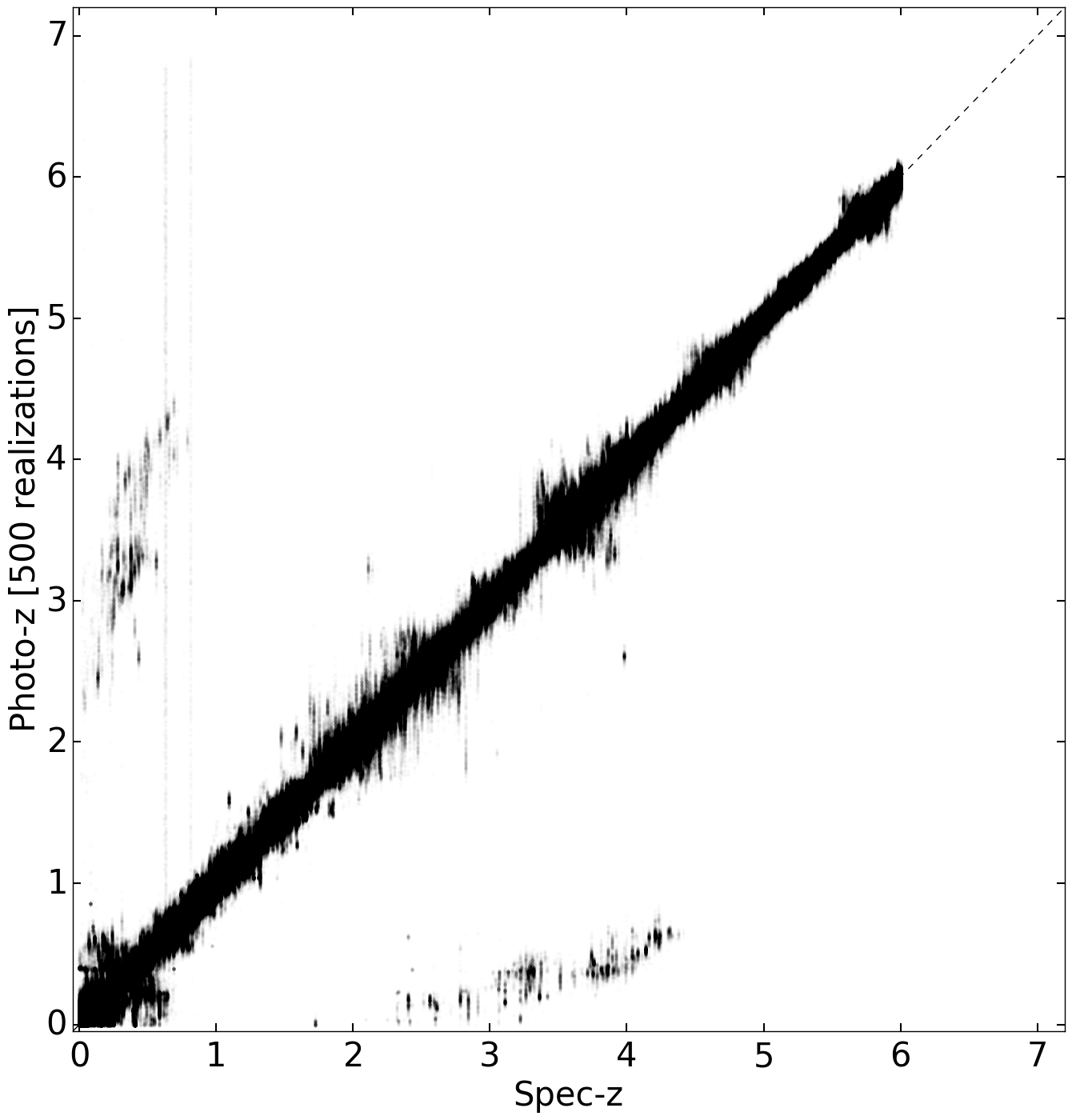}}
	\qquad
	\subfloat[][{\tt{BruteForce\_LinearFuzzy}}]{\includegraphics[scale=0.25]{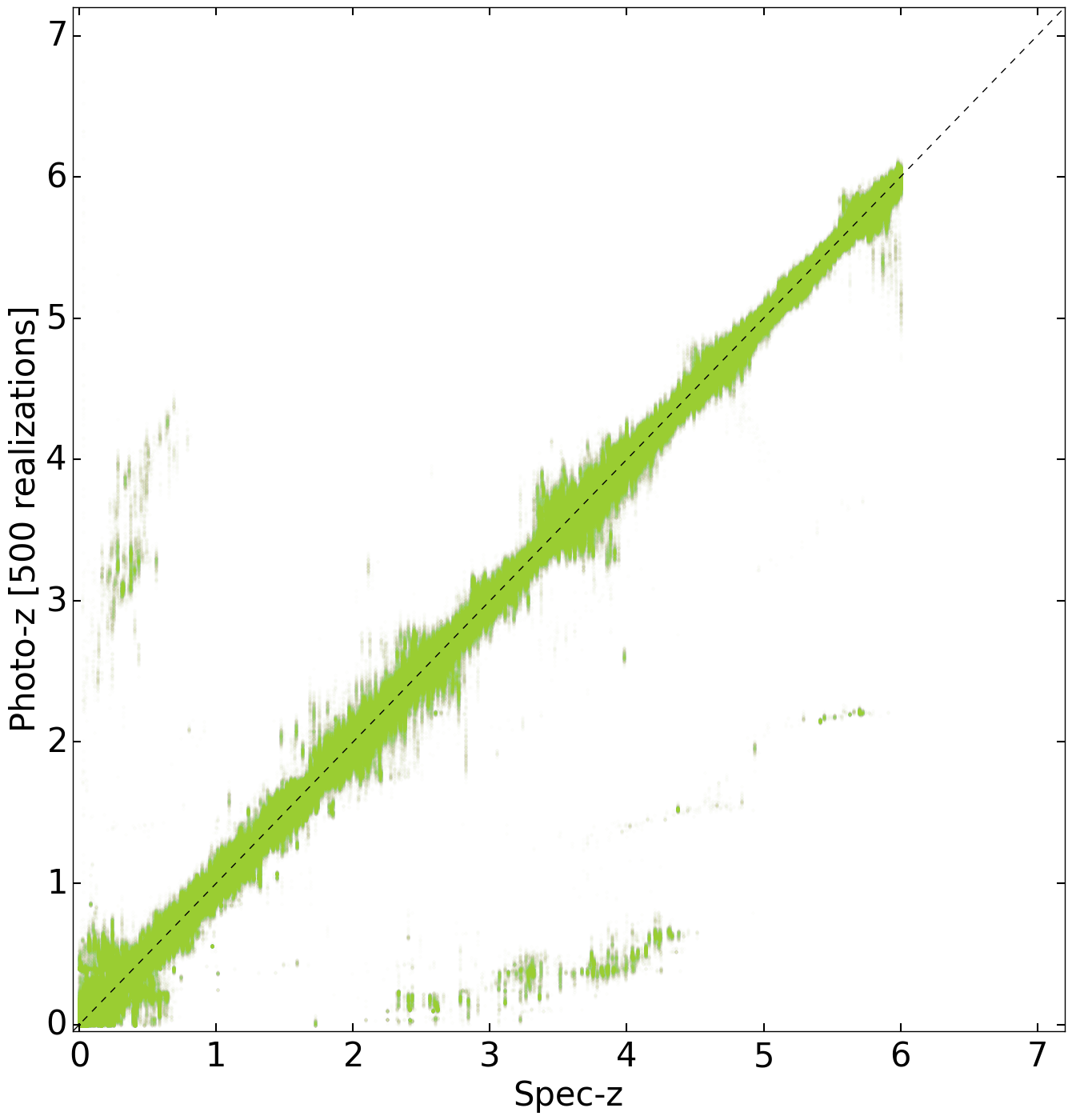}}\\
	\subfloat[][{\tt{SOM\_MCMC\_ObservedFrame}}]{\includegraphics[scale=0.25]{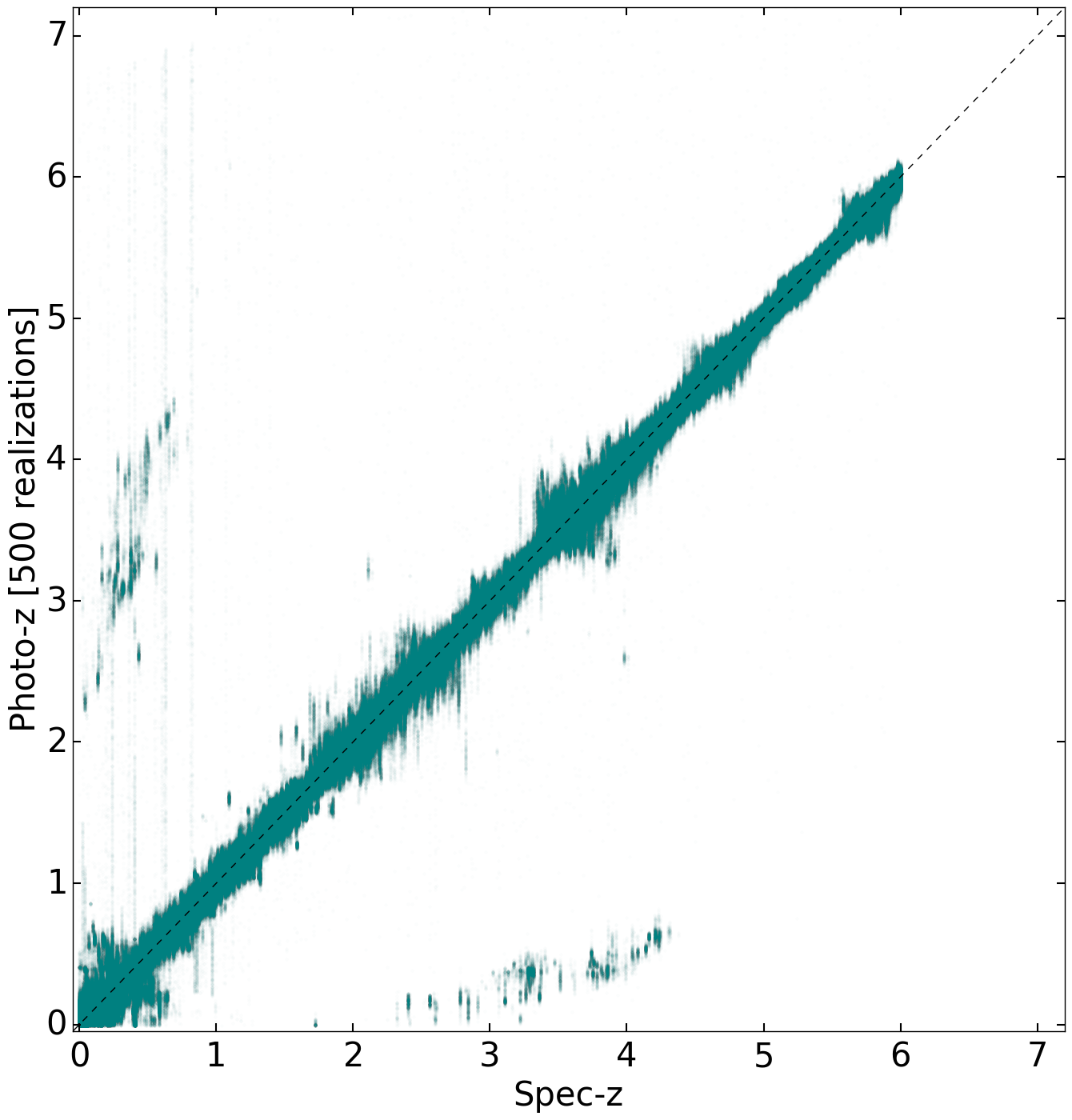}}
	\qquad
	\subfloat[][{\tt{SOM\_MCMC\_RestFrame}}]{\includegraphics[scale=0.25]{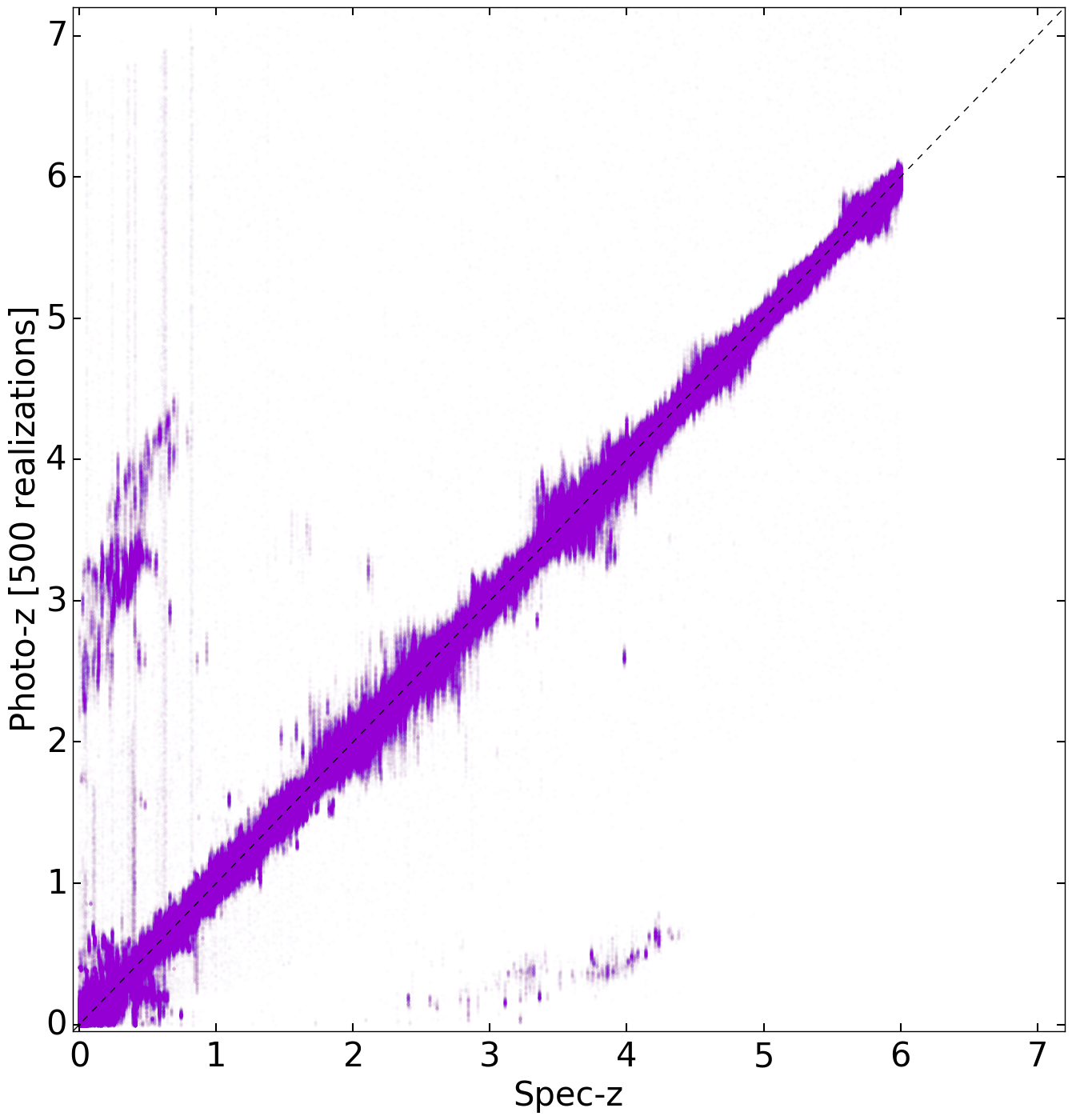}}
	\caption{
		The two-dimensional stacked $P(z|\rom{\mathbf{F}}{obs})$ distributions for our {\tt{BruteForce}} (gold standard), {\tt{BruteForce\_LinearFuzzy}}, {\tt{SOM\_MCMC\_RestFrame}}, and {\tt{SOM\_MCMC\_ObservedFrame}} methods (see \S\ref{sec:methods}) generated using 500 Monte Carlo samples drawn from the redshift PDF of each object and plotted using the same color scheme as Figure~\ref{fig:photz}. 
		See Figure~\ref{fig:photz_pzstack_2} for more details.}
	\label{fig:photz_pzstack}
\end{figure*}

\begin{figure*}
	\centering
	\captionsetup[subfigure]{labelformat=empty}
	\subfloat[][{\tt{SOM\_Hierarchical\_MonteCarlo}}]{\includegraphics[scale=0.25]{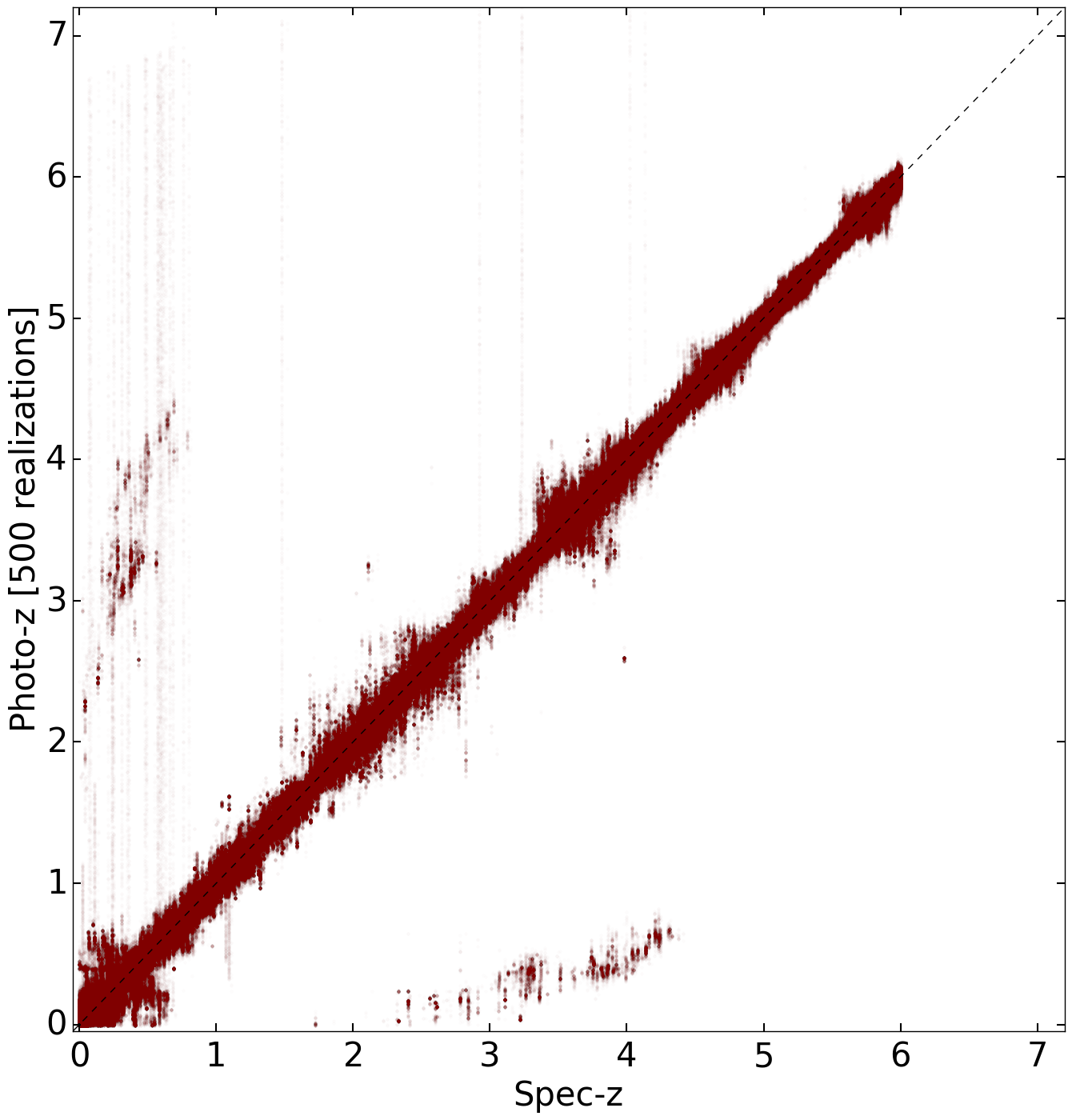}}
	\qquad
	\subfloat[][{\tt{SOM\_Hierarchical\_ImportanceSampling}}]{\includegraphics[scale=0.25]{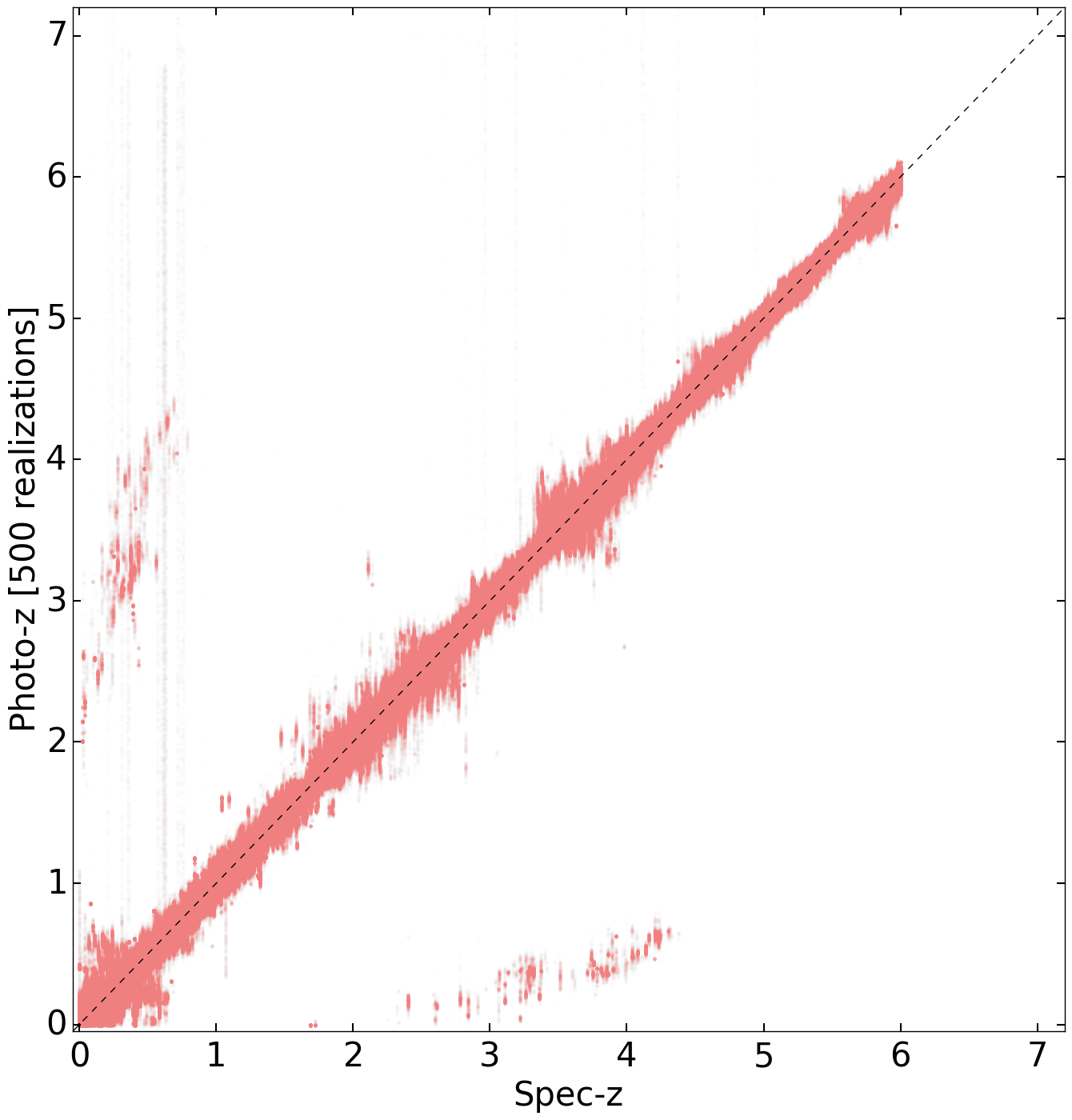}}\\
	\subfloat[][{\tt{SOM\_CellModel\_LimitedSum}}]{\includegraphics[scale=0.25]{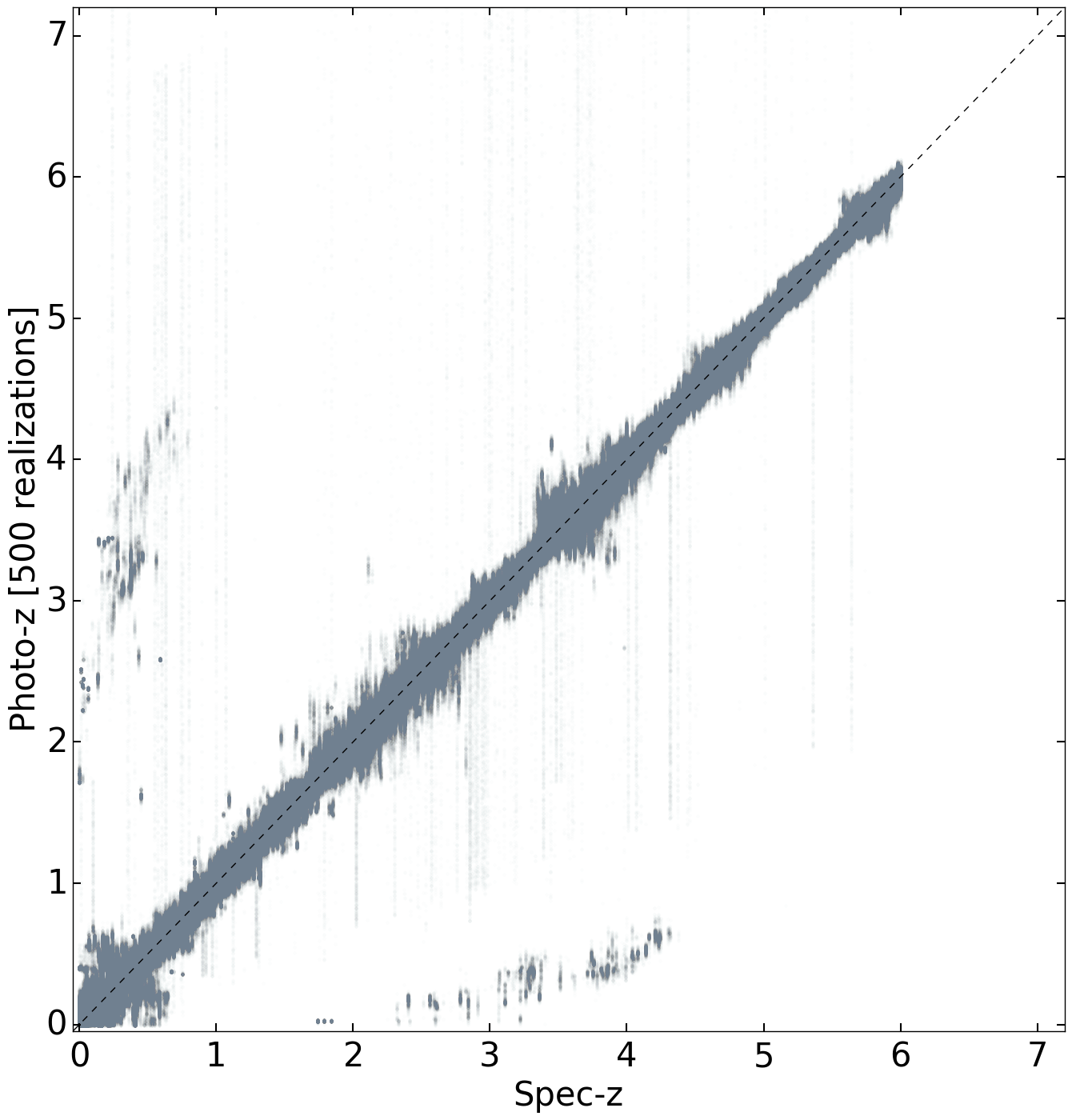}}
	\caption{
		As Figure~\ref{fig:photz_pzstack}, but for our {\tt{SOM\_Hierarchical\_MonteCarlo}}, {\tt{SOM\_Hierarchical\_ImportanceSampling}}, and {\tt{SOM\_CellModel\_LimitedSum}} methods (see \S\ref{sec:methods}). 
		All methods recover the general degeneracies present across the objects in our catalog -- including redshift-reddening degeneracies and confusion between the 1216\,{\AA} and 4000\,{\AA} breaks. We find that the increased flexibility of {\tt{BruteForce\_LinearFuzzy}} not only sharpens degeneracies as compared to {\tt{BruteForce}} but also creates new ones, such as a thin band of probability at lower redshift for sources at $z \sim 4$\,--\,$6$. In addition, we confirm that the large band of flagged outliers at low redshift from {\tt{SOM\_MCMC\_RestFrame}} (see Figure~\ref{fig:photz}) is due to an overabundance of probability at higher redshifts for low-$z$ sources, illustrating the impact of the redshift gradient noted in Speagle \& Eisenstein (2015a). Finally, while {\tt{SOM\_Hierarchical\_MonteCarlo}} gives good overall performance, its PDFs are noticeably ``rougher'' than the other approaches, likely due to sparse sampling in the small fraction of cells where the models with the highest likelihoods are contained.}
	\label{fig:photz_pzstack_2}
\end{figure*}

\begin{figure*}
	\centering
	\captionsetup[subfigure]{labelformat=empty}
	\subfloat[][{\tt{BruteForce}} (gold standard)]{\includegraphics[scale=0.19]{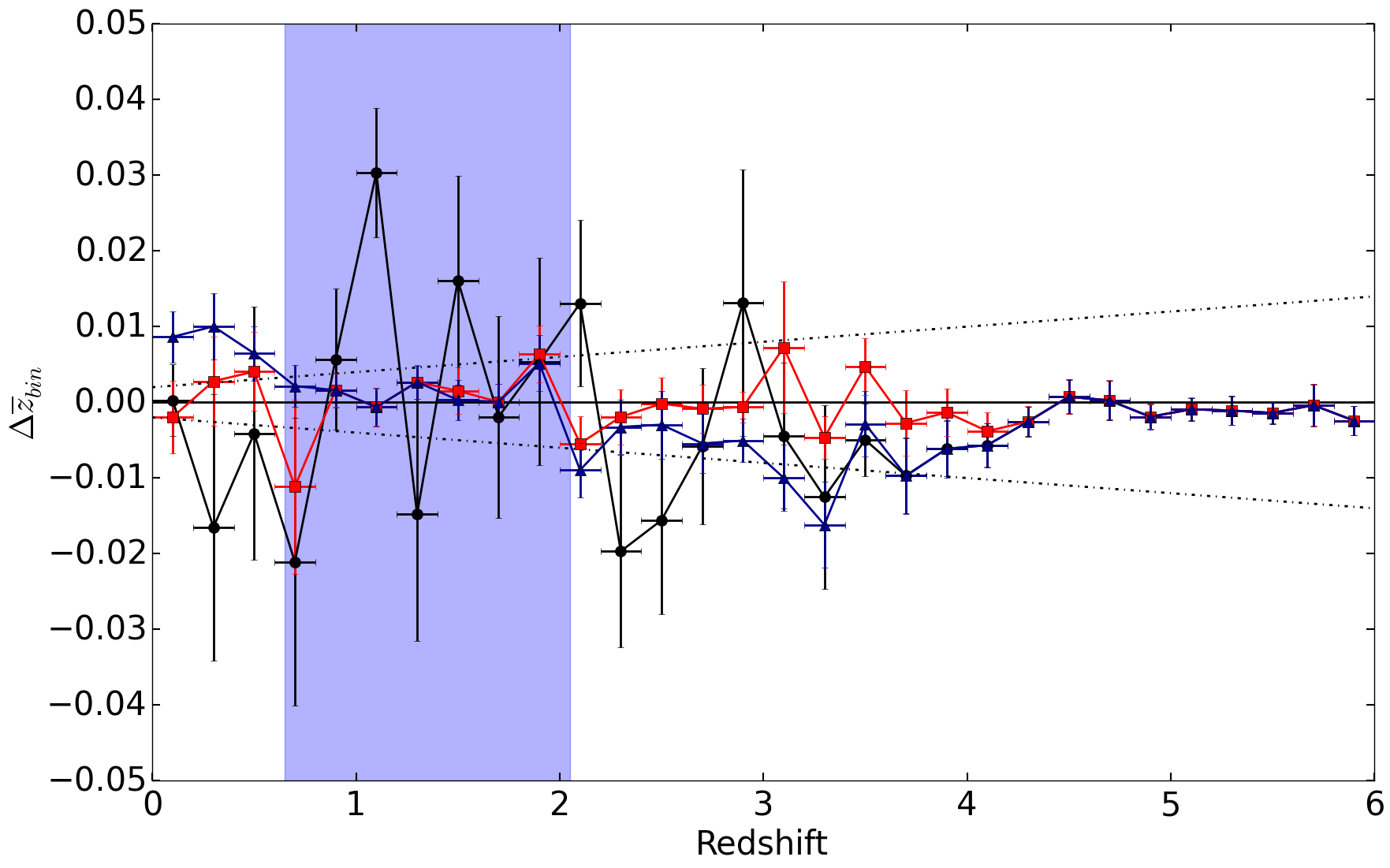}}
	\qquad
	\subfloat[][{\tt{BruteForce\_LinearFuzzy}}]{\includegraphics[scale=0.19]{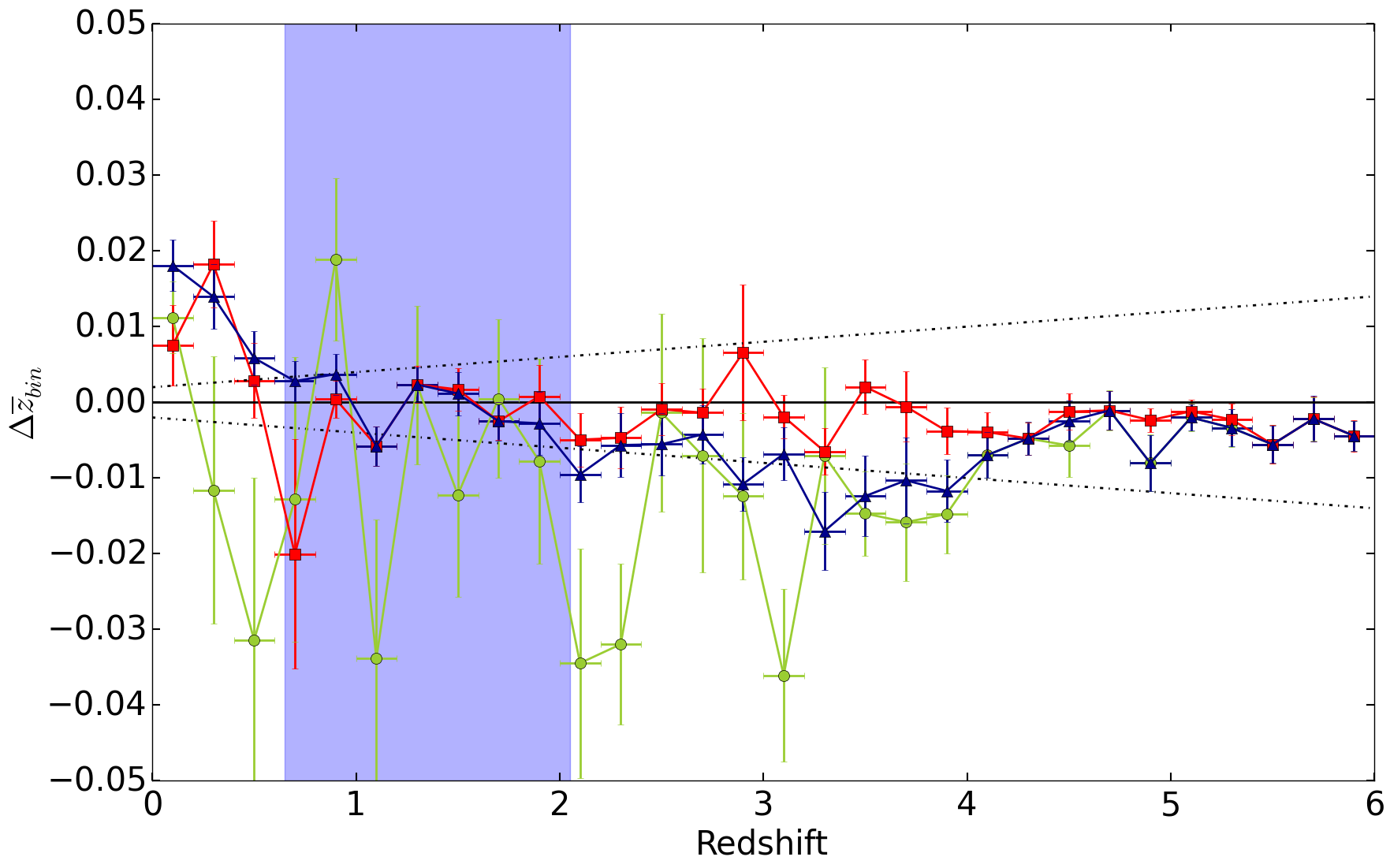}}\\
	\subfloat[][{\tt{SOM\_MCMC\_RestFrame}}]{\includegraphics[scale=0.19]{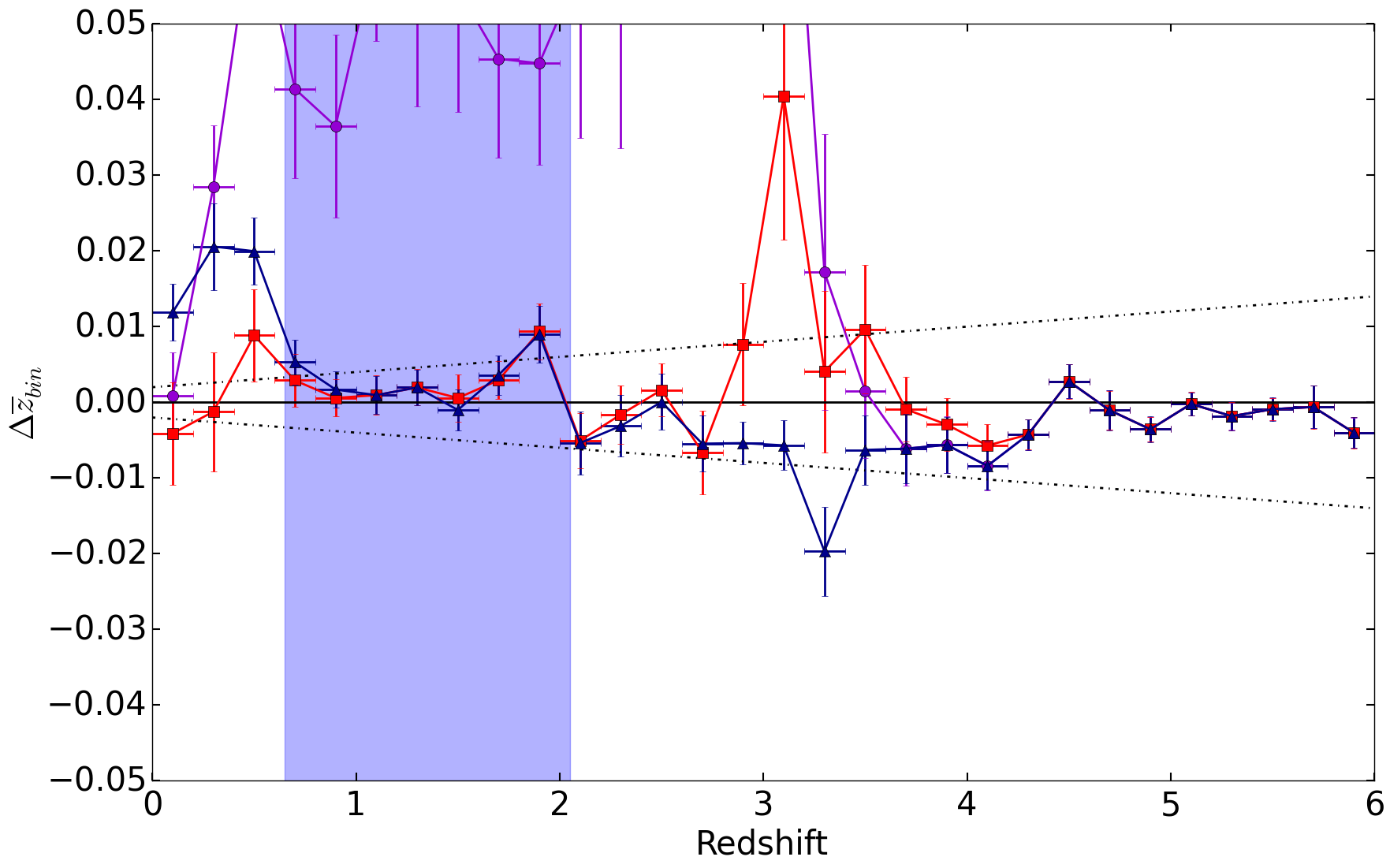}}
	\qquad
	\subfloat[][{\tt{SOM\_MCMC\_ObservedFrame}}]{\includegraphics[scale=0.19]{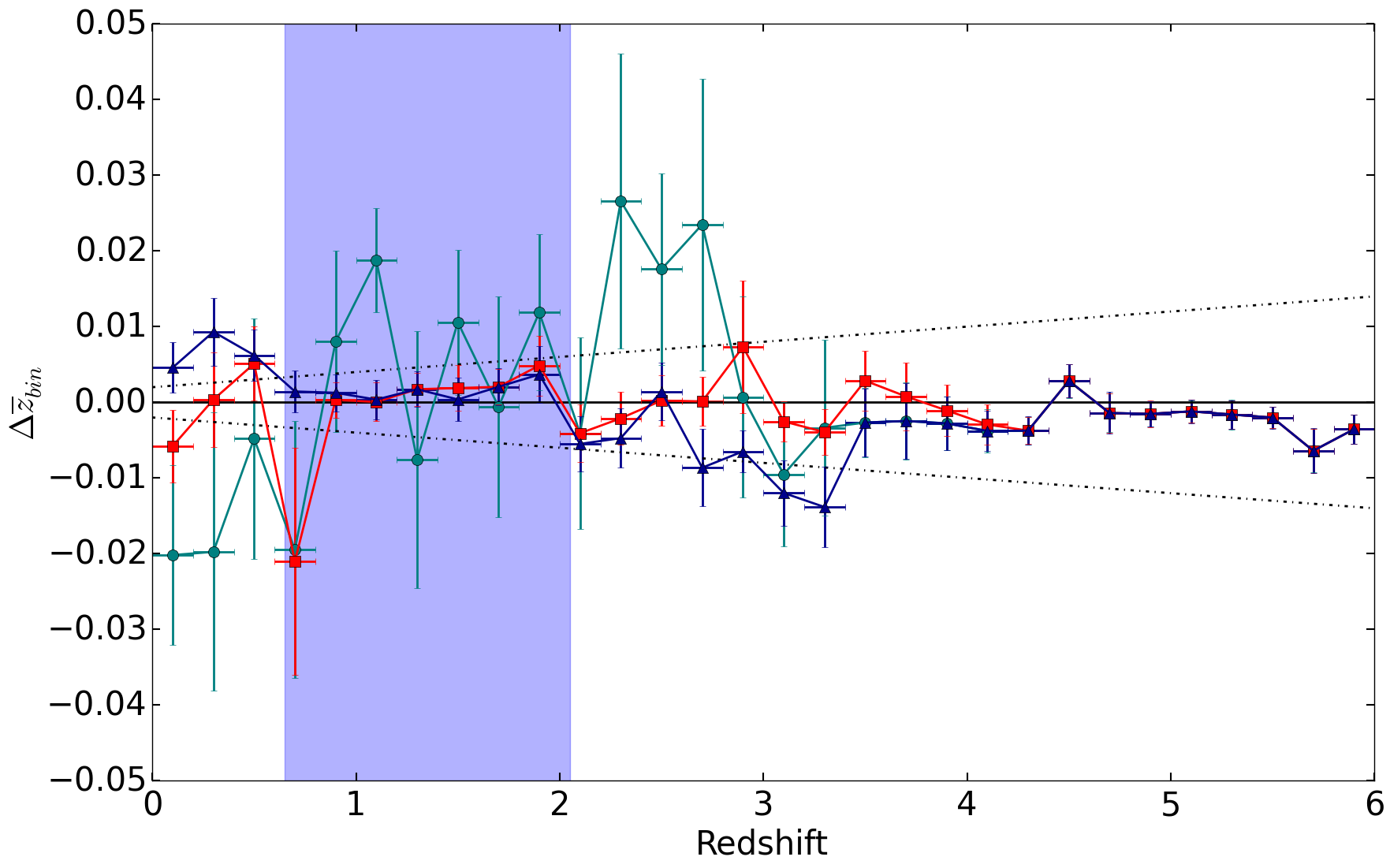}}
	\caption{
		The mean redshift bias $\rom{\Delta \bar{z}}{bin} = \langle\rom{z}{spec}\rangle_{\textrm{bin}} - \langle\rom{\bar{z}}{phot}\rangle_{\textrm{bin}}$ in a given $\rom{\bar{z}}{phot}$-selected redshift tomographic bin, plotted as a function of redshift for our {\tt{BruteForce}} (gold standard), {\tt{BruteForce\_LinearFuzzy}}, {\tt{SOM\_MCMC\_RestFrame}}, and {\tt{SOM\_MCMC\_ObservedFrame}} methods (see \S\ref{sec:methods}). The results for the full sample are colored using the same color scheme as Figure~\ref{fig:photz}, while those from the secure subsample are shown in red. For comparison, the results from a ``cleaned'' subsample where all catastrophic outliers (i.e. objects with $\Delta z^\prime > 0.15$) have been removed are shown in blue. Errors for each sample have been derived through bootstrap resampling, with 1$\sigma$ errors plotted. The \textit{Euclid} photo-z accuracy requirements for weak-lensing of $\Delta \bar{z} < 0.002(1+z)$ from \citet{laureijs+11} are over-plotted as dotted-dashed black lines, with the relevant redshift range highlighted in light blue.
		See Figure~\ref{fig:photz_zbias_2} for more details.}
	\label{fig:photz_zbias}
\end{figure*}

\begin{figure*}
	\centering
	\captionsetup[subfigure]{labelformat=empty}
	\subfloat[][{\tt{SOM\_Hierarchical\_MonteCarlo}}]{\includegraphics[scale=0.19]{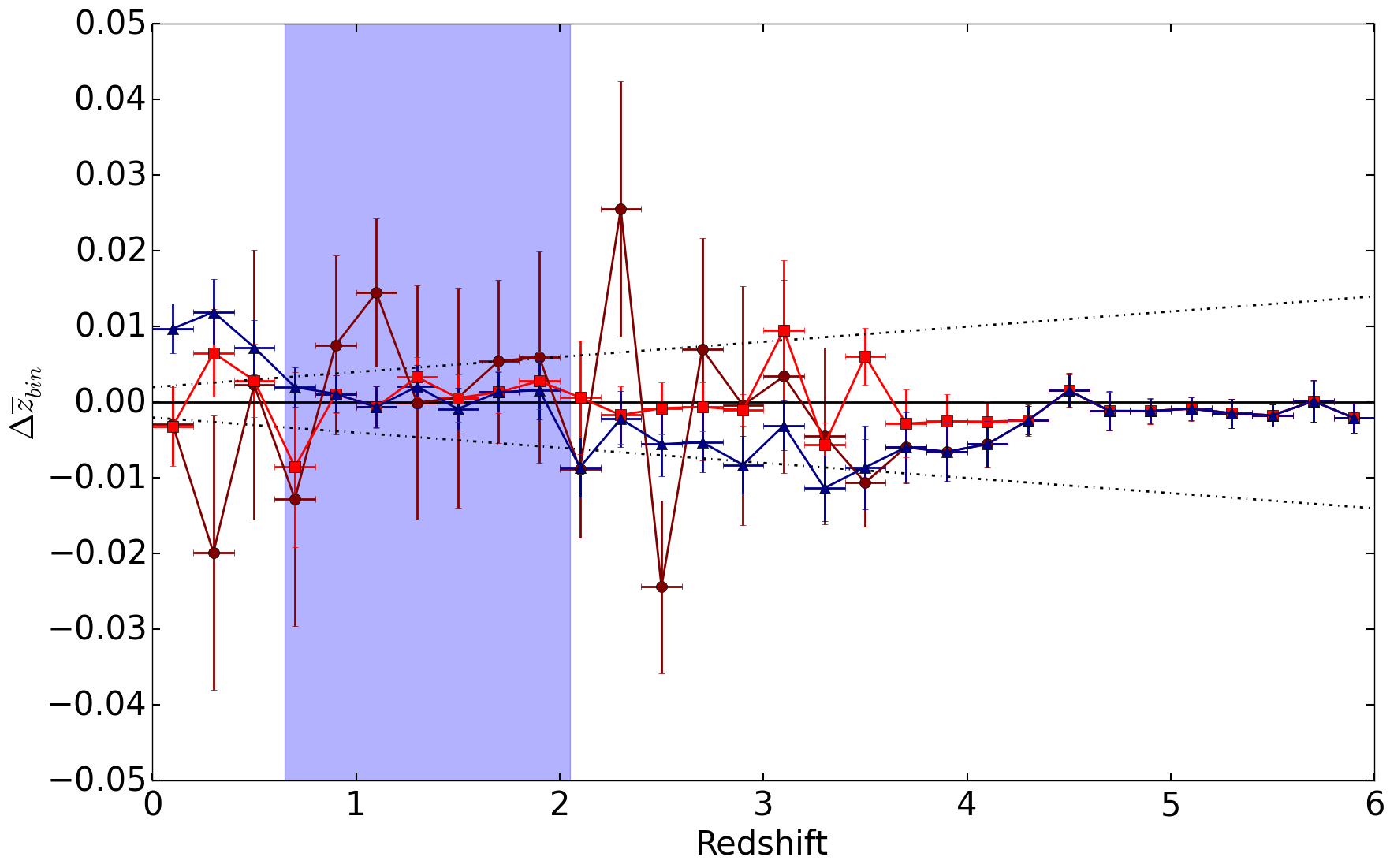}}
	\qquad
	\subfloat[][{\tt{SOM\_Hierarchical\_ImportanceSampling}}]{\includegraphics[scale=0.19]{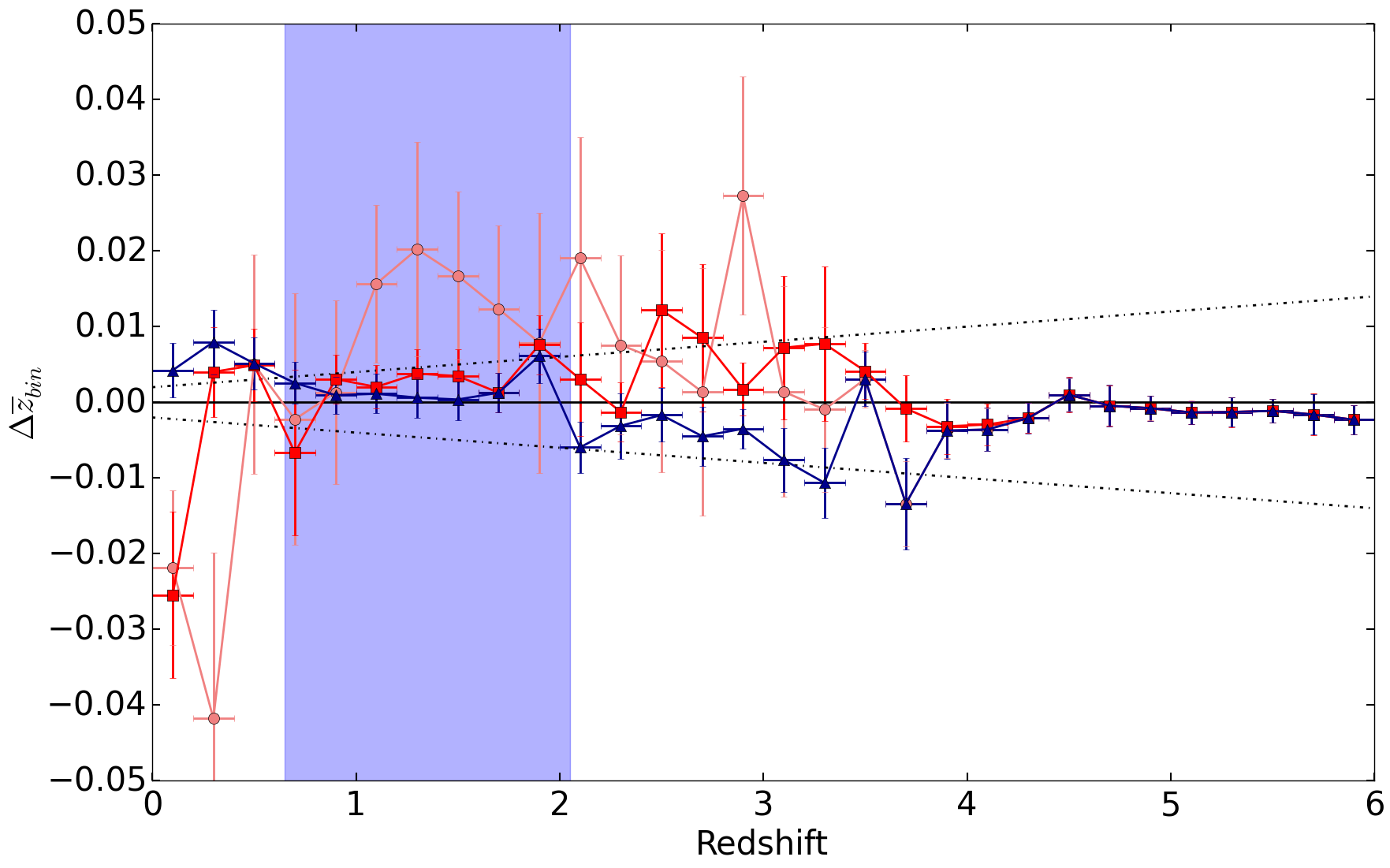}}\\
	\subfloat[][{\tt{SOM\_CellModel\_LimitedSum}}]{\includegraphics[scale=0.19]{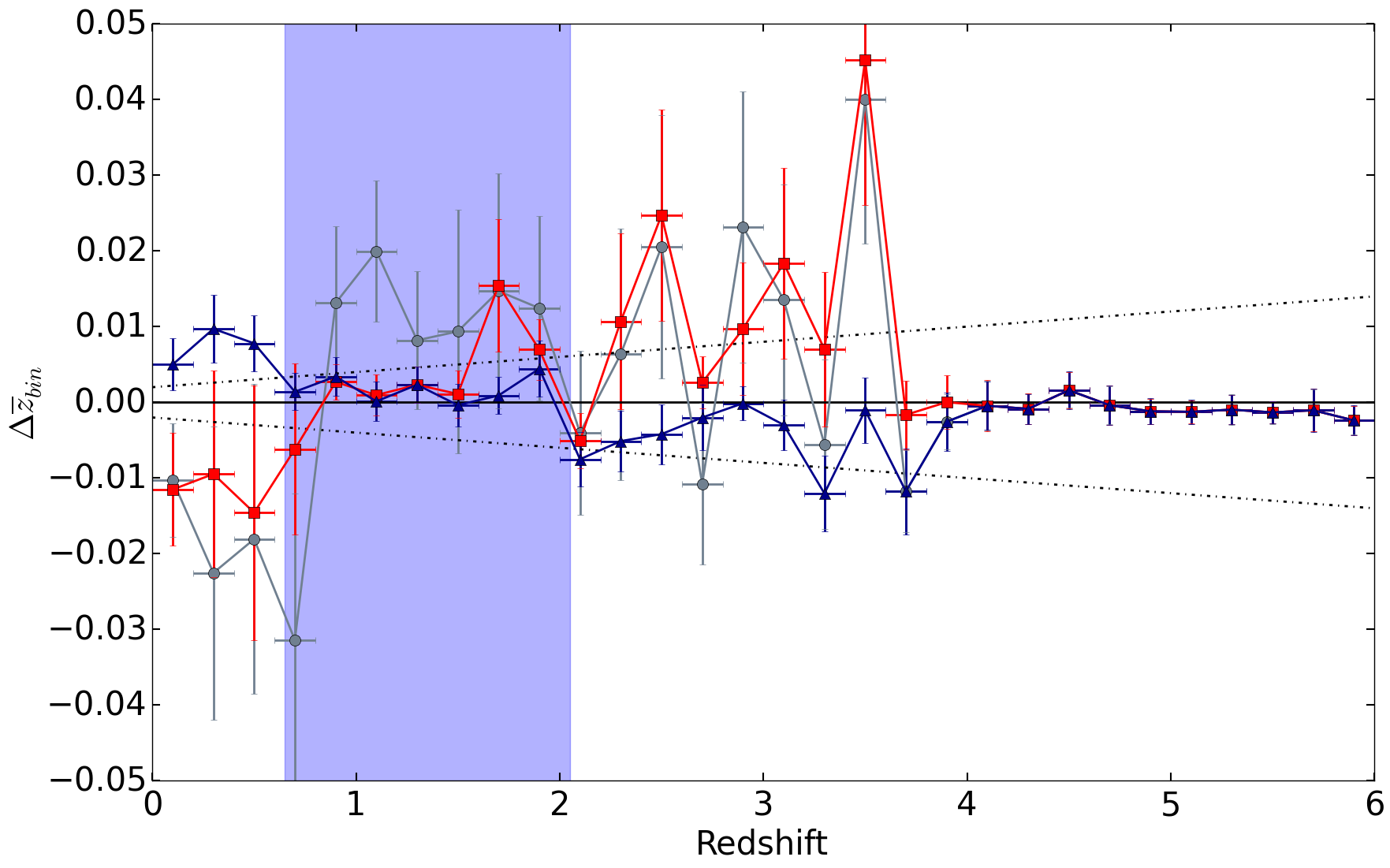}}
	\qquad
	\subfloat[][{\tt{SOM\_CellModel\_Average}}]{\includegraphics[scale=0.19]{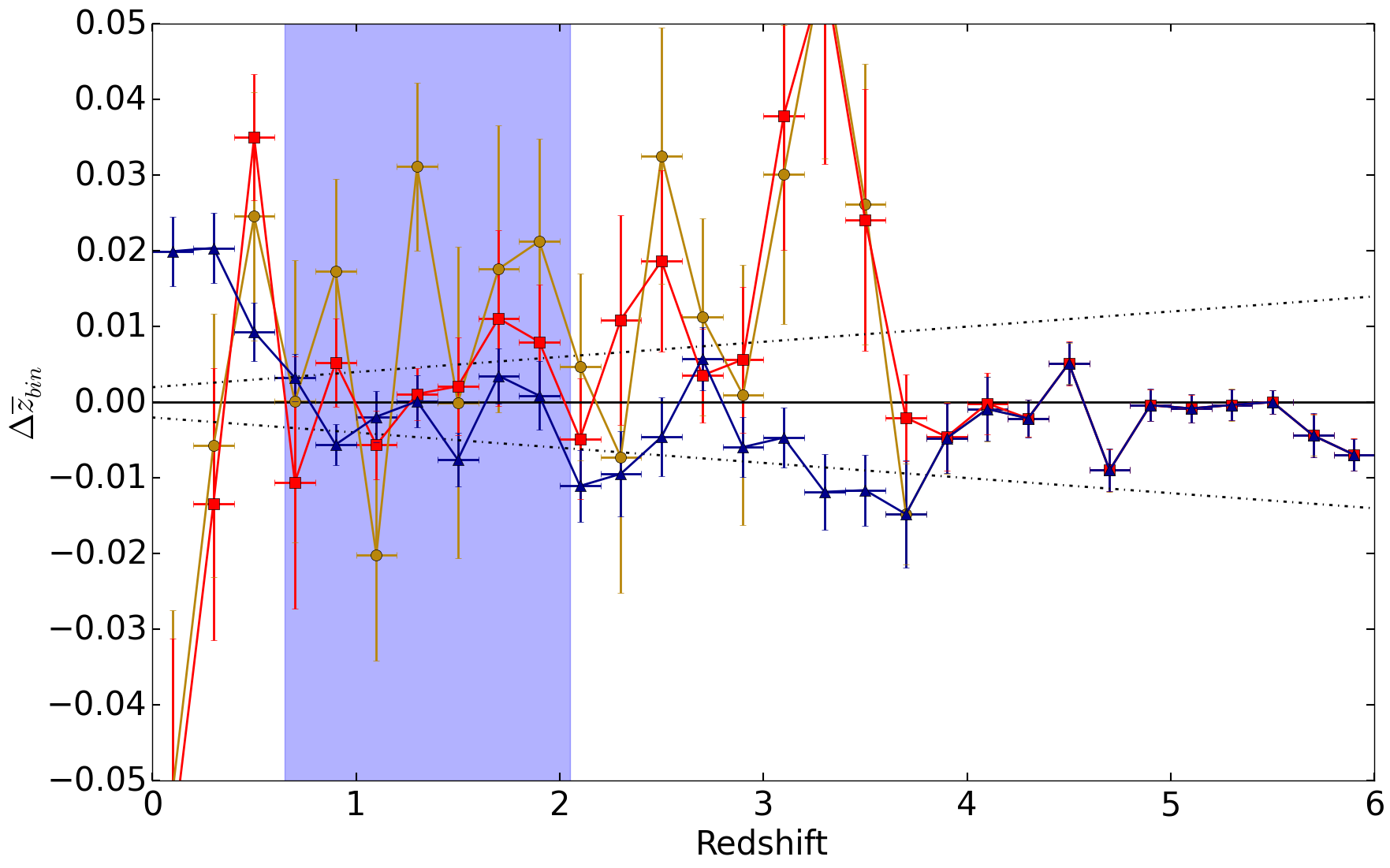}}
	\caption{
		As Figure~\ref{fig:photz_zbias}, but for our {\tt{SOM\_Hierarchical\_MonteCarlo}}, {\tt{SOM\_Hierarchical\_ImportanceSampling}}, {\tt{SOM\_CellModel\_LimitedSum}}, and {\tt{SOM\_CellModel\_Average}} methods (see \S\ref{sec:methods}). 
		For the full sample, numerous outliers lead to significant biases for most redshift bins below $z \lesssim 4$. While variation of $\rom{\Delta \bar{z}}{bin}(\rom{z}{bin})$ appears to be symmetrical for the {\tt{BruteForce}} approach and the sparse-sampling {\tt{SOM\_Hierarchical\_MonteCarlo}} approaches, in the majority of cases the overabundance of up-scattered redshifts tend to drive variation in $\rom{\Delta \bar{z}}{bin}(\rom{z}{bin})$ to be positive (with the exception of {\tt{BruteForce\_LinearFuzzy}}, which introduces new low-$z$ degeneracies; see Figure~\ref{fig:photz_pzstack}). After restricting to secure photo-z's only, the bias decreases substantially (except for {\tt{SOM\_CellModel}}-based approaches; see Figure~\ref{fig:photz_2}), with the majority of tomographic bins for the secure subsample meeting the \textit{Euclid} photo-z accuracy requirements above $z \sim 0.8$.}
	\label{fig:photz_zbias_2}
\end{figure*}

\begin{figure*}
	\centering
	\captionsetup[subfigure]{labelformat=empty}
	\subfloat[][{\tt{BruteForce}} (gold standard)]{\includegraphics[scale=0.19]{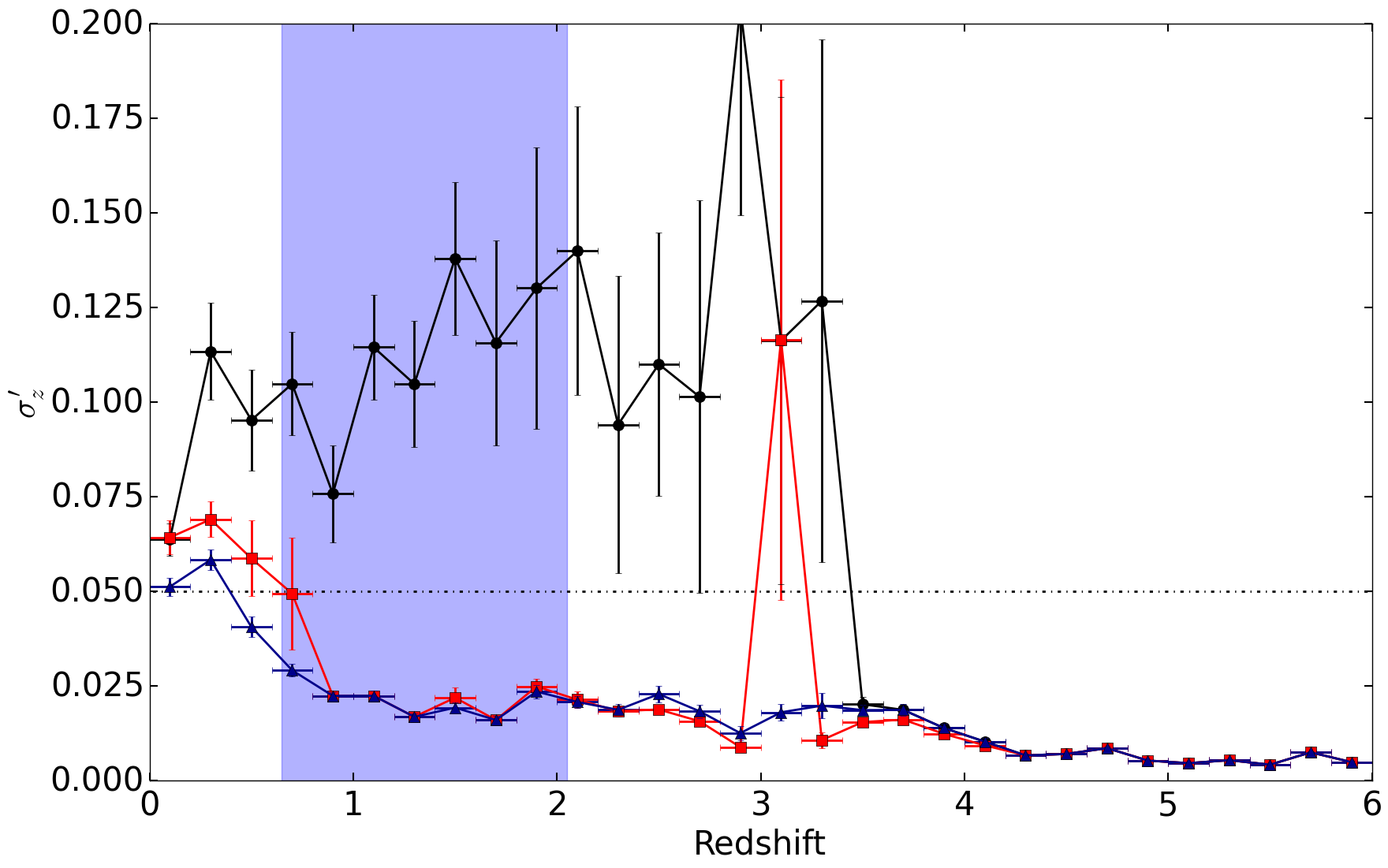}}
	\qquad
	\subfloat[][{\tt{BruteForce\_LinearFuzzy}}]{\includegraphics[scale=0.19]{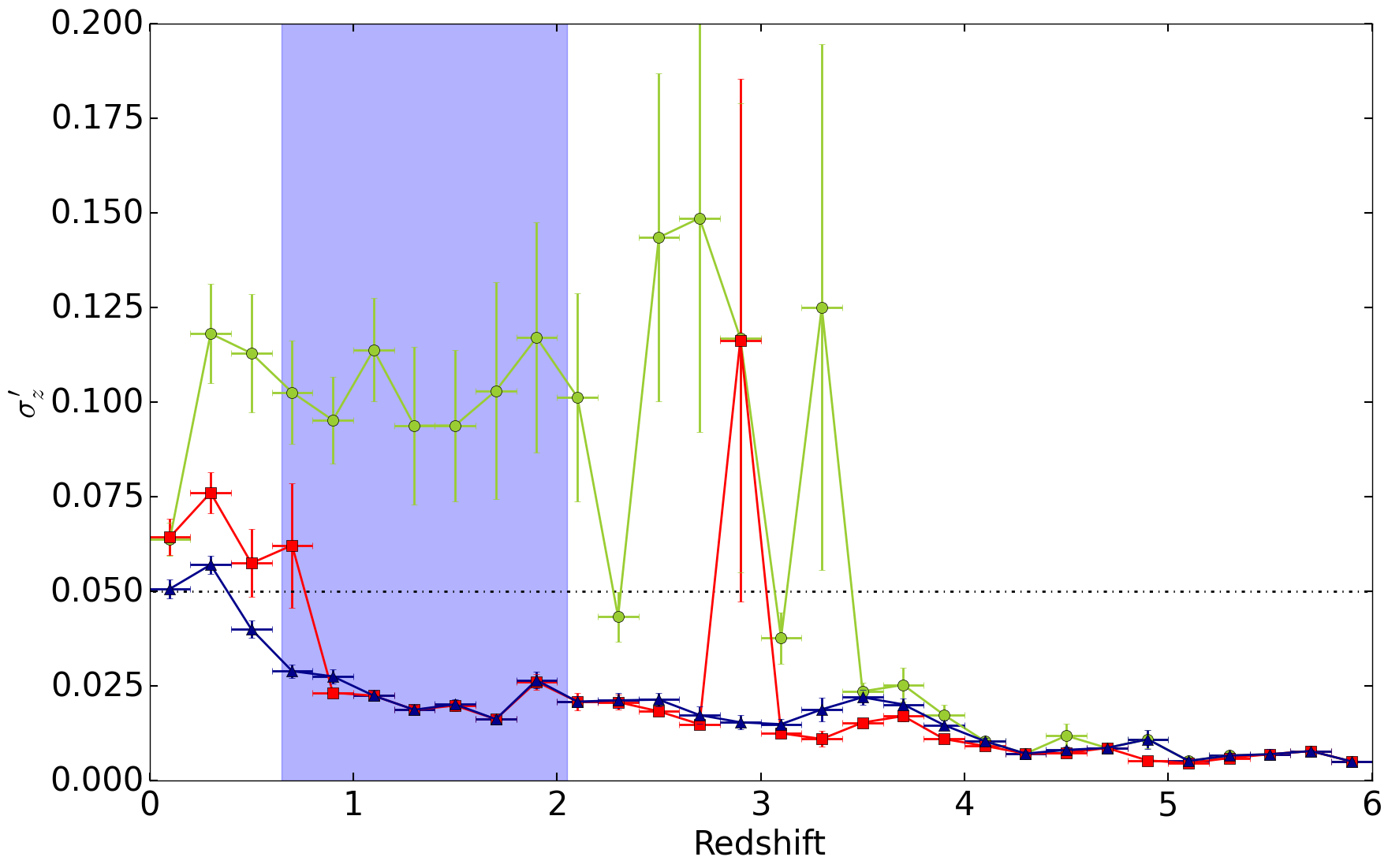}}\\
	\subfloat[][{\tt{SOM\_MCMC\_RestFrame}}]{\includegraphics[scale=0.19]{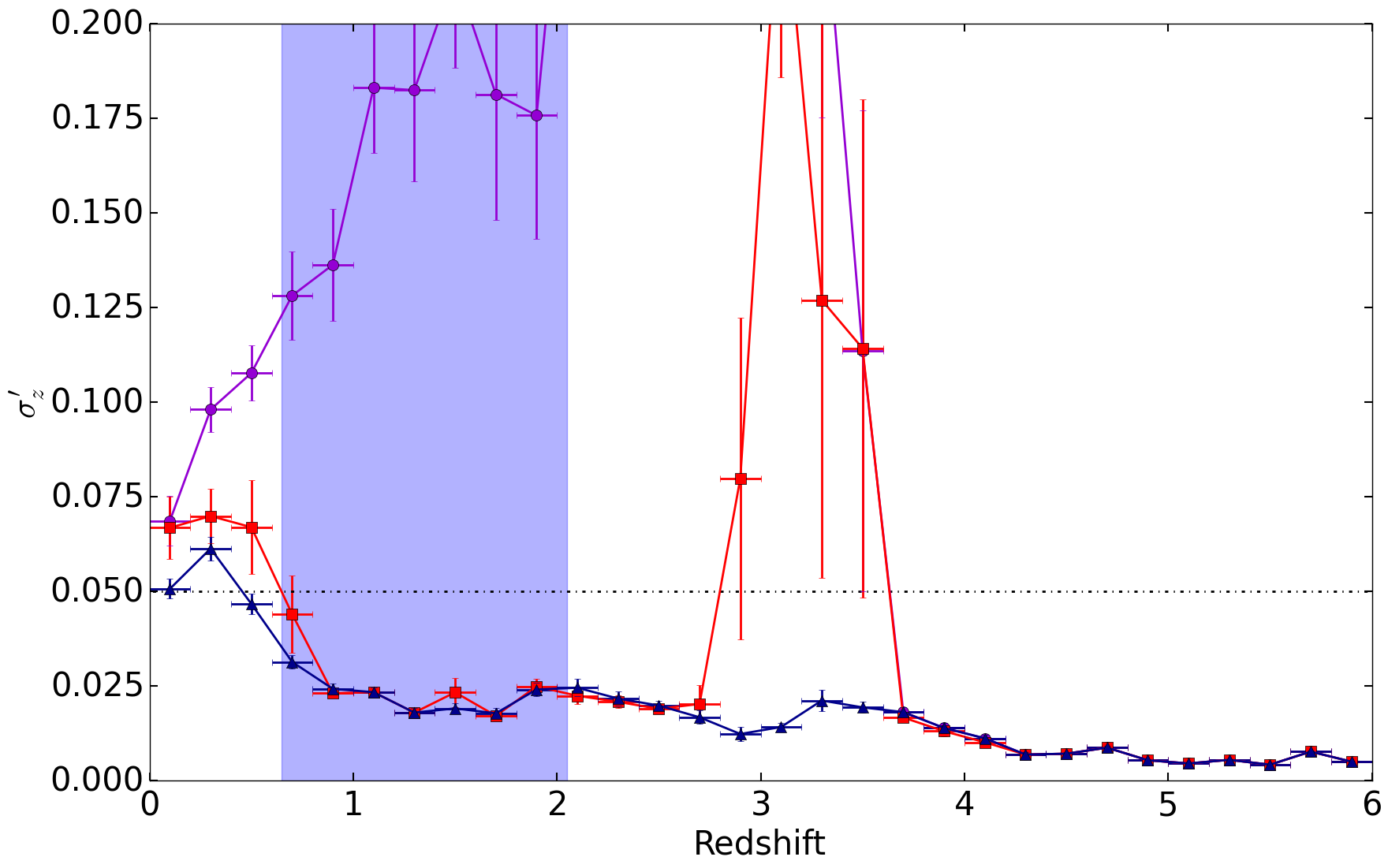}}
	\qquad
	\subfloat[][{\tt{SOM\_MCMC\_ObservedFrame}}]{\includegraphics[scale=0.19]{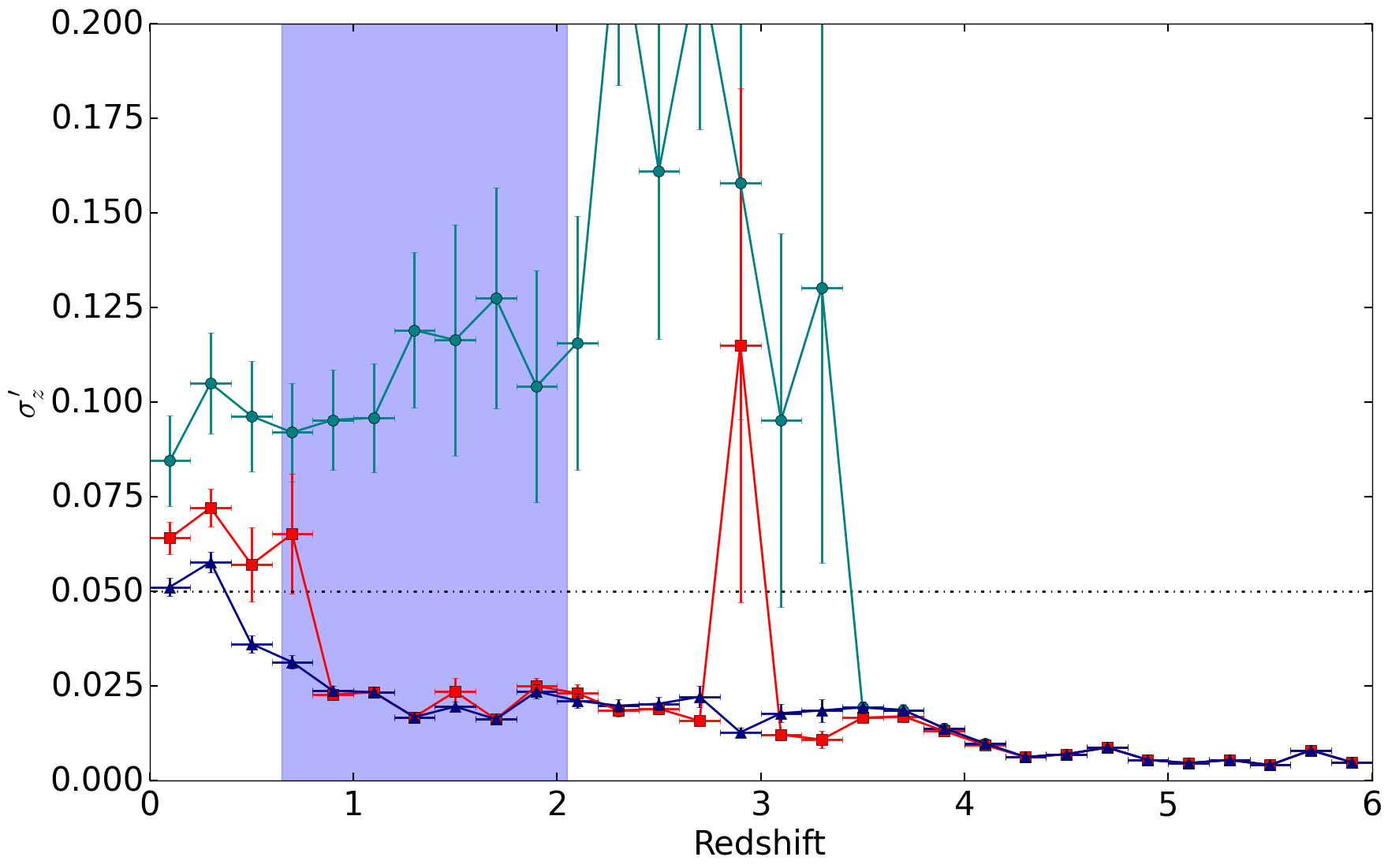}}
	\caption{
		As Figure~\ref{fig:photz_zbias}, but for the redshift-normalized mean scatter $\sigma(\Delta z^\prime)$ for our {\tt{BruteForce}}, {\tt{BruteForce\_LinearFuzzy}}, {\tt{SOM\_MCMC\_RestFrame}}, and {\tt{SOM\_MCMC\_ObservedFrame}} methods (see \S\ref{sec:methods}). The \textit{Euclid} photo-z accuracy requirements for weak-lensing of $\sigma_z^\prime < 0.05$ from \citet{laureijs+11} are over-plotted as dotted-dashed black lines, with the relevant redshift range highlighted in light blue.
		See Figure~\ref{fig:photz_etastd_2} for more details.}
	\label{fig:photz_etastd}
\end{figure*}

\begin{figure*}
	\centering
	\captionsetup[subfigure]{labelformat=empty}
	\subfloat[][{\tt{SOM\_Hierarchical\_MonteCarlo}}]{\includegraphics[scale=0.19]{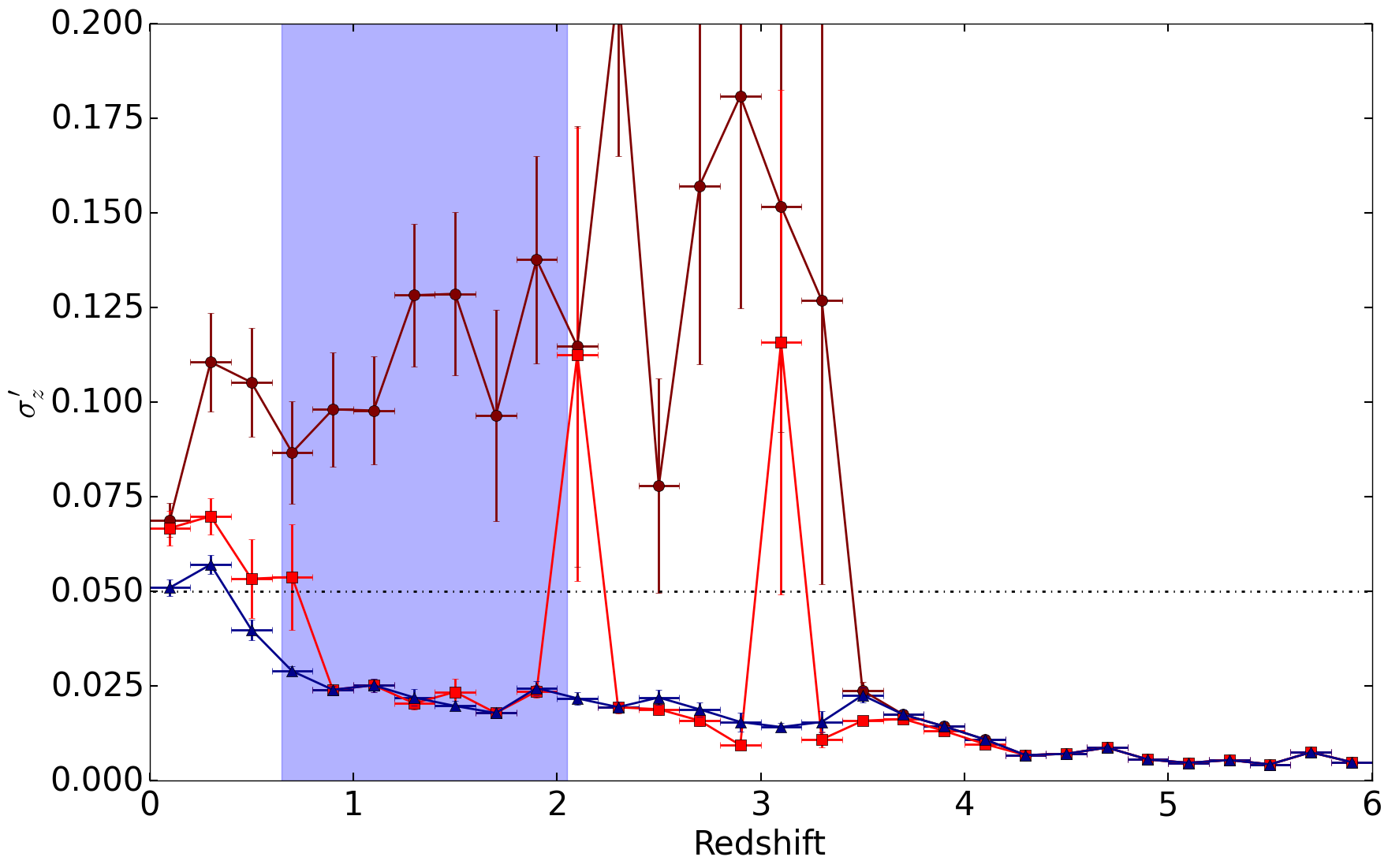}}
	\qquad
	\subfloat[][{\tt{SOM\_Hierarchical\_ImportanceSampling}}]{\includegraphics[scale=0.19]{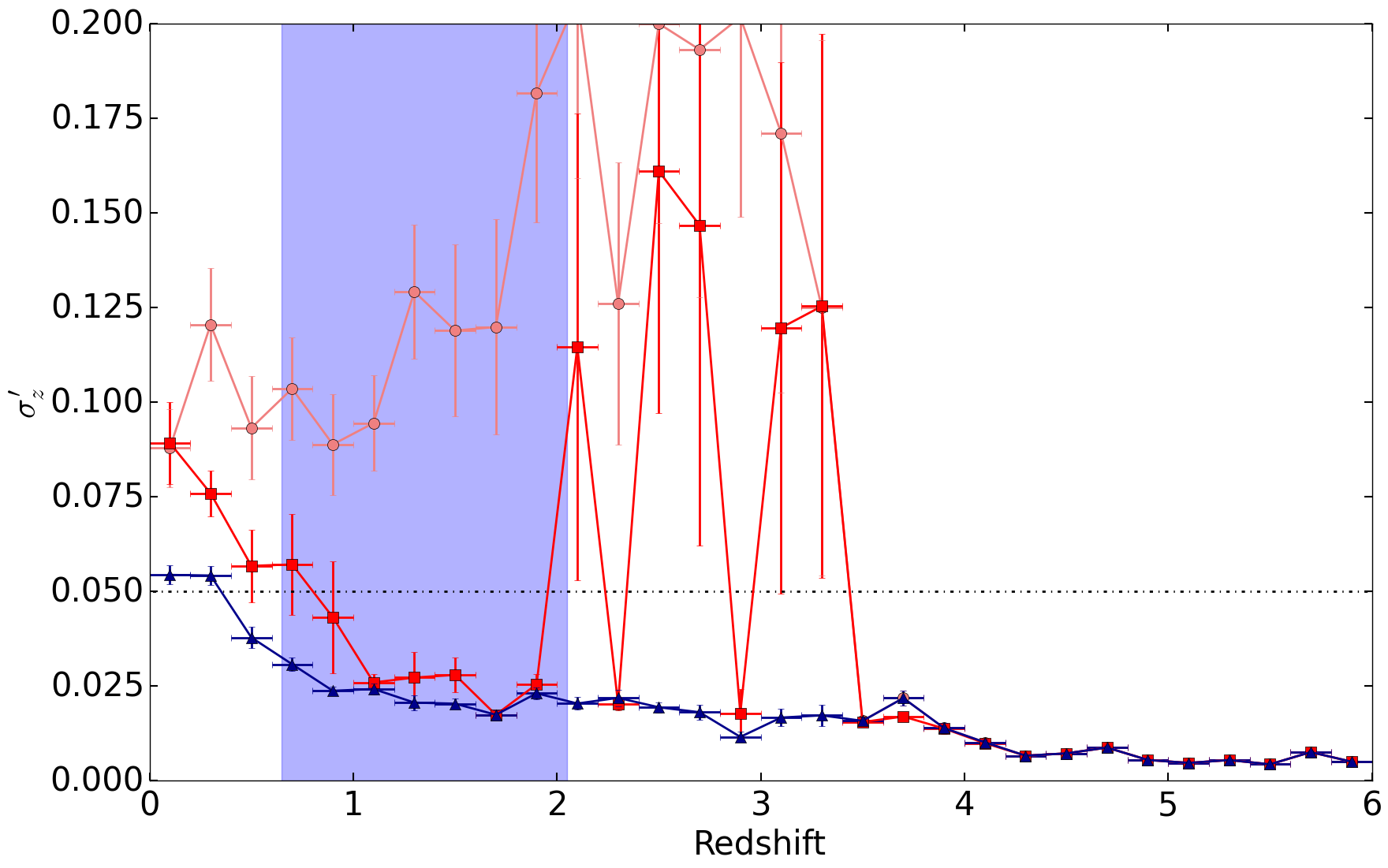}}\\
	\subfloat[][{\tt{SOM\_CellModel\_LimitedSum}}]{\includegraphics[scale=0.19]{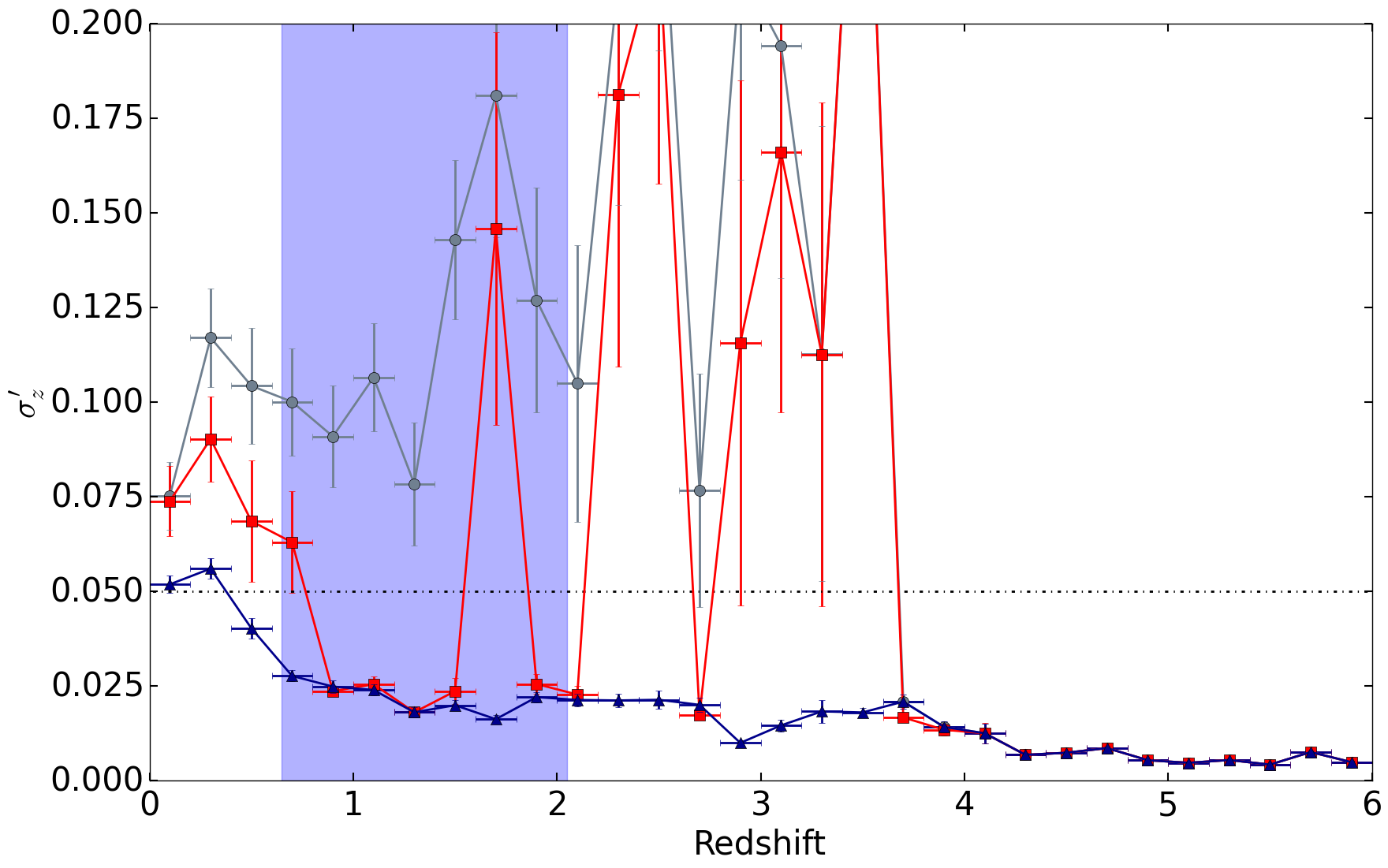}}
	\qquad
	\subfloat[][{\tt{SOM\_CellModel\_Average}}]{\includegraphics[scale=0.19]{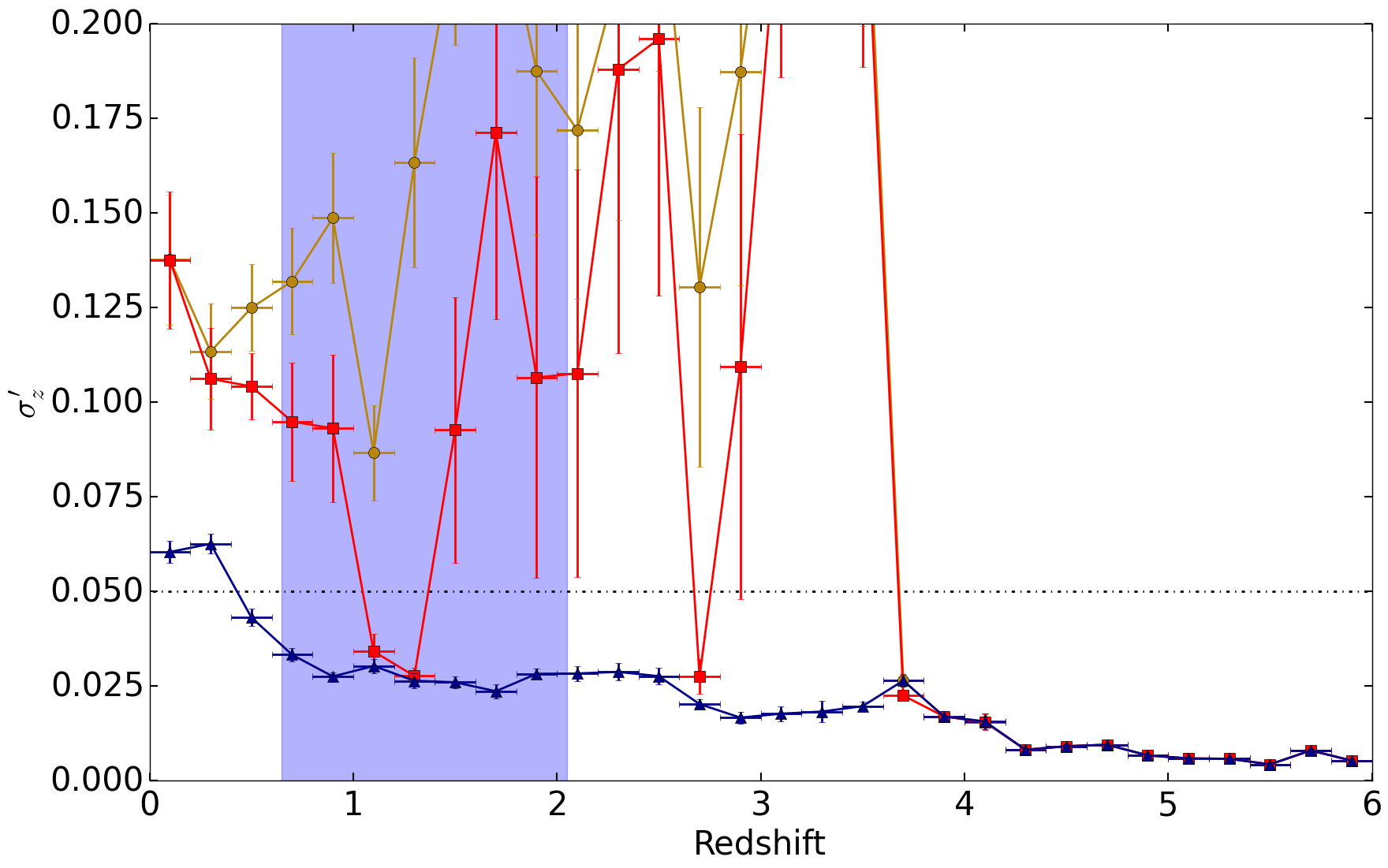}}
	\caption{
		As Figure~\ref{fig:photz_etastd}, but for our {\tt{SOM\_Hierarchical\_MonteCarlo}}, {\tt{SOM\_Hierarchical\_ImportanceSampling}}, {\tt{SOM\_CellModel\_LimitedSum}}, and {\tt{SOM\_CellModel\_Average}} methods (see \S\ref{sec:methods}).
		Similar to Figure~\ref{fig:photz_zbias}, numerous outliers present in the full sample lead to significant biases for most redshift bins below $z \lesssim 3.6$, while the majority of tomographic bins for our secure photo-z's above $z \sim 0.8$ meet the accuracy requirements (with the exception of the {\tt{SOM\_CellModel}}-based approaches). Due to our small sample size and the sensitivity of $\sigma(\Delta z^\prime)$ to outliers, isolated failures in tomographic bins at $z>0.8$ are often due to only a small number of outliers.}
	\label{fig:photz_etastd_2}
\end{figure*}

\section{Performance Tests Against Mock LSST and \textit{Euclid} Data}
\label{sec:results}


Although we would ideally like to test each of these approaches using real data, preliminary tests with a number of individual objects taken from Cosmological Origins Survey \citep[COSMOS;][]{scoville+07} data indicate that the limited range spanned by portions of the \citet{brown+14} archetypes are unrepresentative of a number of actively star-forming galaxies at higher redshift. As noted in Paper I, similar issues exist among all currently used template sets \citep{coleman+80,kinney+96,polletta+07}. Rather than supplementing the \citet{brown+14} galaxies with synthetic spectra created from stellar population synthesis models, we instead use a mock catalog drawn from our fuzzy templates to serve as performance benchmarks for our approach. While this does not give us a fully representative idea of how each of our individual approaches work in practice, it does give us an idealized comparison as to their \textit{intrinsic effectiveness} assuming no template mismatch and similar modeling assumptions, which can help to identify the most useful approaches out of the eight proposed here.

We create a mock catalog of 10,000 objects scaled to $24$\,mag in the LSST $Y$-band to examine performance at a reasonable given magnitude. We generate each object by sampling uniformly from $\boldsymbol{\theta}$ over the redshift interval $z=0$\,--\,$6$ ($\Delta z = 0.01$) and the given set of priors $P(\boldsymbol{\phi})$ on our fiducial set of templates (see \S\ref{subsec:fuzzy}). Each object is then assigned Gaussian errors (added in quadrature) according to the estimated calibration uncertainty (1\%) and observational errors based on the expected imaging depths, and the photometry is jittered according the error bars.

We use each of the methods detailed in \S\ref{sec:methods} to derive photo-z's to the entire mock catalog. Although we save the full $P(z|\rom{\mathbf{F}}{obs})$ for each object when available (i.e. excluding {\tt{SOM\_CellModel\_Average}}), we choose to use the zeroth moment $\rom{\bar{z}}{phot}$ and the square root of the first moment $\sigma_z$ as our fiducial photo-z point estimates and their associated errors.

We report five summary statistics for each sample: the redshift-normalized \textit{mean bias},
\begin{equation}
\bar{\Delta z}^\prime = \left\langle\frac{|\rom{z}{spec} - \rom{\bar{z}}{phot}|}{1+\rom{z}{spec}}\right\rangle,
\end{equation}
the redshift-normalized \textit{median bias},
\begin{equation}
{\Delta z}_{50}^\prime = \textrm{median}\left(\frac{|\rom{z}{spec} - \rom{\bar{z}}{phot}|}{1+\rom{z}{spec}}\right),
\end{equation}
the redshift-normalized \textit{mean scatter},
\begin{equation}
\sigma(\Delta z^\prime) = \left[\textrm{Var}\left(\frac{|\rom{z}{spec} - \rom{\bar{z}}{phot}|}{1+\rom{z}{spec}}\right)\right]^{-\frac{1}{2}}
\end{equation}
the $1\sigma$ \textit{median absolute deviation} (MAD),
\begin{equation}
\sigma_{z,\textrm{MAD}}^\prime = \textrm{68.3 percentile}\left(\left|\frac{|\rom{z}{spec} - \rom{\bar{z}}{phot}|}{1+\rom{z}{spec}} - {\Delta z}_{50}^\prime\right|\right),
\end{equation}
and the \textit{catastrophic outlier fraction},
\begin{equation}
\rom{f}{cat}=N(\Delta z^\prime > 0.15)/N.
\end{equation}
$\rom{z}{spec}$ in all cases is taken to be the input redshift, while $\rom{\bar{z}}{phot}$ is the estimated redshift from each of the different methods outlined in \S\ref{sec:methods}.

In order to remove multimodal and/or unconstrained fits, we classify our objects into ``secure'' and ``insecure'' redshifts based on their mean redshift-normalized \textit{standard deviation},
\begin{equation}
\sigma_{z}^\prime = \sigma_z/(1+\bar{z}).
\end{equation}
Objects are flagged as insecure if $\sigma_z^\prime > 0.15$ (i.e. if the standard deviation of the redshift PDF is larger than the catastrophic error boundaries), and otherwise are classified as secure. These are then used to construct ``full'' and ``secure'' samples. This is done to ensure that we have an unbiased view of the effectiveness of our redshift methods to not only derive accurate photo-z's but also to flag outliers. As weak lensing approaches are extremely sensitive to such outliers \citep{cunha+14}, being able to accurately detect and remove badly constrained photo-z's are key to their successful application.\footnote{We note that median-based point estimates $z_{50,\textrm{phot}}$ give comparable results to $\rom{\bar{z}}{phot}$ in terms of overall accuracy. However, median-based estimates of the relative \textit{spread} $\sigma_{z,\textrm{MAD}}$ are less effective at flagging possible catastrophic errors because the majority of the multi-modal PDFs tends to be concentrated near one mode.}

The generally quality of the photo-z point estimates derived from each method is illustrated in Figures~\ref{fig:photz} and~\ref{fig:photz_2}, along with several summary statistics. We find that all of the methods are able to recover the general one-to-one relation between the input and output redshift relatively well with $\sigma_{z,\textrm{MAD}}^\prime\lesssim1.15$\%. However, we find that both our {\tt{SOM\_CellModel\_LimitedSum}} and {\tt{SOM\_CellModel\_Average}} methods are unable to consistently flag insecure photo-z's, impeding their ability to remove catastrophic outliers that can severely bias weak lensing analyses. We find this is mainly due to the assumption that the SOM cell models are a reliable proxy for the data clustered around them: while this assumption is true in general, it introduces a color-dependent bias during the sampling process that distorts the derived redshift PDF. Even in the worst case, however, we find that our catastrophic outlier fraction is $\lesssim 3.5$\%.

After limiting each of our other six methods to their secure subsamples, we find their accuracy noticeably improves, with redshift-normalized mean biases $\lesssim 0.1$\% and catastrophic outlier fractions of $\sim 0.35$\%, a reduction of a factor of $\sim$\,5\,--\,10. While both {\tt{SOM\_CellModel\_LimitedSum}} and {\tt{SOM\_CellModel\_Average}} also display improved performance, they are only able to reduce their catastrophic outlier fractions by a factor of 2\,--\,3.

Among our ensemble of methods, we find that {\tt{BruteForce\_LinearFuzzy}}, {\tt{SOM\_MCMC\_ObservedFrame}}, and {\tt{SOM\_Hierarchical\_MonteCarlo}} give the best results relative to {\tt{BruteForce}} before and after the removal of insecure redshifts. In addition, although our {\tt{SOM\_Hierarchical\_ImportanceSampling}} approach gives a slightly higher catastrophic outlier fraction after removing photo-z's flagged as insecure, we find its general overall performance is also similar to {\tt{BruteForce}}. Finally, although our {\tt{SOM\_MCMC\_RestFrame}} run displays a noticeably high concentration of misfit objects scattered from low to high redshift, when comparing only the secure subsamples we find its performance is nearly identical to the previous two methods. We thus find both our MCMC-driven and hierarchical sampling methods can give comparable results relative to a brute-force search and are effective at locating and characterizing $P(z|\rom{\mathbf{F}}{obs})$.


To examine the redshift PDFs from our individual methods more closely, in Figure~\ref{fig:photz_pzstack} we plot their two-dimensional stacked $P(z|\rom{\mathbf{F}}{obs})$ distributions using 500 Monte Carlo realizations drawn from the redshift PDF of each object. As can be clearly seen, all of our approaches recover the broad features and degeneracies observed in the {\tt{BruteForce}} run, including the low-$z$ redshift-reddening degeneracy (as well as several others at higher redshifts) as well as confusion between the 1216\,{\AA} and 4000\,{\AA} breaks. We confirm that the {\tt{SOM\_MCMC\_RestFrame}} case displays an overabundance of objects and corresponding probability at higher redshifts, which is likely due to the impact of the redshift gradient observed in first noted in Paper I, which tends to increase the likelihood that chains converge to higher redshift solutions. In addition, we find that while our {\tt{BruteForce\_LinearFuzzy}} run performs comparably to {\tt{BruteForce}} run, the increased flexibility actually causes existing degeneracies to sharpen while simultaneously creating new degeneracies not present in any of the other runs. This indicates that while marginalizing over the linear parameters of our fuzzy templates is generally effective, it occasionally can enhance the likelihoods of degenerate solutions.

In order to measure the dark energy equation-of-state, LSST and \textit{Euclid} will rely on weak lensing to determine the cosmic shear in given redshift tomographic bins to sub-percent accuracy \citep{ivezic+08,laureijs+11}. This approach depends on having extremely accurate measurements of the mean redshift of a given bin. To illustrate the accuracy of each approach more clearly in the context of measuring cosmic shear, we plot the evolution in mean bias $\Delta\bar{z}=\langle\rom{z}{spec}\rangle_{\textrm{bin}}-\langle\rom{\bar{z}}{phot}\rangle_{\textrm{bin}}$ and the normalized standard deviation $\sigma_z^\prime$ in fixed $\Delta z = 0.2$ redshift tomographic bins (selected using $\rom{z}{phot}$) from $z=0$\,--\,$6$ in Figures~\ref{fig:photz_zbias},~\ref{fig:photz_zbias_2},~\ref{fig:photz_etastd}, and~\ref{fig:photz_etastd_2}. To establish an appropriate baseline comparison for each method, we compare both the full and secure samples to an explicitly ``cleaned'' sample where all catastrophic outliers (i.e. objects with $\Delta z^\prime > 0.15$) have been removed.

We find that for all of our methods (excluding our {\tt{SOM\_CellModel}}-based approaches), the secure subsamples are able to meet the \textit{Euclid} mean photo-z bias accuracy requirements for weak-lensing of $\Delta\bar{z} < 0.002(1+z)$ for most redshift bins above $z \sim 0.8$. Without removing insecure objects, however, we find the full sample fails to meet the accuracy requirements for almost all redshift bins where $z \lesssim 4$. After this point, the redshift identification for most objects becomes more secure due to the prominence of the 1216\,{\AA} break combined with \textit{Euclid}'s NIR coverage in the $YJH$ bands.

Likewise, the observed redshift-normalized scatter $\sigma_z^\prime$ for the secure subsample successfully meets the \textit{Euclid} photo-z accuracy requirements of $\sigma_z^\prime < 0.05$ for most redshift bins above $z \sim 1$ with a comfortable margin for error (although betwen $z \sim 3$\,--\,$4$ there is some tension seen in our {\tt{SOM\_MCMC\_RestFrame}} and {\tt{SOM\_Hierarchical\_ImportanceSampling}} results), while the full sample fails for $z \lesssim 3.6$, regardless of the method used.

Together, these results illustrate the power of using the clustering methods such as the SOM to enhance photo-z searches, as well as the danger in relying heavily on summary statistics rather than the underlying collection of model photometry. Ultimately, our findings suggest that the intrinsic ability of the SOM to capture variable information content and widely-separated/broad degeneracies in parameter space and project it into a reduced-dimensional manifold in a topologically smooth fashion lends a significant amount of flexibility to the model-fitting process. This feature is clearly apparent in our 3-D {\tt{SOM\_MCMC\_ObservedFrame}} implementation, which displays performance noticeably better performance than the 2+1-D {\tt{SOM\_RestFrame}} counterpart that is on par with our {\tt{BruteForce}} ``gold standard''. In addition, our hierarchical sampling approaches demonstrate the more generally ability granted by clustering methods to partition the large amount of model photometry into useful partitions in color space that can be quickly explored.

However, while our results emphasize the SOM is a useful tool for clustering and re-organizing the space, using simple summary statistics that describe individual SOM cells to derive photo-z estimates is inherently risky. As individual cell models (and cell means) can display significant variation from the model data assigned to them, either due to high intrinsic variability, sparse sampling, or both (Paper I), relying on them directly will almost guarantee that certain regions of color space will become distorted during the sampling process. This can lead to occasional biases in the final photo-z estimates that make it more difficult to estimate the quality of the final results. While these approaches are reasonable for deriving quick photo-z estimates, they should not be used in place of more rigorous approaches if identifying and removing catastrophic errors from the overall sample are important. It thus remains important to use these approaches judiciously, especially when abstracting away from the underlying set of models.

In addition, it is important to note that all the results derived here have utilized the same set of templates and modeling assumptions, and thus have avoided possible systematic errors due to template mismatches with observed photometry. This will most likely increase the mean bias, scatter, and catastrophic outlier fraction within a given tomographic redshift bin in a redshift-dependent manner. It is thus important to recognize these results as emulating the best-case performance of each of these sampling techniques as applied to upcoming large-scale surveys.

\section{Conclusion}
\label{sec:conc}

Due to the extremely difficult task of obtaining spectroscopic redshifts (spec-z's) to the billions of objects that will be observed in future large-scale photometric surveys, fast and accurate photometric redshifts (photo-z's) remain a necessary tool to enable ``big data'' science. Although significant progress has been made using machine learning to derive photo-z's \citep{carrascokindbrunner13,carrascokindbrunner14,sadeh+15,bonnett15,almosallam+15,hoyle15}, improvements in template-fitting methods have in general lagged behind, with codes today still suffering from many of the same model deficiencies and computational issues that plagued them years ago. While template-fitting methods remain effective (see \citealt{hildebrandt+10,dahlen+13,sanchez+14} for recent comparisons and \citealt{cool+13,ilbert+13,tomczak+14} for recent applications), they are in general under-equipped to handle the enormous number of objects LSST, \textit{Euclid}, and other similar surveys are expected to observed. Because template-fitting methods will likely continue to serve a necessary function in extragalactic science for the upcoming decade, this issue is worrisome: while machine learning approaches are effective within the bounds of their training sets \citep{sanchez+14}, the limited training set provided by spec-z's today \citep{masters+15} and the difficulty of getting reliable spectra to higher redshift sources essentially guarantees that template-based photo-z's will continue to serve a crucial role in high-$z$ science.

We have attempted to test the performance of the new hybrid template-fitting and machine-learning framework outlined in Paper I that alleviates both of these issues by allowing a large set of  ``fuzzy archetypes'' to be fit \textit{en masse} in approximately constant time. While the observational requirements are more intensive than those used by current techniques, our method significantly reduces existing model dependencies and takes advantage of machine learning techniques such as Self Organizing Maps (SOMs) to improve the scope of parameter space that can be quickly yet robustly explored by different sampling techniques. 

Using a mock set of galaxies with LSST $ugrizY$ and \textit{Euclid} $YJH$ photometry with associated calibration uncertainties and observational errors, we quantify the performance of eight different approaches to deriving photo-z's -- a standard brute-force approach ({\tt{BruteForce}}), a fuzzy template-based brute-force approach ({\tt{BruteForce\_LinearFuzzy}}), two approaches based on Markov Chain Monte Carlo (MCMC) sampling over the SOM ({\tt{SOM\_MCMC\_ObservedFrame}} and {\tt{SOM\_MCMC\_RestFrame}}), two approaches based on hierarchical sampling over the SOM ({\tt{SOM\_Hierarchical\_MonteCarlo}} and {\tt{SOM\_Hierarchical\_ImportanceSampling}}), and two ``quick estimate'' approaches based on summary statistics derived from the SOM ({\tt{SOM\_CellModel\_LimitedSum}} and {\tt{SOM\_CellModel\_Average}}). We find that all of our approaches give good photo-z estimates with a redshift-normalized 1$\sigma$ median absolute deviation (MAD) of $\sigma_{z,\textrm{MAD}}\approx 1.1$\% and small catastrophic outlier fractions of $\sim$\,$2\%$. Although we find our {\tt{SOM\_CellModel}}-based approaches are not able to consistently flag insecure photo-z's, our remaining MCMC-driven methods are able to robustly flag outliers and recover the underlying PDF relative to our brute-force approaches, with the results of {\tt{SOM\_MCMC\_ObservedFrame}} and {\tt{SOM\_Hierarchical\_MonteCarlo}} virtually indistinguishable from our {\tt{BruteForce}} ``gold standard''. 

In addition, we find our methods are able to meet the stringent \textit{Euclid} photo-z accuracy requirements for the weak lensing analysis of cosmic shear in the majority of redshift bins in our sample over $z = 0$\,--\,$6$, both in terms of the mean redshift bias as well as the general accuracy over the corresponding ensemble of objects. We thus confirm that the fundamental framework is applicable to future photo-z applications to wide-field surveys and can generate accurate predictions assuming a representative sample of underlying fuzzy archetypes can be obtained/generated.

There are a number of future applications of this work, some of which have been suggested in Paper I and others which have been partially explored by previous authors. In particular, while our {\tt{SOM\_CellModel}}-based  approaches can be computed extremely rapidly, their dependence on individual SOM cell models tends to bias the results for a number of obejcts. Some of this dependence might be substantially reducd or even removed with a recourse to a ``Random Atlas'' of SOMs \citep{carrascokindbrunner14}, which we could use to derive a combined ensemble prediction as outlined in \citet{carrascokindbrunner14}. Alternately, we could move away from discrete mappings of objects onto their best-matching cells onto the SOM (\S\ref{subsec:som}) in favor of general PDFs over the entire collection of SOM cells. These could (along with some set of summary statistics) then be used as a type of non-linear dimensionality reduction whose components would be fed into other machine-learning architectures in an attempt to predict the color-redshift relationship at a given location.

Finally, one could investigate taking the general framework outlined and tested here and applying it to the large regions of parameter space spanned by stellar population synthesis models in an effort to achieve similar results. As the large numbers of parameters used in these models render direct Monte Carlo sampling methods intractable, such an approach would have to involve iterative training steps where a small amount of objects (fit using more computationally expensive methods) would be used to approximate the manifold (and degeneracies) where galaxies live, which would followed by successive SOM training until the entire relevant regions of model space have been subsequently mapped using the minimum number of objects. This approach would then serve as a way of interpolating between sparse sampling of the relevant manifold, which could be used to improve our understanding of galaxy properties and accelerate SED fitting techniques. We are currently looking into a number of these possibilities both as they apply to improving photometric redshifts as well as SED fitting more generally.

\section*{Acknowledgements}

JSS would like to thank Michael Brown, Peter Capak, and Daniel Masters for insightful discussions as well as Charles Alcock for supervising the senior thesis course where a portion of this work was completed. JSS is grateful for financial support from the CREST program, which is funded by the Japan Science and Technology (JST) Agency. This work has benefited extensively from access to Harvard University's Odyssey computing cluster, which is supported by the FAS Division of Science's Research Computing Group.

\appendix
\section{Linearized Fuzzy Templates}
\label{app:linear_fuzzy}

Ignoring rest-frame dust attenuation, the ``baseline'' model photometry $\rom{\mathbf{F}}{base,gal}$ for a galaxy at a given redshift can be written as a linear combination of the photometric fluxes derived from the underlying galaxy and emission line templates,
\begin{equation}
(\rom{\mathbf{F}}{base,gal})(z) = \sum_{\textrm{gal}} c_{\textrm{gal}} \rom{\mathbf{F}}{gal}(z) + \sum_{\textrm{lines},\textrm{gal}} \Delta \textrm{EW}_{\textrm{lines}}^\textrm{gal} \rom{\mathbf{F}}{lines}^\textrm{gal}(z).
\end{equation}
where
\begin{equation*}
\rom{\mathbf{F}}{gal}(z)=\int_{\nu_z} S_{\nu,\textrm{gal}}(\nu)10^{-\rom{A}{IGM}(\nu,z)/2.5} \left[\mathbf{w}(z) \odot \mathbf{T}(\nu)\right]\nu^{-1}d\nu
\end{equation*}
\begin{equation*}
\rom{\mathbf{F}}{lines}^\textrm{gal}(z)=\int_{\nu_z}  S_{\nu,\textrm{lines}}^{\textrm{gal}}(\nu)10^{-\rom{A}{IGM}(\nu,z)/2.5} \left[\mathbf{w}(z) \odot \mathbf{T}(\nu)\right] \nu^{-1}d\nu
\end{equation*}
are photometric fluxes (including dust attenuation from the IGM) for each galaxy and its corresponding set of emission line templates $S_{\nu,\textrm{lines}}^{\textrm{gal}}$, $\odot$ indicates element-wise multiplication, and $\mathbf{w}(z)=1/\int_{\nu_z} \mathbf{T}(\nu) \nu^{-1} d\nu$ is the associated set of \textit{inverse filter weights}. 

As a large set of archetypes should be a representative sample of the corresponding regions of color space, we can ignore linear combinations of galaxy templates, leaving us with the simplified expression
\begin{equation}
s\rom{\mathbf{F}}{base,gal}(z)=s\rom{\mathbf{F}}{gal}(z) + \sum_{\textrm{lines}} \Delta \textrm{EW}_{\textrm{lines}}^\textrm{gal} \,s\rom{\mathbf{F}}{lines}^\textrm{gal}(z),
\end{equation}
where we have included the general scaling factor $s$ to emphasize that our photometry must be properly scaled before comparing with observed photometry $\rom{\mathbf{F}}{obs}$.

Although the impact of attenuation from rest-frame dust screen for a given template is wavelength-dependent, its impact in each band can be approximated as $\exp\left[{\rom{\mathbfcal{R}}{dust}(z)\Delta\ebv}\right]$, where $\rom{\mathbfcal{R}}{dust}(z)$ is a collection of negative numbers determined by the given dust template $\rom{k}{dust}(\nu)$ and the corresponding filter set $\mathbf{T}(\nu)$ derived at a given redshift. As we are only considering the perturbative impact of dust on the few-percent level, we choose to Taylor-expand this expression about $0$. Ignoring all $\mathcal{O}(> 1)$ terms (including cross-products), we get
\begin{eqnarray}
s\rom{\mathbf{F}}{model}(z)= s\rom{\mathbf{F}}{gal}(z) + \Delta\ebv \,s\rom{\mathbf{F}}{dust}(z) \nonumber \\
+  \sum_{\textrm{lines}} \Delta \textrm{EW}_{\textrm{lines}}^\textrm{gal}\,s\rom{\mathbf{F}}{lines}^\textrm{gal}(z),
\end{eqnarray}
where $\rom{\mathbf{F}}{dust}(z) = \rom{\mathbfcal{R}}{dust}(z)\odot\rom{\mathbf{F}}{gal}(z)$ is our new ``dust photometry'' term. 

As our first-order expansion translates dust components that are linear in \textit{magnitudes} to be linear in \textit{flux}, we can use it to probe additional additive dust components that might improve our fitting routine. In particular, we consider a modified dust template of the form 
\begin{equation}
\rom{k}{dust}^\prime(x)=\rom{k}{dust}(x)+c_b\rom{k}{bump}(x|x_0,\gamma),
\end{equation}
where $x\equiv 1/\lambda$ is measured in \ium and
\begin{equation}
\rom{k}{bump}(x|x_0,\gamma)=\frac{x^2}{(x^2-x_0^2)^2+x^2\gamma^2},
\end{equation}
is a Lorentzian-like function where $x_0$ and $\gamma$ the central position and width of the Lorentzian feature, respectively. This ``bump'', most often located around 2175\,{\AA}, is a common observed feature among both Galactic \citep{fitzpatrickmassa07} and extragalactic \citep{kriekconroy13,scoville+15} sources. As it's amplitude $c_b$ can vary widely, ignoring it's impact on observed photometry can significantly bias quantities derived through SED fitting \citep{kriekconroy13}.

Inserting our new dust curve into our original expression for $\rom{\mathbf{F}}{model}$, we get the final expression
\begin{eqnarray}
s\rom{\mathbf{F}}{model}(z)= s\rom{\mathbf{F}}{gal}(z) + \Delta\ebv \,s\rom{\mathbf{F}}{dust}(z) \nonumber \\
+ \Delta c_b^\prime \,s\rom{\mathbf{F}}{bump}(z) + \sum_{\textrm{lines}} \Delta \textrm{EW}_{\textrm{lines}}^\textrm{gal} \,s\rom{\mathbf{F}}{lines}^{\textrm{gal}}(z),
\end{eqnarray}
where $c_b^\prime=\Delta \ebv \times \Delta c_b$ and $\rom{\mathbf{F}}{bump}(z) = \rom{\mathbfcal{R}}{bump}(z)\odot\rom{\mathbf{F}}{gal}(z)$. While this decouples the two dust terms if they are treated independently, it is still an improvement over excluding the feature entirely. 

The corresponding log-likelihood when fitting normally distributed data with a series of independent Gaussian priors is equivalent to a modified $\chi^2$ metric,
\begin{equation}
\rom{\chi^2}{mod}(\boldsymbol{\theta},\boldsymbol{\phi},s) \equiv \sum_i \sigma_{i}^{-2} \left[\Delta F_i(\boldsymbol{\theta},\boldsymbol{\phi},s)\right]^2 + \sum_j \left(\frac{\phi_j}{\sigma_{\phi_j}(\boldsymbol{\theta})}\right)^2
\end{equation}
where $\sigma_i^2=\sigma_{\textrm{obs},i}^2+\sigma_{\textrm{model},i}^2$ is the $i$th component of the total variance,\footnote{For the remainder of the paper, we will assume $\rom{\boldsymbol{\sigma}}{model}=0$ such that $\boldsymbol{\sigma}=\rom{\boldsymbol{\sigma}}{obs}$.} $\Delta F_i(\boldsymbol{\theta},\boldsymbol{\phi},s)=F_{\textrm{obs},i}-sF_{\textrm{model},i}(\boldsymbol{\theta},\boldsymbol{\phi})$ is the $i$th flux residual, $\boldsymbol{\theta}=\left\lbrace z,\textrm{gal},\textrm{dust} \right\rbrace$ contains the non-linear parameters of interest, $\boldsymbol{\phi}=\left\lbrace \ebv, c_b^\prime, \lbrace\Delta \textrm{EW}\rbrace_{\textrm{lines}} \right\rbrace$ contains the linear nuisance parameters, $\sigma_{\phi_j}(\boldsymbol{\theta})$ is the standard deviation of the corresponding Gaussian prior for a given $\boldsymbol{\theta}$, the sum over $i$ is taken over all observed bands, and the sum over $j$ is taken over all relevant nuisance parameters.

For $\boldsymbol{\theta}$ and $\boldsymbol{\phi}$ fixed, we can marginalize over $s$ to minimize $\rom{\chi^2}{mod}(s|\boldsymbol{\theta},\boldsymbol{\phi})$, giving us
\begin{equation}
s = \left. {\sum_i {\sigma_i^{-2}}} F_{\textrm{obs},i}F_{\textrm{model},i} \middle/ {\sum_i \sigma_i^{-2} F_{\textrm{model},i}F_{\textrm{model},i}} \right. ,
\end{equation}
which can be calculated prior to computing the actual $\rom{\chi^2}{mod}$ value.

While marginalizing over $s$ is a simple one-step process, marginalizing over the linear nuisance parameters $\boldsymbol{\phi}$ for fixed $s$ and $\boldsymbol{\theta}$ is slightly more computationally demanding. We first re-write $sF_{\textrm{model},i}(\boldsymbol{\theta},\boldsymbol{\phi})$ as
\begin{equation}
sF_{\textrm{model},i}(\boldsymbol{\theta},\boldsymbol{\phi})=sF_{\textrm{gal},i}(\boldsymbol{\theta})+\sum_j s\mathbf{X}_{ij}(\boldsymbol{\theta}) \boldsymbol{\phi}_j,
\end{equation}
where
\begingroup\makeatletter\def\f@size{8}\check@mathfonts
\begin{equation*}
\mathbf{X}(\boldsymbol{\theta})=
\begin{matrix}
{\rom{F}{$1$,dust}}(z) &  {\rom{F}{$1$,bump}}(z)  &  {\rom{F}{$1$,line-$1$}^{\textrm{gal}}}(z) & \ldots & {\rom{F}{1,line-$m$}^{\textrm{gal}}}(z) \\
\vdots &  \vdots  &  \vdots & \ddots & \vdots \\
{\rom{F}{$n$,dust}}(z) &  {\rom{F}{$n$,bump}}(z)  &  {\rom{F}{$n$,line-$1$}^{\textrm{gal}}}(z) & \ldots & {\rom{F}{$n$,line-$m$}^{\textrm{gal}}}(z) \\
\end{matrix}
\end{equation*}
\endgroup
is the $n \times (2+m) = \rom{N}{filt} \times (2+\rom{N}{lines})$ matrix of pre-computed coefficients for a given $\boldsymbol{\theta}$ and $F_i$ is the $i$th component (i.e. filter) of the corresponding vector.

Minimizing $\rom{\chi^2}{mod}(\boldsymbol{\phi}|\boldsymbol{\theta},s)$ involves solving a system of linear equations of the form
\begin{equation}
\left(\mathbf{X}(\boldsymbol{\theta})^T\rom{\mathbf{W}}{obs}\right)\Delta\rom{\mathbf{F}}{gal}(\boldsymbol{\theta})=\left(\mathbf{X}(\boldsymbol{\theta})^{T}\rom{\mathbf{W}}{obs}\mathbf{X}(\boldsymbol{\theta})+\mathbf{W}_\phi(\boldsymbol{\theta})\right)\boldsymbol{\phi},
\end{equation}
where $\Delta\rom{\mathbf{F}}{gal}(\boldsymbol{\theta})=\mathbf{F}_{\textrm{obs}}-s\rom{\mathbf{F}}{gal}(\boldsymbol{\theta})$ is the baseline galaxy flux residual, $\rom{\mathbf{W}}{obs}=\textrm{diag}(\dots,\sigma_i^{-2},\dots)$ is the associated observational weight matrix, $\mathbf{W}_\phi=\textrm{diag}(\dots,\sigma_{\phi_j}^{-2},\dots)$ is the prior weight matrix, and $T$ is the transpose operator. This gives us an simple iterative scheme for minimizing $\rom{\chi^2}{mod}(\boldsymbol{\theta},\boldsymbol{\phi},s)$ with respect to $s$ and $\boldsymbol{\phi}$ for a given choice of $\boldsymbol{\theta}$.

\bibliography{photoz}

\begin{thebibliography}{}
\makeatletter
\relax
\def\mn@urlcharsother{\let\do\@makeother \do\$\do\&\do\#\do\^\do\_\do\%\do\~}
\def\mn@doi{\begingroup\mn@urlcharsother \@ifnextchar [ {\mn@doi@}
  {\mn@doi@[]}}
\def\mn@doi@[#1]#2{\def\@tempa{#1}\ifx\@tempa\@empty \href
  {http://dx.doi.org/#2} {doi:#2}\else \href {http://dx.doi.org/#2} {#1}\fi
  \endgroup}
\def\mn@eprint#1#2{\mn@eprint@#1:#2::\@nil}
\def\mn@eprint@arXiv#1{\href {http://arxiv.org/abs/#1} {{\tt arXiv:#1}}}
\def\mn@eprint@dblp#1{\href {http://dblp.uni-trier.de/rec/bibtex/#1.xml}
  {dblp:#1}}
\def\mn@eprint@#1:#2:#3:#4\@nil{\def\@tempa {#1}\def\@tempb {#2}\def\@tempc
  {#3}\ifx \@tempc \@empty \let \@tempc \@tempb \let \@tempb \@tempa \fi \ifx
  \@tempb \@empty \def\@tempb {arXiv}\fi \@ifundefined
  {mn@eprint@\@tempb}{\@tempb:\@tempc}{\expandafter \expandafter \csname
  mn@eprint@\@tempb\endcsname \expandafter{\@tempc}}}

\bibitem[\protect\citeauthoryear{{Abazajian} et~al.,}{{Abazajian}
  et~al.}{2009}]{abazajian+09}
{Abazajian} K.~N.,  et~al., 2009, \mn@doi [\apjs]
  {10.1088/0067-0049/182/2/543}, \href
  {http://adsabs.harvard.edu/abs/2009ApJS..182..543A} {182, 543}

\bibitem[\protect\citeauthoryear{{Albrecht} et~al.,}{{Albrecht}
  et~al.}{2006}]{albrecht+06}
{Albrecht} A.,  et~al., 2006, ArXiv Astrophysics e-prints, \href
  {http://adsabs.harvard.edu/abs/2006astro.ph..9591A} {}

\bibitem[\protect\citeauthoryear{{Almosallam}, {Lindsay}, {Jarvis}  \&
  {Roberts}}{{Almosallam} et~al.}{2015}]{almosallam+15}
{Almosallam} I.~A.,  {Lindsay} S.~N.,  {Jarvis} M.~J.,   {Roberts} S.~J.,
  2015, preprint, \href {http://adsabs.harvard.edu/abs/2015arXiv150505489A} {}
  (\mn@eprint {arXiv} {1505.05489})

\bibitem[\protect\citeauthoryear{{Ben{\'{\i}}tez}}{{Ben{\'{\i}}tez}}{2000}]{benitez00}
{Ben{\'{\i}}tez} N.,  2000, \mn@doi [\apj] {10.1086/308947}, \href
  {http://adsabs.harvard.edu/abs/2000ApJ...536..571B} {536, 571}

\bibitem[\protect\citeauthoryear{{Bernton}, {Yang}, {Chen}, {Shephard}  \&
  {Liu}}{{Bernton} et~al.}{2015}]{bernton+15}
{Bernton} E.,  {Yang} S.,  {Chen} Y.,  {Shephard} N.,   {Liu} J.~S.,  2015,
  preprint, \href {http://adsabs.harvard.edu/abs/2015arXiv150608852B} {}
  (\mn@eprint {arXiv} {1506.08852})

\bibitem[\protect\citeauthoryear{{Bolzonella}, {Miralles}  \&
  {Pell{\'o}}}{{Bolzonella} et~al.}{2000}]{bolzonella+00}
{Bolzonella} M.,  {Miralles} J.-M.,   {Pell{\'o}} R.,  2000, \aap, \href
  {http://adsabs.harvard.edu/abs/2000A%26A...363..476B} {363, 476}

\bibitem[\protect\citeauthoryear{{Bonnett}}{{Bonnett}}{2015}]{bonnett15}
{Bonnett} C.,  2015, \mn@doi [\mnras] {10.1093/mnras/stv230}, \href
  {http://adsabs.harvard.edu/abs/2015MNRAS.449.1043B} {449, 1043}

\bibitem[\protect\citeauthoryear{{Bordoloi} et~al.,}{{Bordoloi}
  et~al.}{2012}]{bordoloi+12}
{Bordoloi} R.,  et~al., 2012, \mn@doi [\mnras]
  {10.1111/j.1365-2966.2012.20427.x}, \href
  {http://adsabs.harvard.edu/abs/2012MNRAS.421.1671B} {421, 1671}

\bibitem[\protect\citeauthoryear{{Brown} et~al.,}{{Brown}
  et~al.}{2014}]{brown+14}
{Brown} M.~J.~I.,  et~al., 2014, \mn@doi [\apjs] {10.1088/0067-0049/212/2/18},
  \href {http://adsabs.harvard.edu/abs/2014ApJS..212...18B} {212, 18}

\bibitem[\protect\citeauthoryear{{Calzetti}, {Armus}, {Bohlin}, {Kinney},
  {Koornneef}  \& {Storchi-Bergmann}}{{Calzetti} et~al.}{2000}]{calzetti+00}
{Calzetti} D.,  {Armus} L.,  {Bohlin} R.~C.,  {Kinney} A.~L.,  {Koornneef} J.,
   {Storchi-Bergmann} T.,  2000, \mn@doi [\apj] {10.1086/308692}, \href
  {http://adsabs.harvard.edu/abs/2000ApJ...533..682C} {533, 682}

\bibitem[\protect\citeauthoryear{{Carrasco Kind} \& {Brunner}}{{Carrasco Kind}
  \& {Brunner}}{2013}]{carrascokindbrunner13}
{Carrasco Kind} M.,  {Brunner} R.~J.,  2013, \mn@doi [\mnras]
  {10.1093/mnras/stt574}, \href
  {http://adsabs.harvard.edu/abs/2013MNRAS.432.1483C} {432, 1483}

\bibitem[\protect\citeauthoryear{{Carrasco Kind} \& {Brunner}}{{Carrasco Kind}
  \& {Brunner}}{2014a}]{carrascokindbrunner14}
{Carrasco Kind} M.,  {Brunner} R.~J.,  2014a, \mn@doi [\mnras]
  {10.1093/mnras/stt2456}, \href
  {http://adsabs.harvard.edu/abs/2014MNRAS.438.3409C} {438, 3409}

\bibitem[\protect\citeauthoryear{{Carrasco Kind} \& {Brunner}}{{Carrasco Kind}
  \& {Brunner}}{2014b}]{carrascokindbrunner14c}
{Carrasco Kind} M.,  {Brunner} R.~J.,  2014b, \mn@doi [\mnras]
  {10.1093/mnras/stu827}, \href
  {http://adsabs.harvard.edu/abs/2014MNRAS.441.3550C} {441, 3550}

\bibitem[\protect\citeauthoryear{{Coleman}, {Wu}  \& {Weedman}}{{Coleman}
  et~al.}{1980}]{coleman+80}
{Coleman} G.~D.,  {Wu} C.-C.,   {Weedman} D.~W.,  1980, \mn@doi [\apjs]
  {10.1086/190674}, \href {http://adsabs.harvard.edu/abs/1980ApJS...43..393C}
  {43, 393}

\bibitem[\protect\citeauthoryear{{Collister} \& {Lahav}}{{Collister} \&
  {Lahav}}{2004}]{collisterlahav04}
{Collister} A.~A.,  {Lahav} O.,  2004, \mn@doi [\pasp] {10.1086/383254}, \href
  {http://adsabs.harvard.edu/abs/2004PASP..116..345C} {116, 345}

\bibitem[\protect\citeauthoryear{{Cool} et~al.,}{{Cool} et~al.}{2013}]{cool+13}
{Cool} R.~J.,  et~al., 2013, \mn@doi [\apj] {10.1088/0004-637X/767/2/118},
  \href {http://adsabs.harvard.edu/abs/2013ApJ...767..118C} {767, 118}

\bibitem[\protect\citeauthoryear{{Cunha}, {Huterer}, {Lin}, {Busha}  \&
  {Wechsler}}{{Cunha} et~al.}{2014}]{cunha+14}
{Cunha} C.~E.,  {Huterer} D.,  {Lin} H.,  {Busha} M.~T.,   {Wechsler} R.~H.,
  2014, \mn@doi [\mnras] {10.1093/mnras/stu1424}, \href
  {http://adsabs.harvard.edu/abs/2014MNRAS.444..129C} {444, 129}

\bibitem[\protect\citeauthoryear{{Dahlen} et~al.,}{{Dahlen}
  et~al.}{2013}]{dahlen+13}
{Dahlen} T.,  et~al., 2013, \mn@doi [\apj] {10.1088/0004-637X/775/2/93}, \href
  {http://adsabs.harvard.edu/abs/2013ApJ...775...93D} {775, 93}

\bibitem[\protect\citeauthoryear{Doucet, De~Freitas  \& Gordon}{Doucet
  et~al.}{2001}]{doucet+01}
Doucet A.,  De~Freitas N.,   Gordon N.,  2001, Sequential monte carlo methods
  in practice.
Springer-Verlag

\bibitem[\protect\citeauthoryear{{Elliott}, {de Souza}, {Krone-Martins},
  {Cameron}, {Ishida}  \& {Hilbe}}{{Elliott} et~al.}{2015}]{elliott+15}
{Elliott} J.,  {de Souza} R.~S.,  {Krone-Martins} A.,  {Cameron} E.,  {Ishida}
  E.~E.~O.,   {Hilbe} J.,  2015, preprint, \href
  {http://adsabs.harvard.edu/abs/2015arXiv150701293E} {} (\mn@eprint {arXiv}
  {1507.01293})

\bibitem[\protect\citeauthoryear{{Feldmann} et~al.,}{{Feldmann}
  et~al.}{2006}]{feldmann+06}
{Feldmann} R.,  et~al., 2006, \mn@doi [\mnras]
  {10.1111/j.1365-2966.2006.10930.x}, \href
  {http://adsabs.harvard.edu/abs/2006MNRAS.372..565F} {372, 565}

\bibitem[\protect\citeauthoryear{{Fitzpatrick} \& {Massa}}{{Fitzpatrick} \&
  {Massa}}{2007}]{fitzpatrickmassa07}
{Fitzpatrick} E.~L.,  {Massa} D.,  2007, \mn@doi [\apj] {10.1086/518158}, \href
  {http://adsabs.harvard.edu/abs/2007ApJ...663..320F} {663, 320}

\bibitem[\protect\citeauthoryear{{Foreman-Mackey}, {Hogg}, {Lang}  \&
  {Goodman}}{{Foreman-Mackey} et~al.}{2013}]{foremanmackey+13}
{Foreman-Mackey} D.,  {Hogg} D.~W.,  {Lang} D.,   {Goodman} J.,  2013, \mn@doi
  [\pasp] {10.1086/670067}, \href
  {http://adsabs.harvard.edu/abs/2013PASP..125..306F} {125, 306}

\bibitem[\protect\citeauthoryear{{Gerdes}, {Sypniewski}, {McKay}, {Hao},
  {Weis}, {Wechsler}  \& {Busha}}{{Gerdes} et~al.}{2010}]{gerdes+10}
{Gerdes} D.~W.,  {Sypniewski} A.~J.,  {McKay} T.~A.,  {Hao} J.,  {Weis} M.~R.,
  {Wechsler} R.~H.,   {Busha} M.~T.,  2010, \mn@doi [\apj]
  {10.1088/0004-637X/715/2/823}, \href
  {http://adsabs.harvard.edu/abs/2010ApJ...715..823G} {715, 823}

\bibitem[\protect\citeauthoryear{{Goodman} \& {Weare}}{{Goodman} \&
  {Weare}}{2010}]{goodmanweare10}
{Goodman} J.,  {Weare} J.,  2010, \mn@doi [Communications in Applied
  Mathematics and Computer Science] {10.2140/camcos.2010.5.65}, 5, 65

\bibitem[\protect\citeauthoryear{{Hildebrandt} et~al.,}{{Hildebrandt}
  et~al.}{2010}]{hildebrandt+10}
{Hildebrandt} H.,  et~al., 2010, \mn@doi [\aap] {10.1051/0004-6361/201014885},
  \href {http://adsabs.harvard.edu/abs/2010A%26A...523A..31H} {523, A31}

\bibitem[\protect\citeauthoryear{{Hoyle}}{{Hoyle}}{2015}]{hoyle15}
{Hoyle} B.,  2015, preprint, \href
  {http://adsabs.harvard.edu/abs/2015arXiv150407255H} {} (\mn@eprint {arXiv}
  {1504.07255})

\bibitem[\protect\citeauthoryear{{Hoyle}, {Rau}, {Bonnett}, {Seitz}  \&
  {Weller}}{{Hoyle} et~al.}{2015}]{hoyle+15}
{Hoyle} B.,  {Rau} M.~M.,  {Bonnett} C.,  {Seitz} S.,   {Weller} J.,  2015,
  \mn@doi [\mnras] {10.1093/mnras/stv599}, \href
  {http://adsabs.harvard.edu/abs/2015MNRAS.450..305H} {450, 305}

\bibitem[\protect\citeauthoryear{{Ilbert} et~al.,}{{Ilbert}
  et~al.}{2006}]{ilbert+06}
{Ilbert} O.,  et~al., 2006, \mn@doi [\aap] {10.1051/0004-6361:20065138}, \href
  {http://adsabs.harvard.edu/abs/2006A%26A...457..841I} {457, 841}

\bibitem[\protect\citeauthoryear{{Ilbert} et~al.,}{{Ilbert}
  et~al.}{2013}]{ilbert+13}
{Ilbert} O.,  et~al., 2013, \mn@doi [\aap] {10.1051/0004-6361/201321100}, \href
  {http://adsabs.harvard.edu/abs/2013A%26A...556A..55I} {556, A55}

\bibitem[\protect\citeauthoryear{{Ivezic} et~al.,}{{Ivezic}
  et~al.}{2008}]{ivezic+08}
{Ivezic} Z.,  et~al., 2008, preprint, \href
  {http://adsabs.harvard.edu/abs/2008arXiv0805.2366I} {} (\mn@eprint {arXiv}
  {0805.2366})

\bibitem[\protect\citeauthoryear{{Johnson}, {Wilson}, {Tang}  \&
  {Scott}}{{Johnson} et~al.}{2013}]{johnson+13}
{Johnson} S.~P.,  {Wilson} G.~W.,  {Tang} Y.,   {Scott} K.~S.,  2013, \mn@doi
  [\mnras] {10.1093/mnras/stt1758}, \href
  {http://adsabs.harvard.edu/abs/2013MNRAS.436.2535J} {436, 2535}

\bibitem[\protect\citeauthoryear{{Kennicutt}}{{Kennicutt}}{1998}]{kennicutt98}
{Kennicutt} Jr. R.~C.,  1998, \mn@doi [\araa] {10.1146/annurev.astro.36.1.189},
  \href {http://adsabs.harvard.edu/abs/1998ARA%26A..36..189K} {36, 189}

\bibitem[\protect\citeauthoryear{{Kennicutt} \& {Evans}}{{Kennicutt} \&
  {Evans}}{2012}]{kennicuttevans12}
{Kennicutt} R.~C.,  {Evans} N.~J.,  2012, \mn@doi [\araa]
  {10.1146/annurev-astro-081811-125610}, \href
  {http://adsabs.harvard.edu/abs/2012ARA%26A..50..531K} {50, 531}

\bibitem[\protect\citeauthoryear{{Kinney}, {Calzetti}, {Bohlin}, {McQuade},
  {Storchi-Bergmann}  \& {Schmitt}}{{Kinney} et~al.}{1996}]{kinney+96}
{Kinney} A.~L.,  {Calzetti} D.,  {Bohlin} R.~C.,  {McQuade} K.,
  {Storchi-Bergmann} T.,   {Schmitt} H.~R.,  1996, \mn@doi [\apj]
  {10.1086/177583}, \href {http://adsabs.harvard.edu/abs/1996ApJ...467...38K}
  {467, 38}

\bibitem[\protect\citeauthoryear{Kohonen}{Kohonen}{1982}]{kohonen82}
Kohonen T.,  1982, \mn@doi [Biological Cybernetics] {10.1007/BF00337288}, 43,
  59

\bibitem[\protect\citeauthoryear{{Kohonen}}{{Kohonen}}{2001}]{kohonen01}
{Kohonen} T.,  2001, Self-Organizing Maps.
Springer, 2001, xx, 501 p.~Springer series in information sciences

\bibitem[\protect\citeauthoryear{{Kriek} \& {Conroy}}{{Kriek} \&
  {Conroy}}{2013}]{kriekconroy13}
{Kriek} M.,  {Conroy} C.,  2013, \mn@doi [\apjl] {10.1088/2041-8205/775/1/L16},
  \href {http://adsabs.harvard.edu/abs/2013ApJ...775L..16K} {775, L16}

\bibitem[\protect\citeauthoryear{{Laureijs} et~al.,}{{Laureijs}
  et~al.}{2011}]{laureijs+11}
{Laureijs} R.,  et~al., 2011, preprint, \href
  {http://adsabs.harvard.edu/abs/2011arXiv1110.3193L} {} (\mn@eprint {arXiv}
  {1110.3193})

\bibitem[\protect\citeauthoryear{{Li} \& {Yee}}{{Li} \& {Yee}}{2008}]{liyee08}
{Li} I.~H.,  {Yee} H.~K.~C.,  2008, \mn@doi [\aj]
  {10.1088/0004-6256/135/3/809}, \href
  {http://adsabs.harvard.edu/abs/2008AJ....135..809L} {135, 809}

\bibitem[\protect\citeauthoryear{{Madau}}{{Madau}}{1995}]{madau95}
{Madau} P.,  1995, \mn@doi [\apj] {10.1086/175332}, \href
  {http://adsabs.harvard.edu/abs/1995ApJ...441...18M} {441, 18}

\bibitem[\protect\citeauthoryear{{Masters} et~al.,}{{Masters}
  et~al.}{2015}]{masters+15}
{Masters} D.,  et~al., 2015, preprint, \href
  {http://adsabs.harvard.edu/abs/2015arXiv150903318M} {} (\mn@eprint {arXiv}
  {1509.03318})

\bibitem[\protect\citeauthoryear{{M{\'e}nard}, {Scranton}, {Schmidt},
  {Morrison}, {Jeong}, {Budavari}  \& {Rahman}}{{M{\'e}nard}
  et~al.}{2013}]{menard+13}
{M{\'e}nard} B.,  {Scranton} R.,  {Schmidt} S.,  {Morrison} C.,  {Jeong} D.,
  {Budavari} T.,   {Rahman} M.,  2013, preprint, \href
  {http://adsabs.harvard.edu/abs/2013arXiv1303.4722M} {} (\mn@eprint {arXiv}
  {1303.4722})

\bibitem[\protect\citeauthoryear{{Newman} et~al.,}{{Newman}
  et~al.}{2015}]{newman+15}
{Newman} J.~A.,  et~al., 2015, \mn@doi [Astroparticle Physics]
  {10.1016/j.astropartphys.2014.06.007}, \href
  {http://adsabs.harvard.edu/abs/2015APh....63...81N} {63, 81}

\bibitem[\protect\citeauthoryear{{Polletta} et~al.,}{{Polletta}
  et~al.}{2007}]{polletta+07}
{Polletta} M.,  et~al., 2007, \mn@doi [\apj] {10.1086/518113}, \href
  {http://adsabs.harvard.edu/abs/2007ApJ...663...81P} {663, 81}

\bibitem[\protect\citeauthoryear{{Prevot}, {Lequeux}, {Prevot}, {Maurice}  \&
  {Rocca-Volmerange}}{{Prevot} et~al.}{1984}]{prevot+84}
{Prevot} M.~L.,  {Lequeux} J.,  {Prevot} L.,  {Maurice} E.,
  {Rocca-Volmerange} B.,  1984, \aap, \href
  {http://adsabs.harvard.edu/abs/1984A%26A...132..389P} {132, 389}

\bibitem[\protect\citeauthoryear{{Sadeh}, {Abdalla}  \& {Lahav}}{{Sadeh}
  et~al.}{2015}]{sadeh+15}
{Sadeh} I.,  {Abdalla} F.~B.,   {Lahav} O.,  2015, preprint, \href
  {http://adsabs.harvard.edu/abs/2015arXiv150700490S} {} (\mn@eprint {arXiv}
  {1507.00490})

\bibitem[\protect\citeauthoryear{{S{\'a}nchez} et~al.,}{{S{\'a}nchez}
  et~al.}{2014}]{sanchez+14}
{S{\'a}nchez} C.,  et~al., 2014, \mn@doi [\mnras] {10.1093/mnras/stu1836},
  \href {http://adsabs.harvard.edu/abs/2014MNRAS.445.1482S} {445, 1482}

\bibitem[\protect\citeauthoryear{{Scoville} et~al.,}{{Scoville}
  et~al.}{2007}]{scoville+07}
{Scoville} N.,  et~al., 2007, \mn@doi [\apjs] {10.1086/516585}, \href
  {http://adsabs.harvard.edu/abs/2007ApJS..172....1S} {172, 1}

\bibitem[\protect\citeauthoryear{{Scoville}, {Faisst}, {Capak}, {Kakazu}, {Li}
  \& {Steinhardt}}{{Scoville} et~al.}{2015}]{scoville+15}
{Scoville} N.,  {Faisst} A.,  {Capak} P.,  {Kakazu} Y.,  {Li} G.,
  {Steinhardt} C.,  2015, \mn@doi [\apj] {10.1088/0004-637X/800/2/108}, \href
  {http://adsabs.harvard.edu/abs/2015ApJ...800..108S} {800, 108}

\bibitem[\protect\citeauthoryear{{Speagle}, {Capak}, {Eisenstein}, {Masters}
  \& {Steinhardt}}{{Speagle} et~al.}{2015}]{speagle+15}
{Speagle} J.~S.,  {Capak} P.~L.,  {Eisenstein} D.~J.,  {Masters} D.~C.,
  {Steinhardt} C.~L.,  2015, preprint, \href
  {http://adsabs.harvard.edu/abs/2015arXiv150802484S} {} (\mn@eprint {arXiv}
  {1508.02484})

\bibitem[\protect\citeauthoryear{{Tomczak} et~al.,}{{Tomczak}
  et~al.}{2014}]{tomczak+14}
{Tomczak} A.~R.,  et~al., 2014, \mn@doi [\apj] {10.1088/0004-637X/783/2/85},
  \href {http://adsabs.harvard.edu/abs/2014ApJ...783...85T} {783, 85}

\bibitem[\protect\citeauthoryear{{Wittman}}{{Wittman}}{2009}]{wittman09}
{Wittman} D.,  2009, \mn@doi [\apjl] {10.1088/0004-637X/700/2/L174}, \href
  {http://adsabs.harvard.edu/abs/2009ApJ...700L.174W} {700, L174}

\makeatother
\end{thebibliography}


\bsp	
\label{lastpage}
\end{document}